\documentclass[11pt, oneside]{article}   	

\usepackage{jheppub}

\usepackage{amsmath}
\usepackage{amssymb}
\usepackage{amsfonts}
\usepackage{amsthm}
\usepackage{amsbsy}
\usepackage{array}

\usepackage{mathtools}
\usepackage[usenames,dvipsnames]{xcolor}

\usepackage{dsdshorthand}

\usepackage[vcentermath]{youngtab}
\newcommand{\myng}[1]{\,{\tiny\yng #1}\,}

\usepackage{tikz}
\usetikzlibrary{trees}
\usetikzlibrary{decorations.pathmorphing}
\usetikzlibrary{decorations.markings}
\usetikzlibrary{shapes.misc}

\tikzset{
  threept/.style={
    circle,
    draw,
    inner sep=2pt,
  },
  twopt/.style={
    circle,
    draw,
    fill=black,
    inner sep=1pt,
    minimum size=1pt
  },
  cross/.style={
    cross out,
    draw=black, 
    minimum size=7pt, 
    inner sep=0pt,
    outer sep=0pt
  },
  scalar/.style={
    thick,
    dashed,
    postaction={
      decorate,
      decoration={
        markings,
        mark=at position 0.5 with {\arrow{>}}
      }
    }
  },
  spinning/.style={
    thick,
    postaction={
      decorate,
      decoration={
        markings,
        mark=at position 0.5 with {\arrow{>}}
      }
    }
  },
  spinning no arrow/.style={
    thick,
  },
  finite with arrow/.style={
    decoration={
      snake,
      amplitude=1pt,
      segment length=6pt,
      post length=2pt
    },
    decorate,
    thick,->
  },
  finite/.style={
    decoration={
      snake,
      amplitude=1pt,
      segment length=6pt,
    },
    decorate,
    thick
  }
}

\newcommand{\diagramEnvelope}[1]{#1}

\newcommand{\point}{{\mathbf p}}
\newcommand{\conj}[1]{{\overline #1}}

\renewcommand\hat[1]{\widehat{#1}}
\renewcommand\vol{\mathop{\mathrm{vol}}}

\makeatletter
\def\@fpheader{\ }
\makeatother

\title{Harmonic Analysis and Mean Field Theory}
\author{Denis Karateev$^{+}$, Petr Kravchuk$^{-,0}$, and David Simmons-Duffin$^{-}$}
\affiliation{
${}^+$
Institute of Physics, EPFL, CH-1015 Lausanne, Switzerland\\
${}^-$Walter Burke Institute for Theoretical Physics, Caltech, Pasadena, California 91125, USA \\
${}^{\,0\,\,}$School of Natural Sciences, Institute for Advanced Study, Princeton, New Jersey 08540, USA
}
\date{}
\abstract{
We review some aspects of harmonic analysis for the Euclidean conformal group, including conformally-invariant pairings, the Plancherel measure, and the shadow transform. We introduce two efficient methods for computing these quantities: one based on weight-shifting operators, and another based on Fourier space. As an application, we give a general formula for OPE coefficients in Mean Field Theory (MFT) for arbitrary spinning operators. We apply this formula to several examples, including MFT for fermions and ``seed" operators in 4d, and MFT for currents and stress-tensors in 3d.
}

\preprint{CALT-TH 2018-036}

\begin{document}

\maketitle

\newpage

\section{Introduction}

Mean Field Theory (MFT) provides some of the simplest examples of crossing-symmetric, conformally-invariant correlation functions. Correlators in MFT are simply sums of products of two-point functions. In theories exhibiting large-$N$ factorization, MFT is the leading contribution at large-$N$. For example, in AdS/CFT, MFT is the leading contribution to correlators in bulk perturbation theory \cite{Maldacena:1997re,Gubser:1998bc,Witten:1998qj}. In the analytic conformal bootstrap, MFT is the leading contribution to correlators at large spin \cite{Alday:2007mf,Fitzpatrick:2012yx,Komargodski:2012ek,Alday:2015eya,Alday:2015ewa,Alday:2016njk,Simmons-Duffin:2016wlq}. MFT provides crucial example data for the numerical bootstrap \cite{Rattazzi:2008pe}, especially for spinning operators \cite{Iliesiu:2015qra,Iliesiu:2017nrv,Dymarsky:2017xzb,Dymarsky:2017yzx,Karateev:2017}. Furthermore, MFT OPE coefficients form the ``ladder kernel" in SYK-like models \cite{Kitaev,Maldacena:2016hyu,PR,GR,Murugan:2017eto,Giombi:2017dtl,Liu:2018jhs}. Consequently, the OPE data of MFT (i.e.\ the scaling dimensions and OPE coefficients) is the starting point for many computations.

Although correlators in MFT are simple, the OPE data can be nontrivial. OPE coefficients for a four-point function of fundamental scalars in MFT in 2- and 4- dimensions were guessed in \cite{Heemskerk:2009pn}.\footnote{Here, ``fundamental" means not a composite of other MFT operators. In context of large-$N$ theories, fundamental is equivalent to single-trace.} They were subsequently generalized to $d$-dimensions in \cite{Fitzpatrick:2011dm} using a technique dubbed ``conglomeration." In this work, we point out that conglomeration is part of a general toolkit of harmonic analysis for the Euclidean conformal group $\SO(d+1,1)$ \cite{Dobrev:1977qv}. Although harmonic analysis was first applied to CFTs in the 70's, it has played an especially important role in recent developments \cite{Kitaev,Maldacena:2016hyu,PR,GR,Gadde:2016fbj,Caron-Huot:2017vep,Murugan:2017eto,Giombi:2017dtl,Gromov:2017cja,Gromov:2018hut,Kravchuk:2018htv,Liu:2018jhs}. In section~\ref{sec:harmonic_analysis_review}, we give an introduction to harmonic analysis for (Euclidean) CFTs. 

The calculation of \cite{Fitzpatrick:2011dm} can be rephrased in terms of simple ingredients: the Plancherel measure, three-point pairings, and the ``shadow transform" \cite{Ferrara1972,SimmonsDuffin:2012uy}.
In particular, the computation of MFT OPE coefficients factorizes into two independent shadow transforms of three-point functions, which are essentially generalizations of the famous ``star-triangle" integral \cite{Fradkin:1978pp,Vasiliev:1981dg}. Using these observations, we write a general formula for OPE data of fundamental MFT fields in arbitrary Lorentz representations in section~\ref{sec:mftcoefficients}. Along the way, we derive orthogonality relations for conformal partial waves with arbitrary (internal and external) Lorentz representations.

Our derivation essentially uses a ``Euclidean inversion formula" --- a formula that expresses OPE data as an integral of a four-point function over Euclidean space. MFT OPE data can also in principle be computed by applying the Lorentzian inversion formula of Caron-Huot \cite{Caron-Huot:2017vep} (and its generalization to arbitrary spins \cite{Kravchuk:2018htv}) to the unit operator in the crossed-channel. However, the resulting cross-ratio integral is difficult to perform in general spacetime dimensions, where conformal blocks are not known explicitly. (It is however doable in 2 and 4 dimensions \cite{Liu:2018jhs}.) Our calculation ``gauge-fixes" the conformal symmetry in a different way, resulting in a simpler integral.

In section~\ref{sec:harmonicandweightshifting}, we show how the Plancherel measure and shadow transform can be computed efficiently using weight-shifting operators \cite{Karateev:2017jgd}. Weight-shifting operators are conformally-covariant differential operators that allow one to shift dimensions and spins appearing in conformal correlators. Using them, one can derive recursion relations for essentially any quantity in harmonic analysis of $\SO(d+1,1)$.\footnote{Weight-shifting operators are essentially Clebsch-Gordon coefficients for the tensor product of a finite-dimensional representation and a Verma module. We expect that they can be used in a similar way for harmonic analysis of any group.} In \cite{Karateev:2017jgd}, it was shown how to use weight-shifting operators to derive recursion relations for conformal blocks and $6j$ symbols. In this work, we use weight-shifting operators to give an elementary derivation of the Plancherel measure for general representations in section~\ref{sec:generalplancherel}. In section~\ref{eq:weightshiftingshadowcoeffs}, we show how to compute shadow transforms using weight-shifting operators, and then combine these ingredients in section~\ref{sec:mft} to re-derive MFT OPE coefficients for scalars in general $d$, and derive new formulas for ``seed" correlators and fermion four-point functions in 4d. We focus on these cases to compare to previous results/guesses in the literature, but it is straightforward to study other cases using our techniques. 

In section~\ref{eq:shadowfourier}, we develop another approach to computing shadow transforms based on Fourier space. Because the shadow-transform is a translationally-invariant integral transform, it becomes simply multiplication by a finite-dimensional matrix in Fourier space. In section~\ref{sec:shadowfourieralgorithm}, we describe how to compute the Fourier transform of arbitrary 2- and 3-point functions in 3d and use this to compute shadow transforms in 3d CFTs. We also clarify some subtleties of three-point pairings for conserved operators. As an application, in sections~\ref{sec:opecurrent3d} and \ref{sec:mftopecoeffs} we derive the OPE data for four-point functions of currents and stress-tensors in 3d. We expect that this data will prove useful in future analytic bootstrap studies of these correlators.

The appendices collect conventions and helpful calculations. In particular, appendix~\ref{app:details_4d_formalism} gives a complete summary of our formalism for working with conformal correlators and weight-shifting operators in 4 dimensions.\footnote{A {\it Mathematica} notebook with all the 4d computations is included with this paper. To run it one needs to install the ``CFTs4D'' package~\cite{Cuomo:2017wme}.} Appendix~\ref{app:3dconventions} summarizes our conventions in 3 dimensions.

\section{Review of harmonic analysis for $\SO(d+1,1)$}
\label{sec:harmonic_analysis_review}

In this section, we review some aspects of harmonic analysis on the Euclidean conformal group $\SO(d+1,1)$. This subject was given a detailed and rigorous treatment in the 70's by Dobrev et.\ al. \cite{Dobrev:1977qv}. Here, we highlight some key results and generalize some of the discussion to include arbitrary $\SO(d)$ representations. Our focus is on setting up tools for computations.

\subsection{Shadow representations}
\label{sec:shadow_representations}

A primary operator $\cO$ is labeled by a scaling dimension $\De$ and a finite-dimensional irreducible representation $\rho$ of $\SO(d)$. We denote the $\SO(d+1,1)$ representation of $\cO$ as $V_{\De,\rho}$. A few related representations will be important in this work. The reflected representation $\rho^R$ of $\SO(d)$ is defined by
\be\label{eq:reflecteddefinition}
\rho^R(g) &\equiv \rho(R g R^{-1}),\qquad g\in\SO(d),
\ee
where $R\in \mathrm{O}(d)$ is a reflection in any direction.\footnote{In odd-dimensions, $\rho^R$ is equivalent to $\rho$. In even-dimensions, reflection means swapping the weights associated to the two spinor representations. In cases when $\rho^R\simeq \rho$ it is convenient to use the same realization for both $\rho$ and $\rho^R$.}${}^{,}$\footnote{More generally, given a group $G$, the group of outer automorphisms $\mathrm{Out}(G)$ acts on the ring of representations of $G$. Given $\s\in \mathrm{Out}(G)$ and a representation $\rho$, the action is $\s:\rho\mto \rho'$, where $\rho'(g) = \rho(\s^{-1}(g))$. Conjugation by a reflection is an outer automorphism of $\SO(d)$ when $d$ is even.} The operator $\cO$ has a unique two-point structure  (up to an overall coefficient) with an operator $\cO^\dag$ transforming in $V_{\De,\rho^\dag}$, where $\rho^\dag = (\rho^R)^*$ is the dual of $\rho^R$.\footnote{$\rho^\dag$ is the complex conjugate of $\rho$ in Lorentzian signature.} We denote this two-point structure by
\be
\<\cO(x) \cO^\dag(y)\>=
\diagramEnvelope{\begin{tikzpicture}[anchor=base,baseline]
	\node (opO) at (-1,0) [left] {$\cO$};
	\node (opOprime) at (1,0) [right] {$\cO^\dagger$};
	\node (vert) at (0,0) [twopt] {};
	\draw [spinning] (vert) -- (opO);
	\draw [spinning] (vert) -- (opOprime);
	\end{tikzpicture}},
\ee
where we have suppressed $\SO(d)$-indices for brevity. On the right-hand side we use the same diagrammatic notation for two-point functions as in~\cite{Karateev:2017jgd}.

The ``shadow representation" of $V_{\De,\rho}$ is given by $V_{\tl \De,\rho^R}$, where 
\be
\tl \De &\equiv d-\De.
\ee
We denote an operator transforming in $V_{\tl \De,\rho^R}$ as $\tl \cO$. Note that $\tl \cO^\dag$ transforms in the representation $V_{\tl\De,\rho^*}$ where $\rho^*=((\rho^R)^R)^*$ is the dual of $\rho$. Thus, there exists a natural conformally-invariant pairing\footnote{In principle there is only one canonical way to pair $\rho^*$ and $\rho$. However, if $\rho^*=\rho$, then we can choose which operator to treat as transforming in the dual representation, which may lead to sign ambiguities for fermionic representations. In such cases, care should be taken to assign these signs consistently, e.g.\ by tracking the order in which $\tl\cO^\dag$ and $\cO$ are written. See, e.g.,~\eqref{eq:property_4d}.}
\be
\label{eq:naturalpairing}
(\tl \cO^\dag, \cO) &= \int d^d x\, \tl \cO^\dag(x) \cO(x),
\ee
where we implicitly contract the $\SO(d)$-indices of $\tl\cO^\dag$ and $\cO$, since they are in dual representations $\rho$ and $\rho^*$ of $\SO(d)$. The product $\tl \cO^\dag(x) \cO(x)$ has scaling dimension $\De+\tl \De = d$, which cancels against the scaling dimension of the measure $d^d x$, so that the integral is conformally-invariant.

Instead of writing $V_{\De,\rho}$, etc.\ it will often be convenient to use shorthand notation where $\cO$ stands for the conformal representation with weights $(\De,\rho)$. Similarly $\tl \cO$ will be shorthand for the representation with weights $(d-\De,\rho^R)$, $\cO^\dag$ will be shorthand for the representation with weights $(\De,\rho^\dag)$, and $\tl \cO^\dag$ will be shorthand for the representation with weights $(d-\De,\rho^*)$. In particular, $\cO^\dag$ is not necessarily the hermitian conjugate of a physical operator --- it is simply convenient notation for a particular conformal representation. When we write two- and three-point correlators of operators $\cO_i$, we mean a conformally-invariant structure associated to those representations. Correlation functions in physical theories are linear combinations of the possible structures.

To denote shadow representations diagrammatically, we will use the following convention: an arrow in one direction labeled by $\cO$ is the same as an arrow labeled by $\tl\cO^\dagger$ going in the opposite direction, i.e.
\be
\label{eq:switcharrow}
\diagramEnvelope{\begin{tikzpicture}[anchor=base,baseline]
	\node (opO) at (2,0) [left] {$\cO$};
	\node [circle, fill=black!10, draw=black, inner sep = 7] (vert) at (0,0) {};
	\draw [spinning] (vert) -- (opO);
\end{tikzpicture}}
\quad=\quad
\diagramEnvelope{\begin{tikzpicture}[anchor=base,baseline]
	\node (opO) at (2,0) [left] {$\tl\cO^\dagger$};
	\node [circle, fill=black!10, draw=black, inner sep = 7] (vert) at (0,0) {};
	\draw [spinning] (opO) -- (vert);
\end{tikzpicture}}.
\ee
Here, the shaded circle represents some conformally-invariant structure involving $\cO$.
Using this convention, the conformally-invariant pairing~\eqref{eq:naturalpairing} can be written diagrammatically as
\be\label{eq:diagramcontraction}
\p{\diagramEnvelope{\begin{tikzpicture}[anchor=base,baseline]
	\node (opO) at (2,0) [left] {$\cO$};
	\node [circle, fill=black!10, draw=black, inner sep = 7] (vert) at (0,0) {};
	\draw [spinning] (vert) -- (opO);
	\end{tikzpicture}},
\diagramEnvelope{\begin{tikzpicture}[anchor=base,baseline]
	\node (opO) at (-1.5,0) [left] {$\tl\cO^\dagger$};
	\node [circle, fill=black!10, draw=black, inner sep = 7] (vert) at (0,0) {};
	\draw [spinning] (vert) -- (opO);
	\end{tikzpicture}}}
\quad=\quad
\diagramEnvelope{\begin{tikzpicture}[anchor=base,baseline]
	\node (opO) at (1,0) [below] {$\cO$};
	\node [circle, fill=black!10, draw=black, inner sep = 7] (vert) at (0,0) {};
	\node [circle, fill=black!10, draw=black, inner sep = 7] (vert1) at (2,0) {};
	\draw [spinning] (vert) -- (vert1);
	\end{tikzpicture}}\quad.
\ee

\subsection{Principal series representations and the shadow transform}

Our main interest will be unitary principal-series representations $\cE_{\De,\rho}$, where the scaling dimension takes the form
\be
\De &= \frac d 2 + is,\qquad s\in \R.
\ee
Elements of $\cE_{\De,\rho}$ can be thought of as functions $f^a(x)$ on $\R^d$ satisfying the usual transformation law for primary operators (here, $a=1,\cdots,\dim\rho$ is an index for $\rho$). The significance of principal series representations is that $(f^a(x))^*$ transforms in the representation $\cE_{\tl\De,\rho^*}$, so (\ref{eq:naturalpairing}) gives rise to a positive-definite hermitian inner product on $\cE_{\De,\rho}$,
\be
\label{eq:unitarypairing}
\<g,f\> &= \int d^d x\, (g^a(x))^* f^a(x).
\ee

The representation $\cE_{\De,\rho}$ is equivalent to its shadow representation $\cE_{\tl \De,\rho^R}$ via the shadow transform:\footnote{This transform was called $\bS_E$ (``$E$" for ``Euclidean) in \cite{Kravchuk:2018htv} to distinguish it from other conformally-invariant integral transforms in Lorentzian signature. Those other transforms will not play a role in this work, so we omit the ``$E$".}
\be
\label{eq:shadowtransform}
\bS &: \cE_{\De,\rho} \to \cE_{\tl \De,\rho^R} \nn\\
\bS[\cO](x) &= \int d^d y \<\tl \cO(x) \tl \cO^\dag(y)\>\cO(y).
\ee
Again, in (\ref{eq:shadowtransform}), we are implicitly contracting $\SO(d)$-indices between $\tl \cO^\dag(y)$ and $\cO(y)$.
The shadow transform $\bS$ is an example of a Knapp-Stein intertwining operator \cite{KnappStein1}.
Note that the shadow transform is a convolution, which means it simply becomes multiplication in Fourier space. This fact will play an important role in section~\ref{eq:shadowfourier}.

In diagrammatic language, shadow transform can be expressed as a contraction with the shadow two-point function,
\be\label{eq:shadowdiagram}
	\diagramEnvelope{\begin{tikzpicture}[anchor=base,baseline]
	\node (opO) at (-1.5,0) [left] {$\cO$};
	\node [circle, fill=black!10, draw=black, inner sep = 7] (vert) at (0,0) {};
	\draw [spinning] (vert) -- (opO);
	\end{tikzpicture}}
\quad\xrightarrow[]{\quad\bS\quad}\quad
	\diagramEnvelope{\begin{tikzpicture}[anchor=base,baseline]
		\node (opO) at (0.5,0) [below] {$\cO$};
		\node (opOt) at (-1,0) [left] {$\tl\cO$};
		\node [twopt] (vert) at (0,0) {};
		\node [circle, fill=black!10, draw=black, inner sep = 7] (vert1) at (1.5,0) {};
		\draw [spinning] (vert1) -- (vert);
		\draw [spinning] (vert) -- (opOt);
		\end{tikzpicture}}\quad.
\ee

\subsection{Pairings between structures}
\label{sec:pairings_between_structures}
There is a natural conformally-invariant pairing between $n$-point functions of $\cO_i$ and $n$-point functions of $\tl \cO_i^\dag$, given by multiplying and integrating over all points modulo the conformal group,\footnote{Note that since the product of operators in any $n$-point function is bosonic, there are no potential sign ambiguities in the $n$-point pairing.}
\be
\label{eq:structurepairing}
\p{\<O_1\cdots O_n\>,\<\tl O_1^\dag \cdots \tl O_n^\dag\>} &= \int \frac{d^d x_1\cdots d^d x_n}{\vol\SO(d+1,1)}\<O_1\cdots O_n\>\<\tl O_1^\dag \cdots \tl O_n^\dag\>,\\\label{eq:structurepairingdiagram}
\p{
	\diagramEnvelope{\begin{tikzpicture}[anchor=base,baseline]
	\node (opO1) at (1.5,0.6) [right] {$\cO_1$};
	\node (dots) at (1.6,0.1) [right] {$\vdots$};
	\node (opOn) at (1.5,-0.6) [right] {$\cO_n$};
	\node [circle, fill=black!10, draw=black, inner sep = 7] (vert) at (0,0) {};
	\draw [spinning] (vert) to[out=45,in=180] (opO1);
	\draw [spinning] (vert) to[out=-45,in=180] (opOn);
	\draw [spinning] (vert) -- (1.5,0);
	\end{tikzpicture}}	
	,	
	\diagramEnvelope{\begin{tikzpicture}[anchor=base,baseline]
	\node (opO1) at (-1.5,0.6) [left] {$\tl\cO_1^\dagger$};
	\node (dots) at (-1.7,0.1) [left] {$\vdots$};
	\node (opOn) at (-1.5,-0.6) [left] {$\tl\cO_n^\dagger$};
	\node [circle, fill=black!10, draw=black, inner sep = 7] (vert) at (0,0) {};
	\draw [spinning] (vert) to[out=135,in=0] (opO1);
	\draw [spinning] (vert) to[out=-135,in=0] (opOn);
	\draw [spinning] (vert) -- (-1.5,0);
	\end{tikzpicture}}	
}&=
	\diagramEnvelope{\begin{tikzpicture}[anchor=base,baseline]
	\node (opO1) at (1.5,0.6) [above] {$\cO_1$};
	\node (dots) at (1.3,0.1) [right] {$\vdots$};
	\node (opOn) at (1.5,-0.6) [below] {$\cO_n$};
	\node [circle, fill=black!10, draw=black, inner sep = 7] (vert) at (0,0) {};
	\node [circle, fill=black!10, draw=black, inner sep = 7] (vert1) at (3,0) {};
	\draw [spinning] (vert) to[out=45,in=180] (1.5,0.6);
	\draw [spinning] (vert) to[out=-45,in=180] (1.5,-0.6);
	\draw [spinning] (vert) -- (1.3,0);
	\draw [spinning] (1.5,0.6) to[in=135,out=0] (vert1);
	\draw [spinning] (1.5,-0.6) to[in=-135,out=0] (vert1);
	\draw [spinning] (1.7,0) -- (vert1);
	\end{tikzpicture}}	
\quad.
\ee
Here, we essentially use the pairing (\ref{eq:naturalpairing}) for each pair of operators $\cO_i,\tl \cO_i^\dag$.
Dividing by $\vol\SO(d+1,1)$ is necessary because the integrand is conformally-invariant, so the integral would otherwise be infinite. In diagrammatic language, we agree to implicitly divide by volume of $\vol{\SO(d+1,1)}$ whenever we have a connected subdiagram with completely contracted legs. To implement the quotient, we gauge-fix the conformal group and insert the appropriate Fadeev-Popov determinant.

This pairing (\ref{eq:structurepairing}) is particularly simple for three-point structures. In that case, we can use conformal transformations to set $x_1=0$, $x_2=e$, $x_3=\oo$ (with $e$ a unit vector), and no actual integration is necessary. We have simply
\be
\label{eq:threeptpairinggeneral}
\p{\<\cO_1 \cO_2 \cO_3\>,\<\tl \cO_1^\dag \tl \cO_2^\dag \tl \cO_3^\dag\>} &= \frac{1}{2^d\vol\SO(d-1)} \<\cO_1(0)\cO_2(e)\cO_3(\oo)\>\<\tl \cO_1^\dag(0) \tl \cO_2^\dag(e) \tl \cO_3^\dag(\oo)\>.
\ee
The factor of $2^{-d}$ is the Fadeev-Popov determinant for this gauge fixing.\footnote{In \cite{Simmons-Duffin:2017nub}, a different  definition of $\vol \SO(d+1,1)$ was used which omitted the factor of $2^{-d}$. We include the factor because it simplifies the expression for the Plancherel measure in section~\ref{sec:compactexpression}.} Our convention for an operator insertion at $\oo$ is\footnote{In particular, we do not include a reflection in the $e$ direction. Thus, our definition of an insertion at $\oo$ depends on which direction we move the operator towards $\oo$.}
\be
\cO(\oo) &= \lim_{L\to \oo} L^{2\De} \cO(L e).
\ee
The factor $\vol\SO(d-1)$ is the volume of the stabilizer group of three points. Our normalization convention for the measure on $\SO(d)$ is that
\be\label{eq:relation_volumes_SO}
\vol\SO(d) &= \vol S^{d-1} \vol\SO(d-1)=\frac{2\pi^{d/2}}{\G(d/2)} \vol\SO(d-1), \qquad (d\geq 2)
\ee
and $\vol\SO(1)=1$.

As an example, a scalar-scalar-spin-$J$ correlator has a single conformally-invariant tensor structure (up to normalization), given by\footnote{Recall that $2$ or $3$ point correlators in this paper stand for conformally-invariant structures associated to the given representations.}
\be
\label{eq:scalarscalarspinjstructure}
\<\f_1(x_1)\f_2(x_2)\cO_{3,J}^{\mu_1\cdots \mu_J}(x_3)\> &= \frac{Z^{\mu_1} \cdots Z^{\mu_J} - \textrm{traces}}{x_{12}^{\De_1+\De_2-\De_3} x_{23}^{\De_2 + \De_3 - \De_1} x_{31}^{\De_1 +\De_3-\De_2}},  \nn\\
Z^\mu &\equiv \frac{|x_{13}||x_{23}|}{|x_{12}|} \p{\frac{x_{13}^\mu}{x_{13}^2} - \frac{x_{23}^\mu}{x_{23}^2}}.
\ee
The pairing in this case is
\be
\label{eq:scalarscalarspinjpairing}
\p{\<\f_1\f_2\cO_{3,J}\>,\<\tl \f_1\tl \f_2\tl \cO_{3,J}\>} 
&= \frac{1}{2^d\vol\SO(d-1)}(e^{\mu_1}\cdots e^{\mu_J}-\textrm{traces})(e_{\mu_1}\cdots e_{\mu_J} - \textrm{traces}) \nn\\
&= \frac{1}{2^d\vol\SO(d-1)}\hat C_J(1)\nn\\
&= \frac{1}{2^d\vol(\SO(d-1))}\frac{(d-2)_J}{2^J \p{\frac{d-2}{2}}_J},
\ee
where
\be
\label{eq:chat}
\hat C_J(x) &= \frac{\G(J+1)\G(\frac{d-2}{2})}{2^J \Gamma(J+\frac{d-2}{2})} C_J^{\frac{d-2}{2}}(x)=x^J+\cdots,
\ee
and $C_J^\nu(x)$ is a Gegenbauer polynomial.

In general, spinning operators $\cO_i$ admit multiple conformally-invariant three-point structures $\<\cO_1\cO_2\cO_3\>^{(a)}$, $a=1,\cdots,N$. These structures are classified by \cite{Kravchuk:2016qvl}\footnote{Here, we ignore the possibility of permutation symmetries and conservation conditions. We briefly discuss conservation conditions in section~\ref{sec:conservationconditions}.}
\be
\label{eq:spaceofthreept}
(\rho_1 \otimes \rho_2 \otimes \rho_3)^{\SO(d-1)},
\ee
where $(\rho)^H$ denotes the $H$-invariant subspace of $\rho$.
The pairing (\ref{eq:threeptpairinggeneral}) can be thought of as a pairing between the associated spaces of three-point structures
\be
\label{eq:threeptpairing}
 \p{\<O_1O_2O_3\>^{(a)},\<\tl O_1^\dag \tl O_2^\dag \tl O_3^\dag\>^{(b)}}.
\ee
This pairing is nondegenerate because it can be interpreted as a positive-definite hermitian inner product on the space (\ref{eq:spaceofthreept}), since complex conjugation takes $\rho\to \rho^*$ because $\SO(d)$ is compact. We often omit the structure label $(a)$ when there is a unique structure for the given representations.

The three-point pairing (\ref{eq:spaceofthreept}) can be partially diagonalized as follows. Let 
\be
\mathrm{Res}^{\SO(d)}_{\SO(d-1)}\rho_i = \oplus_j \l_{ij}
\ee
be the restriction of $\rho_i$ to a representation of $\SO(d-1)$, together with its decomposition into irreps $\l_{ij}$ of $\SO(d-1)$. (This decomposition is guaranteed to be multiplicity-free \cite{RepresentationsAndSpecialFunctions}.) The space of three-point structures can be written as
\be
\oplus_{j_1,j_2,j_3} (\l_{1j_1} \otimes \l_{2j_2} \otimes \l_{3 j_3})^{\SO(d-1)}.
\ee
 Because the pairing (\ref{eq:spaceofthreept}) is $\SO(d-1)$-invariant, it is nonzero only if $\l_{ij}$ is paired with $\l_{ij}^*$ for each $i=1,2,3$. For example, in three-dimensions, three-point structures are labeled by a collection of $\SO(2)$ weights $[q_1,q_2,q_3]$, such that $q_1+q_2+q_3=0$ \cite{Kravchuk:2016qvl}. The only nonvanishing pairings are $([q_1,q_2,q_3],[-q_1,-q_2,-q_3])\neq 0$.

The shadow transform gives a linear map between spaces of three-point structures,
\be
\label{eq:shadowcoeffdef}
\<\bS[\cO_1]\cO_2 \cO_3\>^{(a)} &= S([\cO_1]\cO_2\cO_3)^a{}_b\<\tl \cO_1 \cO_2 \cO_3\>^{(b)},
\ee
(A sum over three-point structures $b$ is implicit.)
We call the coefficients $S([\cO_1]\cO_2\cO_3)^a{}_b$ ``shadow coefficients." The brackets indicate which operator has been shadow-transformed. We discuss how to compute shadow coefficients in sections~\ref{eq:weightshiftingshadowcoeffs} and \ref{eq:shadowfourier}.

\subsection{The Plancherel measure}

The Plancherel measure for a group $G$ is a measure on the space $\hat G$ of unitary irreducible representations of $G$. For compact groups, the Plancherel measure is discrete, with value $\frac{\dim \pi}{\vol G}$ for each unitary irreducible representation $\pi$. (Here, $\vol G$ is defined using the Haar measure on $G$.) We can also write this as
\be
\label{eq:compactplancherel}
\mu(\pi) &= \frac{\dim\pi}{\vol G} = \frac{\Tr_\pi(\mathbf{1})}{\vol G} \qquad (\textrm{compact $G$}).
\ee

For noncompact groups, both $\dim \pi$ and $\vol G$ can be infinite, and it is useful to think of the Plancherel measure as a kind of regularized quotient. For principal series representations of $\SO(d+1,1)$, we have
\be
\label{eq:formalplanchereldef}
\frac{\mu(\De,\rho)}{\vol \SO(1,1)}=\frac{\Tr_{\cE_{\De,\rho}}(\mathbf{1})}{\vol \SO(d+1,1)} = 
	\,\,\diagramEnvelope{\begin{tikzpicture}[anchor=base,baseline]
		\node at (0,0.4) [above] {$\cO$};
		\draw[spinning] (0,0) circle (13pt);
	\end{tikzpicture}}\,,
\ee
where $\mu(\De,\rho)$ is a finite quantity that we henceforth refer to as the Plancherel measure.
Both the left and right-hand sides above are formal expressions and we will see how they are useful shortly. One way to understand the infinite factor $\vol \SO(1,1)$ on the left-hand side is to imagine computing the trace via an integral. The identity transformation  $\mathbf{1}$ is given by the kernel
\be
\mathbf{1}_{xx'} &= \de_p^m\de(x-x'),
\ee
where $m,p$ are Lorentz indices for $\rho,\rho^*$. Then we have
\be
\frac{\Tr(\mathbf{1})}{\vol \SO(d+1,1)} = \frac{\int d^d x \de^d(x-x) \de^m_m}{\vol \SO(d+1,1)}
\sim \frac{(\vol \R^d)^2 \dim \rho}{\vol \SO(d+1,1)}.
\ee
The group $\SO(d+1,1)$ has three types of noncompact generators: translations, special conformal generators, and dilatations. The translations and special conformal generators give $(\vol\R^d)^2$, which cancels the infinite factors in the numerator. The remaining infinite factor in the denominator is the volume of dilatations, which is $\vol \SO(1,1)$. This argument makes it hard to see what the remaining finite factors are, but we will show a way to determine them in the next subsection.

The Plancherel measure $\mu(\De,\rho)$ is known in great generality \cite{Dobrev:1977qv}. We will rederive it in section~\ref{sec:generalplancherel}. 

\subsection{Relation to the shadow transform}

Note that the square of the shadow transform $\mathbf{S}^2$ takes the representation $\cE_{\De,\rho}$ to itself. By irreducibility, it must be proportional to the identity,
\be
\label{eq:squareofshadow}
\mathbf{S}^2 &= \cN(\De,\rho) \mathbf{1},
\ee
where $\cN(\De,\rho)$ is some constant that depends on the representation $\mathbf{S}^2$ acts on. $\cN(\De,\rho)$ also depends implicitly on a choice of normalization of the two-point structure used to define $\bS$.

The coefficient $\cN(\De,\rho)$ is closely related to the Plancherel measure \cite{KnappStein1}. To see how, let us take the trace on both sides of (\ref{eq:squareofshadow}) and divide by the volume of the conformal group. Writing out the shadow transforms explicitly, we have 
\be
\cN(\De,\rho) \frac{\mu(\De,\rho)}{\vol\SO(1,1)} &= \int \frac{d^d x d^d y}{\vol \SO(d+1,1)} \<\cO^a(x) \cO^{\dag b}(y)\>\<\tl \cO_b(y)\tl\cO^\dag_a(x)\> \nn\\
&= \frac{1}{2^d \vol \SO(d) \vol \SO(1,1)} \<\cO^a(0) \cO^{\dag b}(\oo)\>\<\tl \cO_b(\oo)\tl\cO^\dag_a(0)\>.
\label{eq:plancherel_and_shadow}
\ee
Here we have gauge-fixed $x=0$ and $y=\oo$ in the integral. The factors in the denominator are the volume of the stabilizer group of two points, $\SO(1,1)\x\SO(d)$, together with a Fadeev-Popov determinant. Thus, we conclude
\be
\mu(\De,\rho) &= \frac{1}{\cN(\De,\rho)} \frac{\<\cO^a(0) \cO^{\dag b}(\oo)\>\<\tl \cO_b(\oo)\tl\cO^\dag_a(0)\>}{2^d\vol \SO(d)}.
\label{eq:relationbetweenshadowsqandplancherel}
\ee
Note that while $\cN(\De,\rho)$ depends on a choice of two-point structures, $\mu(\De,\rho)$ is independent of this choice because any change in convention cancels between $\<\cdots\>\<\cdots\>$ in the numerator and $\cN(\De,\rho)$ in the denominator.

Diagrammatically, we can understand the calculation above as
\be
	\p{\diagramEnvelope{\begin{tikzpicture}[anchor=base,baseline]
			\node (opO) at (-0.5,0) [left] {$\cO$};
			\node (opOprime) at (0.5,0) [right] {$\cO^\dagger$};
			\node (vert) at (0,0) [twopt] {};
			\draw [spinning] (vert) -- (opO);
			\draw [spinning] (vert) -- (opOprime);
			\end{tikzpicture}}
	,
		\diagramEnvelope{\begin{tikzpicture}[anchor=base,baseline]
			\node (opO) at (-0.5,0) [left] {$\tl\cO$};
			\node (opOprime) at (0.5,0) [right] {$\tl\cO^\dagger$};
			\node (vert) at (0,0) [twopt] {};
			\draw [spinning] (vert) -- (opO);
			\draw [spinning] (vert) -- (opOprime);
			\end{tikzpicture}}
	}=
	\diagramEnvelope{\begin{tikzpicture}[anchor=base,baseline]
		\node (opO) at (0.5,0.4) [above] {$\cO$};
		\node (opOprime) at (0.5,-0.4) [below] {$\tl\cO$};
		\node (vert) at (0,0) [twopt] {};
		\node (vert1) at (1,0) [twopt] {};
		\draw [spinning] (vert) to[out=90,in=90] (vert1);
		\draw [spinning] (vert1) to[out=-90,in=-90] (vert);
		\end{tikzpicture}}
	=
	\diagramEnvelope{\begin{tikzpicture}[anchor=base,baseline]
		\node (opO) at (0.5,0.4) [above] {$\cO$};
		\node (opOprime) at (0.5,-0.4) [below] {$\tl\cO$};
		\node (vert) at (0.3,-0.3) [twopt] {};
		\node (vert1) at (0.7,-0.3) [twopt] {};
		\draw [spinning no arrow,->] (vert) to[out=135,in=180] (0.5,0.4);
		\draw [spinning no arrow] (0.5,0.4) to[out=0, in=45] (vert1);
		\draw [spinning] (vert1) to[out=-135, in=-45] (vert);
		\end{tikzpicture}}
	=\cN(\De,\rho)\,\,\diagramEnvelope{\begin{tikzpicture}[anchor=base,baseline]
		\node at (0,0.4) [above] {$\cO$};
		\draw[spinning] (0,0) circle (13pt);
		\end{tikzpicture}}\,\,.
\ee

\subsection{Bubble diagrams}

Suppose $\cO,\cO'$ are both principal series representations, with dimensions $\De=\frac d 2+is$, $\De'=\frac d 2 + is'$ with $s,s'>0$.
A ``bubble" integral of two three-point structures is given by
\be
\label{eq:bubbleintegral}
\int d^dx_1 d^dx_2 \< \cO_1 \cO_2 \cO(x) \> \<\tl \cO^\dag_1 \tl \cO^\dag_2 \tl \cO^{'\dag}(x') \> 
&=
B\, \mathbf{1}_{xx'} \de_{\cO\cO'}, \nn\\
\de_{\cO\cO'} &\equiv 2\pi \de(s-s') \de_{\rho \rho'}.
\ee
For brevity, we use the convention that operators with subscript $i$ are at position $x_i$ (unless otherwise specified).
By irreducibility and inequivalence of principal series representations, the integral on the left-hand side can only be nonzero if the representations $\cO$ and $\cO'$ are the same, in which case it must be proportional to the identity transformation. This explains the factors $\mathbf{1}_{xx'} \de_{\cO\cO'}$.\footnote{The above formula is sometimes written with two terms on the right-hand side, one proportional to $\mathbf{1}\,\de(s-s')$, and another proportional to $\bS\, \de(s+s')$. Here, we have only a single term because we have chosen to restrict $s,s'>0$, which is a fundamental domain under the affine Weyl transformation $\De\to d-\De$. The cases where $s,s'$ are not both positive can be obtained by composing with shadow transforms.} In terms of diagrams we have
\be
	\diagramEnvelope{\begin{tikzpicture}[anchor=base,baseline]
		\node[threept] (Os) at (0,0) {};
		\node[threept] (Ots) at (1,0) {};
		\node (Oout) at (-1,0) [left] {$\cO$};
		\node (Oin) at (+2,0) [right] {$\cO'$};
		\node at (0.5,0.3) [above] {$\cO_1$};
		\node at (0.5,-0.3) [below] {$\cO_2$};
		\draw [spinning] (Oin) -- (Ots);
		\draw [spinning] (Os)  -- (Oout);
		\draw [spinning] (Os) to[in=135, out=45] (Ots);
		\draw [spinning] (Os) to[in=-135, out=-45] (Ots);
	\end{tikzpicture}}
	=
	B\delta_{\cO\cO'}
	\diagramEnvelope{\begin{tikzpicture}[anchor=base,baseline]
		\node (Oout) at (0,0) [left] {$\cO$};
		\node (Oin) at (1,0) [right] {$\cO'$};
		\draw [spinning] (Oin) -- (Oout);
	\end{tikzpicture}}\,\,.
\ee

To figure out the constant $B$, we can set $\cO=\cO'$ and take the trace of both sides, dividing by the volume of the conformal group. This gives
\be
\p{\< \cO_1 \cO_2 \cO \>,\<\tl \cO^\dag_1 \tl \cO^\dag_2 \tl \cO^{\dag} \>}
&=
B\, 2\pi \de(0) \frac{\mu(\De,\rho)}{\vol \SO(1,1)}.
\ee
The infinite factors $\de(0)$ and $\vol \SO(1,1)$ will cancel, leaving a finite coefficient. To determine the correct coefficient, we can compute an example bubble integral, for example in the case where all operators are scalars. The result is $\vol \SO(1,1)=2\pi \de(0)$, so we find
\be
B &= \frac{1}{\mu(\De,\rho)} \p{\< \cO_1 \cO_2 \cO \>,\<\tl \cO^\dag_1 \tl \cO^\dag_2 \tl \cO^{\dag} \>}.
\ee

\subsection{Conformal partial waves}
\label{sec:conformal_partial_waves_and_MFT}

A conformal partial wave is a conformally-invariant integral of two three-point structures,\footnote{
The term ``conformal partial wave" (CPW) has disparate meanings in the literature.
Several recent works on CFT from the past 20 years \cite{Dolan:2000ut,Dolan:2003hv,Dolan:2004iy,SimmonsDuffin:2012uy} use the term CPW for what we call a conformal block --- namely the contribution of an individual conformal multiplet to a four-point function in radial quantization. In those works, ``conformal block" refers to the dimensionless function of $z,\bar z$ obtained by multiplying by standard dimensionful factors. We do not find it helpful to use a separate term for the dimensionless function of cross-ratios, especially because there is no canonical convention for this function in the case of spinning operators. Thus, we follow the terminology of \cite{Dobrev:1977qv} and reserve CPW for the integral of two three-point functions and ``conformal block" for the contribution of a conformal multiplet to a four-point function.
}
\be
\label{eq:partialwavedefinition}
\Psi^{\cO_i(ab)}_\cO(x_i) &= \int d^d x \<\cO_1 \cO_2 \cO(x)\>^{(a)} \< \cO_3 \cO_4 \tl \cO^\dag(x)\>^{(b)}\nn\\
&=\diagramEnvelope{\begin{tikzpicture}[anchor=base,baseline]
	\node (opO1) at (-2,1) [left] {$\cO_1$};
	\node (opO2) at (-2,-1) [left] {$\cO_2$};
	\node (opO3) at (2,1) [right] {$\cO_3$};
	\node (opO4) at (2,-1) [right] {$\cO_4$};
	\node (opO)  at (0,0) [below] {$\cO$};
	\node [threept] (left3pt) at (-1,0) {$a$};
	\node [threept] (right3pt) at (1,-0.05) {$b$};
	\draw [spinning] (left3pt) -- (opO1);
	\draw [spinning] (left3pt) -- (opO2);
	\draw [spinning] (left3pt) -- (right3pt);
	\draw [spinning] (right3pt) -- (opO3);
	\draw [spinning] (right3pt) -- (opO4);
	\end{tikzpicture}}.
\ee
Again, we use the convention that $\cO_i$ is at position $x_i$, so that $\Psi^{\cO_i(ab)}_\cO$ is a function of the four positions $x_1,\dots,x_4$.
Here, $a,b$ label the possible conformally-invariant structures for their respective three-point functions. As usual, we are implicitly contracting the Lorentz indices of $\cO$ and $\tl\cO^\dag$. The integral (\ref{eq:partialwavedefinition}) converges when all operators $\cO_1,\dots \cO_4,\cO$ transform in principal series representations. We will assume this is the case for most of this work, and obtain results for non-principal series representations by analytic continuation in $\De_i,\De$.

By construction, $\Psi^{\cO_i(ab)}_\cO$ transforms like a conformally-invariant four-point function of $\cO_1,\dots,\cO_4$ and is an eigenvector of the conformal Casimir equations~\cite{DO2}. Furthermore, it is clearly single-valued in Euclidean signature because it is a convergent integral of single-valued functions. Principal series conformal partial waves furnish an ``almost complete" set of single-valued solutions to the Casimir equations. That is, a general four-point function of $\cO_i$'s can be expanded in partial waves as
\be
\label{eq:partialwaveexpansion}
\<\cO_1 \cdots \cO_4\> &= \sum_{\rho} \int_{\frac d 2}^{\frac d 2 + i\oo} \frac{d\De}{2\pi i} I_{ab}(\De,\rho) \Psi^{\cO_i(ab)}_\cO(x_i) + \textrm{discrete}.
\ee
(Repeated $a,b$ indices are summed over.)
Here, ``discrete" represents possible additional isolated contributions. Such contributions are absent when the $\cO_i$ are all scalars on the principal series \cite{Dobrev:1977qv}. When the $\cO_i$ are allowed to be non-principal series representations or when $\<\cO_1\cdots \cO_4 \>$ is non-normalizable in the sense that we will define below, we may need to include isolated terms, for example the unit operator \cite{Caron-Huot:2017vep,Simmons-Duffin:2017nub}. We will discuss these subtleties below in some examples.

Conformal partial waves are orthogonal with respect to the pairing between four-point structures defined in section~\ref{sec:pairings_between_structures}. An inner product of partial waves can be computed on very general grounds without ever computing the partial waves themselves. The argument is as follows. Consider an inner product of partial waves and insert their shadow representations (\ref{eq:partialwavedefinition}),
\be
\p{\Psi^{\cO_i(ab)}_{\cO},\Psi^{\tl \cO_i^\dag(cd)}_{\tl \cO'^\dag}} &= \int\frac{d^dx_1 \cdots d^dx_6}{\vol \SO(d+1,1)}
\<\cO_1\cO_2\cO(x_5)\>^{(a)}\<\cO_3\cO_4\tl \cO^\dag(x_5)\>^{(b)} \nn\\
&\qquad\qquad\qquad\qquad\qquad\x\<\tl \cO_1^\dag\tl \cO_2^\dag\tl \cO'^\dag(x_6)\>^{(c)}\<\tl \cO^\dag_3 \tl \cO^\dag_4 \cO'(x_6)\>^{(d)}.
\ee
Using our bubble integral formula for the $x_3,x_4$ integrals, the result is a three-point pairing
\be
&= 
\de_{\cO\cO'} \frac{\p{\<\cO_3 \cO_4\tl \cO^\dag \>^{(b)},\<\tl \cO_3^\dag \tl \cO_4^\dag \cO\>^{(d)}}}{\mu(\Delta,\rho)} \int\frac{d^dx_1 d^dx_2 d^dx_5 d^d x_6}{\vol\SO(d+1,1)}
\<\cO_1\cO_2\cO(x_5)\>^{(a)} \mathbf{1}_{56} \<\tl \cO_1\tl \cO_2\tl \cO'^\dag(x_6)\>^{(c)} \nn\\
&= \frac{\p{\<\cO_1\cO_2 \cO\>^{(a)},\<\tl \cO_1^\dag \tl \cO_2^\dag \tl \cO^\dag \>^{(c)}}\p{\<\cO_3 \cO_4 \tl \cO^\dag \>^{(b)},\<\tl \cO_3^\dag \tl \cO_4^\dag \cO\>^{(d)}}}{\mu(\De,\rho)} 2\pi \de(s-s') \de_{\rho\rho'}.
\label{eq:innerproductcpw}
\ee
It is also possible to derive this expression for the inner product between partial waves by gauge-fixing the integral to cross-ratio space and examining the OPE limit, as we show in appendix~\ref{eq:explicitinnerproduct}. The approach here is much simpler.

It follows from (\ref{eq:innerproductcpw}) that $I_{ab}(\De,\rho)$ is determined by the ``Euclidean inversion formula"
\be
\label{eq:euclideaninversion}
\frac{I_{ab}(\De,\rho)}{\mu(\De,\rho)} \p{\<\cO_1\cO_2 \cO\>^{(a)},\<\tl \cO_1^\dag \tl \cO_2^\dag \tl \cO^\dag \>^{(c)}}\p{\<\cO_3 \cO_4 \tl \cO^\dag \>^{(b)},\<\tl \cO_3^\dag \tl \cO_4^\dag \cO\>^{(d)}} &= \p{\<\cO_1\cdots\cO_4\>,\Psi^{\tl \cO_i^\dag(cd)}_{\tl \cO^\dag}}.
\ee
One can gauge fix the integral (\ref{eq:euclideaninversion}) and express it as an integral over cross-ratios, as described in \cite{Simmons-Duffin:2017nub} and appendix~\ref{eq:explicitinnerproduct}. However this will not be useful for our purposes.

Note that the right-hand side of the Euclidean inversion formula is also described by the following diagram
\be\label{eq:euclideaninversionrhsdiagram}
	\diagramEnvelope{\begin{tikzpicture}[anchor=base,baseline]
	\node (opO1) at (-1,0.4) [above] {$\cO_1$};
	\node (opO2) at (-1,-0.3) [below] {$\cO_2$};
	\node (opO3) at (1,0.4) [above] {$\cO_3$};
	\node (opO4) at (1,-0.3) [below] {$\cO_4$};
	\node (opO)  at (0,-1) [below] {$\cO$};
	\node [threept] (left3pt) at (-1.5,0) {$c$};
	\node [threept] (right3pt) at (1.5,0) {$d$};
	\node [circle, fill=black!10, draw=black, inner sep = 7] (4pt) at (0,0) {};
	\draw [spinning] (4pt) to[out=135,in=45] (left3pt);
	\draw [spinning] (4pt) to[out=-135,in=-45] (left3pt);
	\draw [spinning] (4pt) to[out=45,in=135] (right3pt);
	\draw [spinning] (4pt) to[out=-45,in=-135] (right3pt);
	\draw [spinning] (right3pt) to[out=0,in=0] (+0.7,-1) to[out=180,in=180] (-0.7,-1) to[out=180,in=180] (left3pt);
	\end{tikzpicture}}.
\ee

\subsubsection{Relation to conformal blocks}

Conformal partial waves are a linear combination of a conformal block for the exchange of $\cO$ and a block for the exchange of $\tl\cO$. The coefficients can be determined as follows. Consider the integral
\be
\Psi^{(ab)}_\cO &= \int d^dx \<\cO_1 \cO_2 \cO(x)\>^{(a)}\<\cO_3 \cO_4 \tl \cO^\dag(x) \>^{(b)}.
\ee
(For brevity, we will henceforth leave the dependence of $\Psi$ on the external operators $\cO_i$ implicit.)
Since this is a solution to the Casimir equation, it is uniquely determined by its singularities in the OPE limit $x_1\to x_2$ (equivalently $x_3\to x_4$). It suffices to estimate the integral in this limit.

We cannot simply take the limit $x_1\to x_2$ inside the integrand because $x$ probes the neighborhood near $x_1,x_2$, where the OPE is invalid. However, this is a valid procedure for determining the contribution to the integral from outside this neighborhood. Thus, let us make the replacement
\be
\<\cO_1 \cO_2 \cO(x)\>^{(a)} &\to C^{(a)}_{12\cO} \<\cO^\dag(x_2) \cO(x)\>,
\ee
where $C^{(a)}_{12\cO}$ encodes the leading term in the OPE $\cO_1 \x \cO_2$. (For example, when all the operators are scalars, we have $C_{12\cO} = x_{12}^{\De_\cO-\De_1-\De_2}$. More generally, $C_{12\cO}$ can have indices which are contracted with those of $\cO^\dagger$.)
The integral over $x$ then becomes a shadow transform of $\tl \cO^\dag$,
\be
\Psi^{(ab)}_\cO &\sim C^{(a)}_{12\cO} \<\cO_3 \cO_4 \bS[\tl \cO^\dag] \>^{(b)} = 
S(\cO_3\cO_4[\tl \cO^\dag])^{b}{}_{c}
 C^{(a)}_{12\cO} \<\cO_3 \cO_4 \cO^\dag \>^{(c)},
\label{eq:onelimitofpsi}
\ee
where we are using the notation for shadow coefficients introduced in (\ref{eq:shadowcoeffdef}).

We define a conformal block $G^{(ab)}_{\cO}$ as the solution to the conformal Casimir equations that behaves in the OPE limit as
\be
G^{(ab)}_{\cO} &\sim C_{12\cO}^{(a)}\<\cO_3 \cO_4 \cO^\dag \>^{(b)} \qquad (x_1 \to x_2).
\ee
(Like partial waves, conformal blocks also depend on the external operators, but here we leave this dependence implicit.) Note that in our conventions, $\cO^\dag$ appears in the $\cO_1\x\cO_2$ OPE and $\cO$ appears in the $\cO_3\x\cO_4$ OPE. In the notation of \cite{Kravchuk:2018htv}, the block is given by
\be
G^{(ab)}_\cO &= \frac{\<\cO_1 \cO_2 \cO\>^{(a)}\<\cO_3 \cO_4 \cO^\dag\>^{(b)}}{\<\cO^\dag \cO\>},
\ee
where the three-point structures in the numerator should be merged by summing over descendants. Concretely, the leading term in the OPE limit is given in (\ref{eq:blockopelimit}).
Thus, (\ref{eq:onelimitofpsi}) shows that $\Psi_\cO^{(ab)}$ contains a term
\be
\Psi^{(ab)}_\cO &\supset S(\cO_3\cO_4[\tl \cO^\dag])^{b}{}_{c} G_\cO^{(ac)}.
\ee
To compute the term in $\Psi^{(ab)}_\cO$ proportional to $G_{\tl \cO}$, we repeat the above argument using the $\cO_3 \x \cO_4$ OPE. This gives
\be
\Psi^{(ab)}_\cO & \sim C^{(b)}_{34\tl \cO^\dag} \<\cO_1 \cO_2 \bS[\cO]\>^a \sim S(\cO_1\cO_2[\cO])^a{}_c G^{(cb)}_{\tl \cO}.
\ee
Putting everything together, we find\footnote{In case of fermionic $\cO$, and depending on ordering conventions, there may be an additional $(-1)$ factor in the second term, which is inconsequential for our purposes. For simplicity of discussion, we ignore this subtlety.}
\be
\label{eq:expressionforpartialwave}
\Psi^{(ab)}_\cO &= S(\cO_3\cO_4[\tl \cO^\dag])^{b}{}_{c} G^{(ac)}_{\cO} + S(\cO_1\cO_2[\cO])^a{}_c G^{(cb)}_{\tl \cO}.
\ee

We can now relate the conformal partial wave decomposition to the usual conformal block decomposition. We plug (\ref{eq:expressionforpartialwave}) into (\ref{eq:partialwaveexpansion}) to obtain
\be
\<\cO_1\cdots\cO_4\> &=
\sum_\rho \int_{\frac d 2}^{\frac d 2 + i\oo} \frac{d\De}{2\pi i} I_{ab}(\De,\rho) \p{S(\cO_3\cO_4[\tl \cO^\dag])^{b}{}_{c} G^{(ac)}_{\cO} + S(\cO_1\cO_2[\cO])^a{}_c G^{(cb)}_{\tl \cO}}.
\ee
Using the inversion formula (\ref{eq:euclideaninversion}), one can show that the second term above gives the same contribution as we get by extending the range of integration to the whole imaginary axis. Thus, we have
\be\label{eq:contour_integral}
\<\cO_1\cdots\cO_4\> &= \sum_\rho \int_{\frac d 2-i\oo}^{\frac d 2 + i\oo} \frac{d\De}{2\pi i} C_{ac}(\De,\rho) G^{(ac)}_{\cO}(x_i),\nn\\
C_{ac}(\De,\rho) &\equiv I_{ab}(\De,\rho)S(\cO_3\cO_4[\tl \cO^\dag])^{b}{}_{c}.
\ee
The block $G^{(ac)}_{\cO}$ now decays exponentially in the right half-$\De$ plane, so we can deform the contour in that direction. If $C_{ac}(\De,\rho)$ is meromorphic in this half-plane, with only simple poles on the positive real axis, then we obtain a discrete sum of conformal blocks
\be\label{eq:physical spectrum_from_contour_integral}
\<\cO_1\cdots\cO_4\> &= \sum_{\De_*,\rho_*} P_{ac}(\De_*,\rho_*)  G^{(ac)}_{\cO_*}(x_i),\nn\\
P_{ac}(\De_*,\rho_*) &= -\mathrm{Res}_{\De\to \De_*} C_{ac}(\De,\rho_*).
\ee
In deforming the contour from the principal series, one encounters spurious poles that cancel in various ways \cite{Caron-Huot:2017vep,Murugan:2017eto,Simmons-Duffin:2017nub}. We will see some explicit examples below.

\subsection{Mean Field Theory}
\label{sec:mftcoefficients}

Having developed this technology, it is almost trivial to decompose a Mean Field Theory (MFT)\footnote{AKA Generalized Free Field Theory.} four-point function into conformal partial waves. The four-point function and its partial wave decomposition are given by\footnote{When the operators $\cO_1,\cO_2$ are identical, there is an additional contribution to the four-point function with $3\leftrightarrow 4$. This is straightforward to deal with in the same way. If instead $\cO_1=\cO_2^\dagger$ then there is a contribution with $2\leftrightarrow 3$, which simply gives the contribution of the identity operator in $\cO_1\cO_2$ OPE. If $\cO_1=\cO_2=\cO_2^\dagger$, then both contribution are present.}${}^{,}$\footnote{For fermions there can be extra $(-)$ signs.}
\be
\label{eq:basicgff}
\<\cO_1(x_1) \cO_2(x_2) \cO_1^\dag(x_3) \cO_2^\dag(x_4)\>&=\<\cO_1(x_1) \cO_1^\dag(x_3)\> \<\cO_2(x_2) \cO_2^\dag(x_4)\>\nn\\ &= \sum_{\rho} \int_{\frac d 2}^{\frac d 2 + i\oo} \frac{d\De}{2\pi i} I^\mathrm{MFT}_{ab}(\De,\rho) \Psi^{(ab)}_{\De,\rho}(x_i).
\ee
Applying the Euclidean inversion formula~\eqref{eq:euclideaninversion}, we have
\be
&\frac{I^\mathrm{MFT}_{ab}(\De,\rho)}{\mu(\De,\rho)} \p{\<\cO_1\cO_2 \cO\>^{(a)},\<\tl \cO_1^\dag \tl \cO_2^\dag \tl \cO^\dag \>^{(c)}}\p{\<\cO_1^\dagger \cO_2^\dagger \tl \cO^\dag \>^{(b)},\<\tl \cO_1 \tl \cO_2 \cO\>^{(d)}}\nn\\
 &= \p{\<\cO_1(x_1) \cO_1^\dag(x_3)\> \<\cO_2(x_2) \cO_2^\dag(x_4)\>,\Psi_{\tl \cO^\dag}^{(cd)}} \nn\\
 &=
 \int \frac{d^dx_1\cdots d^dx_5}{\vol \SO(d+1,1)} 
 \<\cO_1(x_1) \cO_1^\dag(x_3)\> \<\cO_2(x_2) \cO_2^\dag(x_4)\> \nn\\
 &\qquad \qquad \x\<\tl \cO_1^\dag(x_1)\tl \cO_2^\dag(x_2) \tl \cO^\dag(x_5)\>^{(c)}\<\tl \cO_1(x_3)\tl \cO_2(x_4) \cO(x_5)\>^{(d)} \nn\\
 &= 
 \int \frac{d^dx_1 d^dx_2 d^dx_5}{\vol\SO(d+1,1)}
 \<\tl \cO_1^\dag(x_1)\tl \cO_2^\dag(x_2) \tl \cO^\dag(x_5)\>^{(c)}\<\bS[\tl \cO_1](x_1)\bS[\tl \cO_2](x_2) \cO(x_5)\>^{(d)} \nn\\
 &= \p{
 \<\tl \cO_1^\dag\tl \cO_2^\dag \tl \cO^\dag\>^{(c)},
 \<\bS[\tl \cO_1]\bS[\tl \cO_2] \cO\>^{(d)}
 }
.
\label{eq:derivation_mftcoeffs}
\ee
Above, we recognize the integrals over $x_3,x_4$ as shadow transforms and the remaining integral as a three-point pairing. (We can act with the shadow transform on either $\tl \cO_1^\dag$ or $\tl \cO_1$, and similarly for $\cO_2$. Above, we made an arbitrary choice.) Diagrammatically we have, starting from~\eqref{eq:euclideaninversionrhsdiagram},
\be
	\diagramEnvelope{\begin{tikzpicture}[anchor=base,baseline]
	\node (opO1) at (-1.5,0.4) [above] {$\cO_1$};
	\node (opO2) at (-1.5,-0.3) [below] {$\cO_2$};
	\node (opO3) at (1.5,0.4) [above] {$\cO_3$};
	\node (opO4) at (1.5,-0.3) [below] {$\cO_4$};
	\node [circle, fill=black!10, draw=black, inner sep = 7] (4pt) at (0,0) {};
	\draw [spinning] (4pt) -- (opO1);
	\draw [spinning] (4pt) -- (opO2);
	\draw [spinning] (4pt) -- (opO3);
	\draw [spinning] (4pt) -- (opO4);
	\end{tikzpicture}}&=
	\diagramEnvelope{\begin{tikzpicture}[anchor=base,baseline]
		\node (opO1) at (-1.5,0.4) [above] {$\cO_1$};
		\node (opO2) at (-1.5,-0.3) [below] {$\cO_2$};
		\node (opO3) at (1.5,0.4) [above] {$\cO_1^\dagger$};
		\node (opO4) at (1.5,-0.3) [below] {$\cO_2^\dagger$};
		\node[twopt] (up2pt) at (0,0.6) {};
		\node[twopt] (dn2pt) at (0,-0.5) {};
		\draw[spinning] (up2pt) -- (opO1);
		\draw[spinning] (up2pt) -- (opO3);
		\draw[spinning] (dn2pt) -- (opO2);
		\draw[spinning] (dn2pt) -- (opO4);
	\end{tikzpicture}},\nn\\
	\diagramEnvelope{\begin{tikzpicture}[anchor=base,baseline]
	\node (opO1) at (-1,0.4) [above] {$\cO_1$};
	\node (opO2) at (-1,-0.3) [below] {$\cO_2$};
	\node (opO3) at (1,0.4) [above] {$\cO_1^\dagger$};
	\node (opO4) at (1,-0.3) [below] {$\cO_2^\dagger$};
	\node (opO)  at (0,-1) [below] {$\cO$};
	\node [threept] (left3pt) at (-1.5,0) {$c$};
	\node [threept] (right3pt) at (1.5,0) {$d$};
	\node [circle, fill=black!10, draw=black, inner sep = 7] (4pt) at (0,0) {};
	\draw [spinning] (4pt) to[out=135,in=45] (left3pt);
	\draw [spinning] (4pt) to[out=-135,in=-45] (left3pt);
	\draw [spinning] (4pt) to[out=45,in=135] (right3pt);
	\draw [spinning] (4pt) to[out=-45,in=-135] (right3pt);
	\draw [spinning] (right3pt) to[out=0,in=0] (+0.7,-1) to[out=180,in=180] (-0.7,-1) to[out=180,in=180] (left3pt);
	\end{tikzpicture}}&=
	\diagramEnvelope{\begin{tikzpicture}[anchor=base,baseline]
	\node (opO1) at (-1,0.4) [above] {$\cO_1$};
	\node (opO2) at (-1,-0.3) [below] {$\cO_2$};
	\node (opO3) at (0.8,0.4) [above] {$\cO_1^\dagger$};
	\node (opO4) at (0.8,-0.3) [below] {$\cO_2^\dagger$};
	\node (opO)  at (0,-1) [below] {$\cO$};
	\node [threept] (left3pt) at (-1.5,0) {$c$};
	\node [threept] (right3pt) at (1.5,0) {$d$};
	\node[twopt] (up2pt) at (0,0.4) {};
	\node[twopt] (dn2pt) at (0,-0.3) {};
	\draw [spinning] (up2pt) to[out=180,in=45] (left3pt);
	\draw [spinning] (dn2pt) to[out=-180,in=-45] (left3pt);
	\draw [spinning] (up2pt) to[out=0,in=135] (right3pt);
	\draw [spinning] (dn2pt) to[out=0,in=-135] (right3pt);
	\draw [spinning] (right3pt) to[out=0,in=0] (+0.7,-1) to[out=180,in=180] (-0.7,-1) to[out=180,in=180] (left3pt);
	\end{tikzpicture}}=
	\diagramEnvelope{\begin{tikzpicture}[anchor=base,baseline]
	\node (opO1) at (-1,0.4) [above] {$\cO_1$};
	\node (opO2) at (-1,-0.3) [below] {$\cO_2$};
	\node (opO3) at (0.8,0.4) [above] {$\tl\cO_1$};
	\node (opO4) at (0.8,-0.3) [below] {$\tl\cO_2$};
	\node (opO)  at (0,-1) [below] {$\cO$};
	\node [threept] (left3pt) at (-1.5,0) {$c$};
	\node [threept] (right3pt) at (1.5,0) {$d$};
	\node[twopt] (up2pt) at (0,0.4) {};
	\node[twopt] (dn2pt) at (0,-0.3) {};
	\draw [spinning] (up2pt) to[out=180,in=45] (left3pt);
	\draw [spinning] (dn2pt) to[out=-180,in=-45] (left3pt);
	\draw [spinning] (right3pt) to[in=0,out=135] (up2pt);
	\draw [spinning] (right3pt) to[in=0,out=-135] (dn2pt);
	\draw [spinning] (right3pt) to[out=0,in=0] (+0.7,-1) to[out=180,in=180] (-0.7,-1) to[out=180,in=180] (left3pt);
	\end{tikzpicture}}.
\ee
It is indeed easy to identify two shadow transforms~\eqref{eq:shadowdiagram} and a pairing~\eqref{eq:structurepairingdiagram} of three-point functions in the last diagram.

By performing the shadow transforms in some order, we can write the above in terms of shadow coefficients. In doing so, one of the three-point pairings cancels, and we obtain
\be
\frac{I_{ab}^{\mathrm{MFT}}(\De,\rho)}{\mu(\De,\rho)}
\p{\<\cO_1^\dag \cO_2^\dag \tl \cO^\dag \>^{(b)},\<\tl \cO_1 \tl \cO_2 \cO\>^{(d)}} 
&=
S([\tl \cO_1]\tl \cO_2 \cO)^{d}{}_{e} S(\cO_1 [\tl \cO_2]\cO)^e{}_a.
\label{eq:mftcoeffs}
\ee
Because the three-point pairing is nondegenerate, the above formula determines $I^\mathrm{MFT}_{ab}$.
Alternatively, doing the shadow transforms in a different way, we could write
\be
\frac{I_{ab}^{\mathrm{MFT}}(\De,\rho)}{\mu(\De,\rho)}
\p{\<\cO_1 \cO_2 \cO \>^{(a)},\<\tl \cO_1^\dag \tl \cO_2^\dag \tl\cO^\dag\>^{(c)}} 
&=
S([\tl \cO_1^\dag]\tl \cO_2^\dag \tl\cO^\dag)^{c}{}_{e} S(\cO_1^\dag [\tl \cO_2^\dag]\tl\cO^\dag)^e{}_b.
\label{eq:mftcoeffsII}
\ee

Essentially this way of computing MFT OPE coefficients was used (for scalar operators) in \cite{Fitzpatrick:2011dm}, where it was called ``conglomeration." Here, we have generalized this method to arbitrary operator representations and highlighted the role of the Plancherel measure, shadow transform, and three-point pairings, thus making precise contact with harmonic analysis~\cite{Dobrev:1977qv}.\footnote{Our approach also removes the need for prescriptions involving $\G(0)$ which were needed in \cite{Fitzpatrick:2011dm}.} We are still left with the question of determining these quantities, which is what we turn to next.
 
\section{Harmonic analysis and weight-shifting operators}
\label{sec:harmonicandweightshifting}

\subsection{Weight-shifting operators review}

A weight-shifting operator \cite{Karateev:2017jgd} is a conformally-covariant differential operator 
\be
\cD^{A}:[\De,\rho]\to [\De',\rho'].
\ee
Acting on a structure for an operator $\cO$ with weights $[\De,\rho]$, it produces a new structure for an operator $\cO'$ with weights $[\De',\rho']$. It also possesses a free index $A=1,\dots,\dim W$ for a finite-dimensional representation $W$ of the conformal group $\SO(d+1,1)$. In general, a pair of representations $\cO,\cO'$ may be connected by multiple weight-shifting operators with the same $W$, so we use a label $a$ to index the possible choices, $\cD^{(a)A}$. As in~\cite{Karateev:2017jgd}, we will use the following pictorial representation for the action of a weight-shifting differential operator
\begin{equation}
\cD^{(a)A} \quad=\quad
\diagramEnvelope{\begin{tikzpicture}[anchor=base,baseline]
	\node (vert) at (0,0) [threept] {$a$};
	\node (opO) at (-0.5,-1) [below] {$\cO$};
	\node (opOprime) at (-0.5,1) [above] {$\cO'$};
	\node (opFinite) at (1,0) [right] {$W$};	
	\draw [spinning] (vert)-- (opOprime);
	\draw [spinning] (opO) -- (vert);
	\draw [finite with arrow] (vert) -- (opFinite);
	\end{tikzpicture}}.
\label{eq:diffoppicture}
\end{equation}
In diagrams where a weight-shifting operator is placed on a leg which is contracted between two structures (such as right-hand side in~\eqref{eq:diagramcontraction}), we will assume that the operator is acting on the structure from which the arrow is outgoing. For example, if we put a weight-shifting operator on the internal line in the right-hand side of~\eqref{eq:diagramcontraction}, it should be interpreted as acting on the left conformally-invariant structure.

Weight-shifting operators can be used very generally to relate quantities associated to different conformal representations. For example, they can be used to write conformal blocks for general conformal representations as derivatives of conformal blocks for scalars \cite{Karateev:2017jgd}. In this section, we will see that they are an efficient tool for computing quantities in harmonic analysis.

As an example, let us describe  weight-shifting operators for $W=\Box$ (the vector representation of $\SO(d+1,1)$). Using the embedding space formalism of \cite{Costa:2011mg}, we represent a symmetric traceless tensor operator as a homogeneous function $\cO(X,Z)$ of coordinates $X,Z\in \R^{d+1,1}$ such that $X^2 = X\.Z=Z^2 = 0$, with gauge invariance under $Z\to Z+\b X$. It has homogeneity
\be
\cO(\l X,\a Z) = \l^{-\De} \a^J \cO(X,Z),
\ee
where $\De,J$ are the dimension and spin of $\cO$. The dictionary between embedding-space operators and operators on $\R^d$ is
\be
\label{eq:tofromflatdict}
\cO(X,Z) &= (X^+)^{-\De}\cO\p{x=\frac{X}{X^+},z=Z-\frac{Z^+}{X^+}X}, \nn\\
\cO(x,z) &= \cO\p{X=(1,x^2,x),Z=(0,2x\.z,z)}.
\ee
Here, $z \in \C^d$ is an auxiliary null polarization vector, and we are using index-free notation
\be
\label{eq:index_free_vector_formalism}
f(x,z) &= f^{\mu_1\cdots \mu_J}(x) z_{\mu_1}\cdots z_{\mu_J}.
\ee

The simplest weight-shifting operators for $W=\Box$ are
\be\label{eq:vectoroperators}
\cD_m^{-0} &= X_m, \nn\\
\cD_m^{0-} &=
\p{(\De-d + 2 - J)\de_m^n + X_m \pdr{}{X_n}} \p{(d - 4 + 2 J)  \pdr{}{Z^n}
-Z_n \frac{\ptl^2}{\ptl Z^2}},
\nn\\
\cD_m^{0+} &= (J+\De)Z_m + X_m Z\.\pdr{}{X},\nn\\
\cD_m^{+0} &= c_1\pdr{}{X^m} + c_2 X_m \frac{\ptl^2}{\ptl X^2} + c_3 Z_m \frac{\ptl^2}{\ptl Z\.\ptl X} + c_4 Z\.\pdr{}{X} \pdr{}{Z^m}\nn\\
&\quad + c_5 X_m Z\.\pdr{}{X}\frac{\ptl^2}{\ptl Z\.\ptl X} + c_6 Z_m Z\.\pdr{}{X} \frac{\ptl^2}{\ptl Z^2} + c_7 X_m \p{Z\.\pdr{}{X}}^2 \frac{\ptl^2}{\ptl Z^2},
\ee
where the coefficients $c_i$ are given in \cite{Karateev:2017jgd}. Here $m=0,\dots,d+1$ is a vector index for $\SO(d+1,1)$. The superscripts indicate how the operators shift dimension and spin, respectively:
\be
\cD^{\a\b}_m : [\De,J] \to [\De+\a,J+\b].
\ee
For example, $\cD^{-0}$ is a zeroth-order differential operator that simply multiplies by $X_m$. This increases the homogeneity in $X$ by $1$ and does not affect the homogeneity in $Z$. Thus it shifts $[\De,J]\to [\De-1,J]$.
The representation $W=\Box$ possesses other weight-shifting operators that produce non traceless-symmetric representations. These will play a role in section~\ref{sec:generalplancherel}.

In 4d, one uses a specialized embedding formalism \cite{Weinberg:2010fx,SimmonsDuffin:2012uy,Elkhidir:2014woa,Cuomo:2017wme} that efficiently describes all Lorentz representations. 4d operators $\cO$ are encoded into functions on the 6d embedding space denoted by
\begin{equation}
\label{eq:generic_4D_operator}
\cO_\Delta^{(\ell,\bar\ell)}(\point),\quad \point\equiv(X,S,\conj S),
\end{equation}
where $X^m$ ($m=1\ldots 6$) is the 6d embedding space coordinate, and $S_a$ and $\overline S^a$ ($a=1\ldots 4$) are 6d spinors. The lower and upper indices $a$ denote the fundamental representation and its dual of the $\mathrm{SU}(2,2)$ conformal group.\footnote{Note that $\mathrm{SU}(2,2)$ is the real form of the 4d conformal group in Lorentzian signature, whereas most of our discussion has focused on Euclidean signature. The distinction between signatures is not important for weight-shifting operators because they are finite-order differential operators with polynomial coefficients, which can be trivially analytically continued between signatures.} (They should not be confused with the general Lorentz indices or tensor structure labels used in section~\ref{sec:harmonic_analysis_review}.) The operators~\eqref{eq:generic_4D_operator} satisfy the homogeneity property
\be\label{eq:homoginety_degree}
\cO(\l X, \a S, \bar \a S) &= \l^{-\ka} \a^\ell \bar \a^{\bar \ell} \cO(X,S,\bar S), \nn\\
\kappa&\equiv \Delta+\frac{\ell+\bar\ell}{2}.
\ee
Instead of $X^m$ we often use the following antisymmetric objects
\begin{equation}\label{eq:coordinates_SU(2,2)}
\mathbf{X}_{ab}\equiv X_m \Sigma^m_{ab},\qquad
\overline{\mathbf{X}}^{ab}\equiv X_m \overline\Sigma^{m\,ab},
\end{equation}
where the 6d $\Sigma$-matrices are defined for example in appendix B of~\cite{Cuomo:2017wme}. 

In 4d there are two fundamental sets of weight-shifting operators that generate all others~\cite{Karateev:2017jgd}. The first set associated to the fundamental representation of $SU(2,2)$ is
\begin{align}%\label{eq:4Doperators_1}
\cD^a_{-0+} &\equiv \overline S^a,\nn\\
\cD^a_{-0-} &\equiv \overline{\mathbf{X}}^{ab}\partial_{\bar S,b},\nn\\
\cD^a_{++0} &\equiv {\bar a}\overline\partial^{ab}S_{b} +\overline S^a(S\overline\partial\partial_{\bar S}),\nn\\
\label{eq:4Doperators_4}
\cD^a_{+-0} &\equiv {\bar b} c\partial_S^{a}+
{\bar b}\overline S^a(\partial_S \partial_{\bar S})+
c\mathbf{X}_{bc}\overline\partial^{ab}\partial_S^{c} -
\overline S^{a}(\mathbf{X}_{bc}\overline\partial^{bd} \partial_S^{c}\partial_{\bar S,d}).
\end{align}
The second set associated to the dual representation is
\begin{align}%\label{eq:4Doperators_dual_1}
\overline{\cD}_a^{-+0} &\equiv S_{a},\nn\\
\overline{\cD}_a^{--0} &\equiv \mathbf{X}_{ab}\partial_S^b,\nn\\
\overline{\cD}_a^{+0+} &\equiv  a\partial_{ab}\overline S^b +S_{a}(\overline S\partial \partial_S),\nn\\
\label{eq:4Doperators_dual_4}
\overline{\cD}_a^{+0-} &\equiv b c\partial_{\bar S,a} +
bS_{a}(\partial_{\bar S}\partial_S)+
c\overline{\mathbf{X}}^{bc}\partial_{ab}\partial_{\bar S,c}-
S_{a}(\overline{\mathbf{X}}^{bc}\partial_{bd} \partial_{\bar S,c}\partial_S^d).
\end{align}
In the above, we assumed that the weight-shifting differential operators act on a generic operator~\eqref{eq:generic_4D_operator} in the representation $[\De,\ell,\bar\ell]$. We have also used the following short-hand notation for derivatives
\begin{equation}
\partial_{\bar S,a}\equiv \frac{\partial}{\partial \overline S^a},\qquad
\partial_S^a\equiv \frac{\partial}{\partial S_{a}},\qquad
\partial_{ab}\equiv\Sigma^M_{ab}\frac{\partial}{\partial X^M},\qquad
\overline\partial^{ab}\equiv\overline\Sigma^{M\,ab}\frac{\partial}{\partial X^M}.
\end{equation}
The parameters depend on the spins $(\ell, \bar\ell)$ and the homogeneity degree~\eqref{eq:homoginety_degree} as
\begin{equation}\label{eq:definitions_parameters_weight_shifting_operators_4d}
a \equiv 1-\kappa+\ell,\quad
\bar a \equiv 1-\kappa+\bar{\ell},\quad
b = 2\,(\bar \ell+1),\quad
\bar b \equiv 2\,(\ell+1),\quad
c \equiv -2 +\kappa-\ell-\bar{\ell}.
\end{equation}

\subsection{Recursion relations for the Plancherel measure}
\label{sec:plancherelrecursions}

As a first application of weight-shifting operators, let us compute the Plancherel measure for $\SO(d+1,1)$. Our strategy will be to derive a recursion relation for $\mu(\De,\rho)$ that allows us to shift $[\De,\rho]$ by the weights of finite-dimensional representations of $\SO(d+1,1)$.

 As a base case, consider the Plancherel measure for scalars. Via (\ref{eq:scalarplancherel}) it is related to the square of the shadow transform $\cN(\De,0)$. This is particularly simple to compute in Fourier space, where the shadow transform acts by multiplication. Note that
\be
\int d^d x \frac{1}{x^{2\De}} e^{-ipx} &= \frac{\pi^{d/2} 2^{d-2\De} \G(\frac d 2 - \De)}{\G(\De)} p^{2\De-d}.
\ee
Thus, in Fourier space
\be\label{eq:coefficient_square_shadow_transform}
\bS^2 &= \cN(\De,0) =  \frac{\pi^d \G(\frac d 2 - \De)\G(\De-\frac d 2)}{\G(\De)\G(d-\De)}.
\ee
Using (\ref{eq:relationbetweenshadowsqandplancherel}), we have
\be
\label{eq:scalarplancherel}
\mu(\De,0) = \frac{1}{2^d \cN(\De,0) \vol\SO(d)}.
\ee

We can derive a recursion relation for $\mu(\De,\rho)$ as follows. Note first that a weight-shifting operator $\cD$ in the trivial representation $W=\bullet$ must be proportional to the identity when $\De$ is generic. The reason is that $\cD$ is a conformally-invariant map from a generic conformal representation $V_{\De,\rho}$ to itself. The claim follows by irreducibility of $V_{\De,\rho}$ for generic $\De$ and Schur's lemma.

Consider now a pair of weight-shifting operators associated to $W$ and its dual $W^*$ that map between conformal representations $\cO$ and $\cO'$.
\be
\cD^{(a)A}:\cO \to \cO', \nn\\
\cD^{(b)}_A:\cO' \to \cO.
\ee
Taking the product and contracting the $\SO(d+1,1)$ indices, the result is a weight-shifting operator associated to the trivial representation, which must be a constant
\be\label{eq:definition_bubble_constant}
\cD^{(b)}_A \cD^{(a)A} &= \begin{pmatrix}
\cO \\
\cO'\ W
\end{pmatrix}^{ba}.
\ee
We call the quantity on the right-hand side a ``bubble coefficient" because it corresponds to a bubble of $\cO'$ and $W$ in the diagrammatic notation,
\be
	\begin{pmatrix}
		\cO \\
		\cO'\ W
	\end{pmatrix}^{ba}
	\diagramEnvelope{\begin{tikzpicture}[anchor=base,baseline]
	\node (opOin) at (-1,0) [left] {$\cO$};
	\node (opOout) at (1,0) [right] {$\cO$};
	\draw [spinning] (opOin)-- (opOout);
	\end{tikzpicture}}=
	\diagramEnvelope{\begin{tikzpicture}[anchor=base,baseline]
	\node (opOin) at (-2,0) [left] {$\cO$};
	\node (opOout) at (2,0) [right] {$\cO$};
	\node (opFinite) at (0,0.8) [above] {$W$};
	\node (opinternal) at (0,-0.1) [below] {$\cO'$};
	\node [threept] (ws1) at (-0.7,-0.083) {$a$};
	\node [threept] (ws2) at (+0.7,-0.13) {$b$};
	\draw [spinning] (opOin)-- (ws1);
	\draw [spinning] (ws1)-- (ws2);
	\draw [spinning] (ws2)-- (opOout);
	\draw [finite with arrow] (ws1) to[out=90, in=90] (ws2);
	\end{tikzpicture}}.
\ee
Taking the trace of both sides and using cyclicity of the trace, we find
\be
\begin{pmatrix}
\cO \\
\cO'\ W
\end{pmatrix}^{ba}\Tr_\cO(\mathbf{1}) = \Tr_\cO(\cD^{(b)}_A \cD^{(a)A})=\Tr_{\cO'}( \cD^{(a)A}\cD^{(b)}_A) = \begin{pmatrix}
\cO'\\
\cO\ W^*
\end{pmatrix}^{ab} \Tr_{\cO'}(\mathbf{1}).
\ee
Dividing by $\vol \SO(d+1,1)$ and using the definition (\ref{eq:formalplanchereldef}), we conclude
\be
\label{eq:recursionforplancherel}
\begin{pmatrix}
\cO \\
\cO'\ W
\end{pmatrix}^{ba}\mu(\cO)
= \begin{pmatrix}
\cO'\\
\cO\ W^*
\end{pmatrix}^{ab} \mu(\cO').
\ee
Diagrammatically, we interpret this recursion relation as reflecting the consistency of the two possible ways of closing the bubble in the diagram
\be
\diagramEnvelope{\begin{tikzpicture}[anchor=base,baseline]
	\node[threept] (ws2) at (1,0) {$b$};
	\node[threept] (ws1) at (-1,0) {$a$};
	\draw[spinning no arrow] (0,1.2) to[out=0, in=90] (ws2);
	\draw[spinning no arrow,->] (ws2) to[out=-90, in=0] (0,-1);
	\draw[spinning no arrow] (0,-1) to[out=180, in=-90] (ws1);
	\draw[spinning no arrow,->] (ws1) to[out=90, in=180] (0,1.2);
	\node at (0,-1) [below] {$\cO$};
	\node at (0,1.2) [above] {$\cO'$};
	\node at (0,0) [below] {$W$};
	\draw[finite with arrow] (ws1) -- (ws2);
	\end{tikzpicture}}\,.
\ee

The recursion relation (\ref{eq:recursionforplancherel}) is both a practical tool for computing $\mu(\cO)$ and also an interesting statement about the algebra of weight-shifting operators.

\subsubsection{Example: traceless symmetric tensors}

Let us apply the recursion relation (\ref{eq:recursionforplancherel}) to traceless symmetric tensors using the weight-shifting operators (\ref{eq:vectoroperators}).
It is straightforward to compute bubble coefficients by applying $\cD^{(b)}_A \cD^{(a)A}$ to any conformally-invariant structure. For example, we can choose the standard conformally-invariant two-point function, which in the embedding space takes the form
\be
\label{eq:two_point_structure_vector_formalism}
\<\cO(X_1,Z_1) \cO(X_2,Z_2)\> &= 
\frac{H_{12}^J}{X_{12}^{\De+J}},\nn\\
H_{12} &\equiv -2((X_1\.X_2)(Z_1\.Z_2)- (X_1\.Z_2)(X_2\.Z_1)), \nn\\
X_{ij} &\equiv -2X_i\.X_j.
\ee
Using the dictionary (\ref{eq:tofromflatdict}), this reduces to the standard convention in $\R^d$
\be
\label{eq:twoptconvnentionflatspace}
\<\cO^{\mu_1\cdots\mu_J}(x_1) \cO_{\nu_1\cdots\nu_J}(x_2)\> &= \frac{I^{(\mu_1}{}_{(\nu_1}(x_{12})\cdots I^{\mu_J)}{}_{\nu_J)}(x_{12})- \textrm{traces}}{x_{12}^{2\De}} \nn\\
I^\mu{}_\nu(x) &= \de^\mu{}_\nu - 2\frac{x^\mu x_\nu}{x^2}.
\ee

Applying spin-shifting operators $\cD^{0\pm}_m$, we find
\be
\begin{pmatrix}
[\De,J] \\
[\De,J-1]\ \Box
\end{pmatrix}^{(0+)(0-)} &= -J (d+2 J-4) (\Delta +J-2) (d-\Delta +J-2),\nn\\
\begin{pmatrix}
[\De,J-1] \\
[\De,J]\ \Box
\end{pmatrix}^{(0-)(0+)} &= -(d+J-3) (d+2 J-2) (\Delta +J-1) (d-\Delta +J-1),
\ee
which gives the recursion relation
\be\label{eq:musttjrecursion}
\mu(\De,J) &= \frac{(d+J-3) (d+2 J-2) (\Delta +J-1) (d-\Delta +J-1)}{J (d+2 J-4) (\Delta +J-2) (d-\Delta +J-2)} \mu(\De,J-1).
\ee
The solution 
 with base case (\ref{eq:scalarplancherel}) is
\be
\mu(\De,J) &= \frac{1}{\cN(\De,\rho)} \frac{\dim(\rho_J)}{2^d\vol\SO(d)},\nn\\
\dim(\rho_J) &=  \frac{\G(J+d-2)(2J+d-2)}{\G(J+1)\G(d-1)}, \nn\\
\label{eq:normforstts}
\cN(\De,J) &= \frac{\pi^d \G(\De-\frac d 2)\G(\frac d 2-\De)}{\G(\De-1)\G(d-\De-1)(\De+J-1)(d-\De+J-1)}.
\ee
Here, $\dim(\rho_J)$ is the dimension of a spin-$J$ representation of $\SO(d)$. 
Note that in our convention for the two-point function (\ref{eq:twoptconvnentionflatspace}), this is the same as
\be
\label{eq:contractionoftwopt}
\<\cO^{\mu_1\cdots\mu_J}(0) \cO^{\nu_1\cdots\nu_J}(\oo)\>\<\tl \cO_{\nu_1\cdots\nu_J}(\oo)\tl\cO_{\mu_1\cdots\mu_J}(0)\> &= \dim(\rho_J).
\ee
(This follows because the numerator of the two-point function (\ref{eq:twoptconvnentionflatspace}) is simply the action of $I\otimes \cdots \otimes I$ in $\rho_J \subset \mathrm{Sym}^J(\Box)$, where $I$ is a reflection. Equation~(\ref{eq:contractionoftwopt}) follows from $I^2 = 1$.) Thus, $\cN(\De,J)$ in (\ref{eq:normforstts}) is $\bS^2$ acting on a symmetric traceless tensor with dimension $\De$ and spin $J$.

One can additionally derive a recursion relation using dimension-shifting operators and verify that the expression (\ref{eq:normforstts}) obeys it.

\subsubsection{Example: general irreps in 4d}
All the simplest bubble coefficients in 4d are summarized in appendix~\ref{app:weight-shifting-formulas-4d}. We use them and the recursion relation~\eqref{eq:recursionforplancherel} to write
\begin{align}
\label{eq:plancherel_recursionA_4d}
\mu(\De,\; \ell,\; \bar\ell) &= 
\mathcal{B}_4(\De+1/2,\bar\ell-1,\ell)\;\mathcal{B}_1^{-1}(\De,\ell,\bar\ell)
\times \mu(\De+1/2,\; \ell-0,\; \bar\ell-1),\\
\label{eq:plancherel_recursionB_4d}
\mu(\De,\; \ell,\; \bar\ell) &= 
\mathcal{B}_2(\De-1/2,\bar\ell,\ell-1)\;\mathcal{B}_3^{-1}(\De,\ell,\bar\ell)
\times \mu(\De-1/2,\; \ell-1,\; \bar\ell-0).
\end{align}
The solution to the recursion relation~\eqref{eq:plancherel_recursionA_4d} and~\eqref{eq:plancherel_recursionB_4d} with the base case~\eqref{eq:scalarplancherel} is given by
\begin{align}
\label{eq:plancherel_measure_4d}
\mu(\De,\; \ell,\; \bar\ell) &=
\frac{1}{\mathcal{N}(\De,\; \ell,\; \bar\ell)}\times
\frac{(-1)^{\ell+\bar\ell}\times\dim(\ell,\bar\ell)}{16\,\vol\SO(4)},\\
\dim(\ell,\; \bar\ell) &=(1+\ell)(1+\bar\ell),\\
\mathcal{N}(\De,\; \ell,\; \bar\ell) &=\frac{(-1)^{\ell+\bar\ell}\times\pi^4}
{\left(\Delta-2+\frac{\ell-\bar\ell}{2}\right)
	\left(2-\Delta+\frac{\ell-\bar\ell}{2}\right)
	\left(\Delta+\frac{\ell+\bar\ell}{2}-1\right)
	\left(4-\Delta+\frac{\ell+\bar\ell}{2}-1\right)
}.
\label{eq:shadow_square_coefficient_indirect_4d}
\end{align}
In writing these relation we have used~\eqref{eq:plancherel_shadow_4d} and~\eqref{eq:two_point_pairing_4d}. As a consistency check we have also explicitly computed the square shadow coefficient in~\eqref{eq:shadow_square_4d}. The results of indirect~\eqref{eq:shadow_square_coefficient_indirect_4d} and direct~\eqref{eq:shadow_square_4d} computations perfectly coincide. They also coincide with the one in~\eqref{eq:normforstts} for $d=4$ and $\ell=\bar\ell=J$.

\subsection{Plancherel measure for general representations}
\label{sec:generalplancherel}

In this section we use the recursion relation~\eqref{eq:recursionforplancherel} to compute $\mu(\De,\rho)$ for general $\SO(d)$ representations $\rho$. We will use weight-shifting operators that transform in the vector representation of the conformal group and shift only the $\SO(d)$ weights $\rho$. In principle, the Plancherel measure is known in general~\cite{Dobrev:1977qv,KnappStein1}. The purpose of this section is to give an elementary derivation and clarify certain subtleties associated with fermionic representations.

\subsubsection{Weight-shifting operators}
Following the notation of~\cite{Kravchuk:2017dzd}, we represent an $\SO(d)$ weight as $\rho=\bfm_d$, a vector of $n$ integers or half-integers. In even dimensions $d=2n$ and we have
\be
	m_{d,1}\geq m_{d,2}\geq \ldots \geq m_{d,n-1} \geq |m_{d,n-1}|\geq 0,
\ee
while in odd dimensions $d=2n+1$ and
\be
	m_{d,1}\geq m_{d,2}\geq \ldots \geq m_{d,n-1} \geq m_{d,n-1}\geq 0.
\ee
When $\bfm_d$ consists of integers, its entries can be thought of as defining the lengths of the rows in the Young diagram of $\rho$.\footnote{The precise relation of $m_{d,i}$ to Dynkin labels can be found e.g.\ in~\cite{Kravchuk:2017dzd}.} According to the discussion in~\cite{Karateev:2017jgd}, the vector weight-shifting operators can change a single $m_{d,i}$ by $\pm 1$. We denote the resulting representations by $\bfm_d(\pm i)$.

Suppose $\cO$ has scaling dimension $\De$ and spin $\bfm_d$. Then the weight-shifting operator $D^A_{\pm i}$ that turns it into an operator with spin $\bfm_d(\pm i)$ has the form
\be\label{eq:vectorshifting}
	e_A \otimes (D^A_{\pm i} \cO)^a(x)=\pi^a_{b\,\mu}(\ptl^\mu w(x)\otimes \cO^b(x)+\a_{\pm i}(\De,\bfm_d) w(x)\otimes \ptl^\mu\cO^b(x)),
\ee
where $w(x)=w^A(x) e_A$, $e_A$ are basis elements of the vector representation of $\SO(d+1,1)$ and $w^A(x)$ are the conformal Killing scalars of dimension $-1$~\cite{Karateev:2017jgd}.\footnote{In embedding space language, we have $w^A(x)=X^A$.} Moreover, $\pi^a_{b\,\mu}$ gives the Clebsch-Gordan coefficient for $\bfm_d(\pm i)\in \bfm_d\otimes \myng{(1)}$.\footnote{Note that there are no multiplicities in this tensor product, so $\pi$ is determined uniquely up to a phase.} This is the most general expression, up to a multiple, which has the correct scaling and $\SO(d)$ transformation properties, according to the rules described in~\cite{Karateev:2017jgd}. In this expression, we have a single undetermined coefficient $\a_{\pm i}(\De,\bfm_d)$, which is fixed by the requirement that~\eqref{eq:vectorshifting} is primary,
\be
	(K_\nu\otimes 1+1\otimes K_\nu)\.\p{e_A \otimes (D^A_{\pm i} \cO)^a(0)}=0.
\ee
By recalling that $\ptl^\mu w(0) = P^\mu\. w(0)$ and similarly for $\cO$, and using the commutation relations of conformal group\footnote{In this section we use conventions of~\cite{Kravchuk:2017dzd}.} we find
\be
	0&=\pi^a_{b\,\mu}([K_\nu,P^\mu]\. w(0)\otimes \cO^b(0)+\a_{\pm i}(\De,\bfm_d) w(0)\otimes [K_\nu,P^\mu]\.\cO^b(0))\nn\\
	&=2(\a_{\pm i}(\De,\bfm_d)\De-1)\pi^a_{b\,\nu}w(0)\otimes \cO^b(0)+2\a_{\pm i}(\De,\bfm_d)\pi^a_{b\,\mu}(S_{\nu}{}^{\mu})^b{}_c w(0)\otimes \cO^c(0),
\ee
where $(S^{\nu\mu})^{b}{}_c$ are the generators of rotations in the representation $\bfm_d$.
From the index structure, we must have
\be
	\pi^a_{b\,\mu}(S_{\nu}{}^{\mu})^b{}_c=-(\bfm_d \myng{(1)}|\bfm_{d}(\pm i))\,\pi^a_{c\,\nu}
\ee
for some coefficient $(\bfm_d\myng{(1)}|\bfm_{d}(\pm i))$. We thus have
\be
\a_{\pm i}(\De,\bfm_d)=\frac{1}{\De-(\bfm_d\myng{(1)}|\bfm_{d}(\pm i))}.
\ee
It was observed in appendix~B$.$4 of~\cite{Kravchuk:2017dzd} that
\be
	(\bfm_d\myng{(1)}|\bfm_{d}(+i))&=-m_{d,i}+i-1,\\
	(\bfm_d\myng{(1)}|\bfm_{d}(-i))&=m_{d,i}+d-i-1.
\ee

\subsubsection{Bubble coefficients}

Let us now compute the bubble coefficients. First, we compute the composition
\be
	e_B\otimes e_A \otimes (D^B_{\mp i} &D^A_{\pm i} \cO(x))^a=\nn\\
	=\pi^a_{b\, \nu} \pi^b_{c\,\mu}&\left[\ptl^\nu w(x)\otimes \ptl^\mu w(x) \otimes \cO^c(x)+\a_{\pm i}(\De,\bfm_d)\ptl^\nu w(x)\otimes w(x) \otimes \ptl^\mu \cO^c(x)\right.\nn\\
	&+\a_{\mp i}(\De,\bfm_d(\pm i))\p{w(x)\otimes \ptl^\nu\ptl^\mu w(x) \otimes \cO^c(x)+w(x)\otimes \ptl^\mu w(x) \otimes \ptl^\nu\cO^c(x)}\nn\\
	&+\a_{\mp i}(\De,\bfm_d(\pm i))\a_{\pm i}(\De,\bfm_d)w(x)\otimes \ptl^\nu w(x) \otimes \ptl^\mu\cO^c(x)\nn\\
	&\left.+\a_{\mp i}(\De,\bfm_d(\pm i))\a_{\pm i}(\De,\bfm_d)w(x)\otimes  w(x) \otimes \ptl^\nu\ptl^\mu\cO^c(x)\right].
\ee
We now contract the indices on the operators by replacing $e_B\otimes e_A\to \eta_{AB}$, where $\eta_{AB}$ is the metric in $\R^{d+1,1}$. This amounts to the substitution $w(x)\otimes w(y)\to -\half(x-y)^2$ which yields
\be
	(D_{\mp i,A} &D^A_{\pm i} \cO(x))^a=\de^{\mu\nu}\pi^a_{b\, \mu} \pi^b_{c\,\nu}[1-\a_{\mp i}(\De,\bfm_d(\pm i))]\cO^c(x).
\ee
One can check that in the normalization of CG coefficients as in~\cite{Kravchuk:2017dzd} we have
\be
	\de^{\mu\nu}\pi^a_{b\, \mu} \pi^b_{c\,\nu}=\sqrt{\frac{1}{d}\frac{\dim \bfm_d(\pm i)}{\dim \bfm_d}}\delta^a_c.
\ee
This implies that the bubble coefficient is given by
\be
	\begin{pmatrix}
		[\De,\bfm_d]\\
		[\De,\bfm_d(\pm i)]\,\,\myng{(1)}
	\end{pmatrix}=
	\sqrt{\frac{1}{d}\frac{\dim \bfm_d(\pm i)}{\dim \bfm_d}}
	\frac{\De-(\bfm_d(\pm i)\myng{(1)}|\bfm_{d})-1}{\De-(\bfm_d(\pm i)\myng{(1)}|\bfm_{d})}.
\ee

\subsubsection{Recursion relation}
Given the bubble coefficients which we found above, we can conclude from~\eqref{eq:recursionforplancherel}
\be\label{eq:generalrecursionforplancherel}
	\mu(\De,\bfm_d(+i))=&\begin{pmatrix}
		[\De,\bfm_d]\\
		[\De,\bfm_d(+ i)]\,\,\myng{(1)}
	\end{pmatrix}
	\begin{pmatrix}
		[\De,\bfm_d(+i)]\\
		[\De,\bfm_d]\,\,\myng{(1)}
	\end{pmatrix}^{-1}\mu(\De,\bfm_d)\nn\\
=&\frac{\dim \bfm_d(+i)}{\dim \bfm_d}
\frac{d-\De+m_{d,i}-i+1}{d-\De+m_{d,i}-i}\frac{\De+m_{d,i}-i+1}{\De+m_{d,i}-i}\mu(\De,\bfm_d).
\ee
In particular if we set $i=1$ and let $\bfm_d$ be a traceless-symmetric irrep, we find
\be
	\mu(\De,J) = \frac{(J+d-3)(2J+d-2)}{J(2J+d-4)}
	\frac{d-\De+J-1}{d-\De+J-2}\frac{\De+J-1}{\De+J-2}\mu(\De,J-1),
\ee
in full agreement with~\eqref{eq:musttjrecursion}.

For bosonic representations the solution to the general recursion relation is given by
\be
	\mu(\De,\bfm_{d})=\dim \bfm_d \prod_{i=1}^n \frac{(\De+m_{d,i}-i)}{(\De-i)}\frac{(d-\De+m_{d,i}-i)}{(d-\De-i)}\mu(\De,0),
\ee
where $\mu(\De,0)$ is the Plancherel measure for scalar representations given by~\eqref{eq:scalarplancherel}. For fermionic representations we have
\be
	\mu(\De,\bfm_{d})=\frac{\dim \bfm_d}{\dim S} \prod_{i=1}^n \frac{(\De+m_{d,i}-i)}{(\De+\half-i)}\frac{(d-\De+m_{d,i}-i)}{(d-\De+\half-i)}\mu(\De,S),
\ee
where $S$ is the spinor representation with all $m_{d,i}=\half$. We compute $\mu(\De,S)$ in appendix~\ref{app:spinorplancherel}. Let us now substitute $\mu(\De,0)$ and $\mu(\De,S)$ into the above expressions.

In even dimensions, $d=2n$, we have
\be
	\mu(\De,0)&=(-1)^{n}\p{\De-\frac{d}{2}}(\De+1-d)_{d-1}\frac{1}{(2\pi)^d \vol\SO(d)},\\
	\mu(\De,S)&=(-1)^{n}\p{\De+\half-d}_{d}\frac{\dim S}{(2\pi)^d \vol\SO(d)}.
\ee
We then find that the general Plancherel measure for $d=2n$ is given by
\be\label{eq:generalplancherelevend}
	\mu(\De,\bfm_{d})=\frac{\dim \bfm_d}{(2\pi)^d \vol\SO(d)} \prod_{i=1}^n (\De+m_{d,i}-i)(d-\De+m_{d,i}-i).
\ee
In odd dimensions, $d=2n+1$, we have
\be
	\mu(\De,0)&=(-1)^{n}\p{\De-\frac{d}{2}}(\De+1-d)_{d-1}\frac{\cot\pi\De}{(2\pi)^d \vol\SO(d)},\\
	\mu(\De,S)&=(-1)^{n+1}\p{\De+\half-d}_{d}\frac{\dim S\,\tan\pi\De}{(2\pi)^d \vol\SO(d)},
\ee
and we find the general Plancherel measure for $d=2n+1$
\be\label{eq:generalplanchereloddd}
	\mu(\De,\bfm_{d})=\frac{\dim \bfm_d}{(2\pi)^d \vol\SO(d)} \p{\De-\frac{d}{2}}\cot\pi(\De-m_{d,1})\prod_{i=1}^n (\De+m_{d,i}-i)(d-\De+m_{d,i}-i).
\ee
Note that we have $\cot\pi(\De-m_{d,1})=\cot\pi\De$ for bosonic irreps and $\cot\pi(\De-m_{d,1})=-\tan\pi\De$ for fermionic irreps.

In order to compare to the formulas in~\cite{Dobrev:1977qv}, let us recall the expression for $\dim\bfm_d$. For that, define $p_i=m_{d,n-i+1}+i-1$ and $q_i=i-1$ for even $d=2n$, $p_i=m_{d,n-i+1}+i-\half$ and $q_i=i-\half$ for odd $d=2n+1$. We then have
\be\label{eq:irrepdimensionseven}
	\dim \bfm_{2n}&=\prod_{1\leq i<j \leq n}\frac{p_j^2-p_i^2}{q_j^2-q_i^2},\\
	\label{eq:irrepdimensionsodd}
	\dim \bfm_{2n+1}&=\prod_{i=1}^n \frac{p_i}{q_i}\prod_{1\leq i<j \leq n}\frac{p_j^2-p_i^2}{q_j^2-q_i^2}.
\ee
If we write $\De=\frac{d}{2}+c$ we find
\be
	\prod_{i=1}^n (\De+m_{d,i}-i)(d-\De+m_{d,i}-i)=\prod_{i=1}^n (p_i^2-c^2)
\ee
and we can now recognize formulas (8.6) and (8.17b) of~\cite{Dobrev:1977qv}.\footnote{We have $p_i=n_i$ in even $d$ or $p_i=\ell_i+i-\half$ in odd $d$ in their notation. For fermionic representations in odd $d$ (8.17b) of~\cite{Dobrev:1977qv} should be modified by replacing $\tan\pi c$ with $-\cot\pi c$ in order to agree with our~\eqref{eq:generalplanchereloddd}. We believe our formula~\eqref{eq:generalplanchereloddd} is correct in these cases since it is consistent with the 3d results obtained in section~\ref{sec:plancherel_3d} by two independent methods. Also, see (79) in~\cite{Kitaev:2017hnr} for $\mu=0$ and $\mu=\half$.} 

\subsubsection{Compact expression}
\label{sec:compactexpression}

Finally, let us give a compact expression for the Plancherel measure. For that, we first define an $\SO(d+2)$ weight
\be
	\bfm_{d+2}=(-\De,m_{d,1},\ldots,m_{d,n}).
\ee
We then have
\be
\label{eq:finalformulasforplancherel}
	\mu(\De,\bfm_d)&=(-1)^n 2\pi\frac{\dim \bfm_{d+2}}{\vol\SO(d+2)} &(\textrm{$d=2n$}), \nn\\
	\mu(\De,\bfm_d)&=(-1)^{n+1} 2\pi\frac{\dim \bfm_{d+2}}{\vol\SO(d+2)}\cot\pi(\De-m_{d,1}) &(\textrm{$d=2n+1$}).
\ee
In the above formulas $\dim\bfm_{d+2}$ should be interpreted as a formal rational function given by~\eqref{eq:irrepdimensionseven} and~\eqref{eq:irrepdimensionsodd}. Note the similarity between these expressions and the Plancherel measure for a compact group $G$,
\be
	\mu(\pi) = \frac{\dim\pi}{\vol G}.
\ee
It is not hard to understand why the Plancherel measure for $\SO(d+1,1)$  is proportional to $\dim \bfm_{d+2}$. Indeed, we should be able to use weight-shifting operators to compute recursion relations for the Plancherel measure of the compact group $\SO(d+2)$. These recursion relations will be identical to what we have found above. It then follows that the ratio of the Plancherel measures for $\SO(d+2)$ and $\SO(d+1,1)$ should be invariant under shifts by weights of $\mathfrak{so}(d+2)$, which is precisely what we observe. In fact, this reasoning can be used to fix the formulas (\ref{eq:finalformulasforplancherel}) from the answer for scalars and fermions without explicitly computing the weight-shifting operators $D_{\pm i}^A$.

We expect that similar reasoning should hold for the Plancherel measure with respect to other real forms of $\mathfrak{so}(d+2)$, and indeed for different real forms of arbitrary Lie groups. Specifically, the ratio between Plancherel measures for different real forms of $\mathfrak{g}$ (analytically continued appropriately in the weights) should be invariant under shifts by weights of $\mathfrak{g}$.

\subsection{Integration by parts and shadow coefficients}
\label{eq:weightshiftingshadowcoeffs}

To compute shadow coefficients and three-point pairings, we must understand how to integrate weight-shifting operators by parts. Recall the natural pairing between operators 
\be
(\tl \cO^\dag,\cO) &= \int d^d x\,  \tl \cO^\dag(x)\cO(x).
\ee
 Given a weight-shifting operator $\cD:\cO\to \cO'$ in the representation $W$ (we suppress its $W$-index for brevity), there exists an adjoint $\cD^*:\tl \cO'^\dag\to \tl \cO^\dag$ in the same representation $W$ such that
\be
(\tl \cO'^\dag,\cD \cO) &= (\cD^* \tl \cO'^\dag,\cO).
\ee
Note that $\cD$ is always a finite-order differential operator with polynomial coefficients, so $\cD^*$ can be computed straightforwardly by integrating by parts in each term. The fact that $\cD^*$ is a weight-shifting operator follows from conformal invariance of the pairing.

Integration by parts is useful for computing pairings between two- and three-point structures, as we show via examples in appendix~\ref{sec:two_and_three_point_pairings}. Although this is in principle simply a matter of contracting indices, weight-shifting operators can help organize the computation efficiently.

With the ability to integrate by parts, we can also commute weight-shifting operators past shadow transforms. We have
\be
\bS \cD \cO(x) &= \int d^d y \<\tl \cO'(x) \tl \cO'^\dag(y)\> \cD \cO(y) \nn\\
&= \int d^d y \<\tl \cO'(x) (\cD^* \tl \cO'^\dag)(y)\> \cO(y) \nn\\
&= \int d^d y \<(\cD' \tl \cO')(x) \tl \cO^\dag(y)\> \cO(y) \nn\\
&= \cD' \bS \cO(x).
\label{eq:commutewithshadow}
\ee
Here, $\cD'$ is a weight-shifting operator determined by solving the ``crossing equation"
\be
\label{eq:twoptcrossing}
\<\tl \cO'(x) (\cD^* \tl \cO'^\dag)(y)\> &= \<(\cD' \tl \cO')(x) \tl \cO^\dag(y)\>,
\ee
which is a linear equation in a finite-dimensional space of conformally-covariant two-point structures. The coefficients relating $\cD^*$ and $\cD'$ are examples of $6j$ symbols~\cite{Karateev:2017jgd}. They can be computed by simply evaluating both sides of (\ref{eq:twoptcrossing}).

From representation theory point of view, the shadow transform $\bS$ implements a particular affine Weyl reflection of the conformal group $\SO(d+1,1)$ (see~\cite{Kravchuk:2018htv} for a recent discussion in the context of CFT). On the other hand, each weight-shifting operator $\cD$ shifts the weights of the primary it is acting on by some weight $\mu$. It is easy to see that $\cD'$ shifts by $\bS\mu$, where the action of the Weyl group is interpreted in the usual (not affine) sense. This extends to other affine Weyl transforms which exist in Lorentzian signature~\cite{Kravchuk:2018htv}.

Finally, by commuting weight-shifting operators past shadow transforms, we can efficiently derive recursion relations for shadow coefficients. This leads to much simpler calculations than in previous cases, e.g.\ \cite{Dobrev:1977qv,SimmonsDuffin:2012uy,Goncalves:2014rfa,Costa:2014rya}. We will see several examples below.

\subsubsection{Example: traceless symmetric tensors}
\label{sec:shadow_coefficients_example_TS}

\paragraph{Integration by parts}

To integrate the operators (\ref{eq:vectoroperators}) by parts, we translate from the embedding space to $\R^d$ using the dictionary (\ref{eq:tofromflatdict}). Integration by parts proceeds in the usual way for $x$-derivatives. Dealing with the polarization vector $z$ takes more care. A contraction between two traceless symmetric tensors can be written as \cite{Costa:2011mg}
\be
\label{eq:contractionpairing}
f\.g &= f_{\mu_1\cdots\mu_J} g^{\mu_1\cdots\mu_J}= \frac{1}{J!(h-1)_J} f(D) g(z) = \frac{1}{J!(h-1)_J} g(D) f(z).
\ee
Here, we use index-free notation $f(z)=f^{\mu_1\cdots\mu_J} z_{\mu_1}\cdots z_{\mu_J}$ and similarly for $g(z)$. The Thomas/Todorov operator $D$ is given by
\be
D_\mu &= \p{h - 1 + z\.\pdr{}{z}}\pdr{}{z^\mu} - \frac 1 2 z_\mu \pdr{{}^2}{z^2},
\ee
with $h=\frac d 2$. 
Now suppose $f(z)$ has spin-$(J+1)$ and $g(z)$ has spin-$J$. By symmetry of the pairing (\ref{eq:contractionpairing}), we have
\be
f\.(z_\mu g) &= \frac{1}{(J+1)!(h-1)_{J+1}} g(D)D_\mu f(z)\nn\\
&= \frac{1}{(J+1)(h-1+J)} g\.D_\mu f,
\ee
(Here we have also used that $[D_\mu,D_\nu]=0$.) Thus, the adjoint of $z$ under the pairing (\ref{eq:contractionpairing}) is
\be
z^* |_{J} &= \frac{1}{(J+1)(h-1+J)} D_z|_{J+1},
\ee
where $|_J$ indicates that the operator acts on a spin-$J$ representation.

Applying these results to the weight-shifting operators (\ref{eq:vectoroperators}), we find
\be
(\cD^{-0}|_{\De,J})^* &= \cD^{-0}|_{d-\De+1,J}, \nn\\
(\cD^{0-}|_{\De,J})^* &= -2J(h-2+J) \cD^{0+}|_{d-\De,J-1}, \nn\\
(\cD^{0+}|_{\De,J})^* &= -\frac{1}{2(J+1)(h-1+J)} \cD^{0-}|_{d-\De,J+1}, \nn\\
(\cD^{+0}|_{\De,J})^* &= \cD^{+0}|_{d-\De-1,J},
\ee
where $\cD|_{\De,J}$ indicates that $\cD$ acts on a multiplet with weights $\De,J$. 
Following the computation (\ref{eq:commutewithshadow}), we find
\be
\label{eq:commuteshadowandd}
\mathbf{S} \cD^{0+} |_{[\De,J]} &= \frac{\Delta +J-1}{d-\De +J} \cD^{0+} \mathbf{S}|_{[\De,J]}.
\ee

\paragraph{Shadow coefficients for scalar-scalar-spin-$J$}

Let us apply these results to compute shadow coefficients for the scalar-scalar-spin-$J$ three-point structure (\ref{eq:scalarscalarspinjstructure}). In the embedding space, this structure is given by
\be
\<\f_{\De_1}(X_1) \f_{\De_2}(X_2) \cO_{\De_3,J}(X_3,Z_3)\> &= \frac{V_{3,12}^J}{X_{12}^{\frac{\De_1+\De_2-\De_3}{2}} X_{23}^{\frac{\De_2+\De_3-\De_1}{2}} X_{13}^{\frac{\De_1+\De_3-\De_2}{2}}},\nn\\
V_{3,12} &= -2\frac{(Z_3\.X_1)(X_2\.X_3) - (Z_3\.X_2)(X_1\.X_3)}{X_{12}^{1/2} X_{23}^{1/2} X_{13}^{1/2}}.
\ee
In the case $J=0$, the shadow transform of any of the operators is given by the classic star-triangle formula \cite{Fradkin:1978pp,Vasiliev:1981dg}
\be
\label{eq:scalartransform}
\<\f_{\De_1} \f_{\De_2} \bS[\f_{\De_3}]\> &= \int d^dx_3 \frac{1}{(x_3'-x_3)^{2(d-\De_3)}} \frac{1}{x_{12}^{\De_1+\De_2-\De_3} x_{23}^{\De_2+\De_3-\De_1} x_{31}^{\De_3+\De_1-\De_2}} \nn\\
&= \frac{\pi^{\frac d 2} \G(\De_3-\frac d 2)\G(\frac{\tl\De_3+\De_1-\De_2}{2})\G(\frac{\tl\De_3+\De_2-\De_1}{2})}{ \G(d-\De_3) \G(\frac{\De_3+\De_1-\De_2}{2})\G(\frac{\De_3+\De_2-\De_1}{2})} \<\f_{\De_1} \f_{\De_2} \f_{\tl \De_3}\>.
\ee
To study nonzero $J$, we use weight-shifting operators to relate three-point structures with different spins,\footnote{The trick of using differential operators to produce spinning three-point structures from scalar structures was originally described in \cite{Costa:2011dw}.}
\be
\label{eq:relationbetweenstructs}
\<(\cD^{-0} \f_{\De_1}) \f_{\De_2} (\cD^{0+} \cO_{\De_3,J})\> &= -\frac 1 2 (\De_1-J-\De_2-\De_3)
\<\f_{\De_1-1} \f_{\De_2} \cO_{\De_3,J+1}\>.
\ee
Here, we have chosen to act with a particular combination of weight-shifting operators, but other choices are possible. The operators $\cD^{-0}$ and $\cD^{0+}$ each have a free vector index for $\SO(d+1,1)$, and we are implicitly contracting those indices. From (\ref{eq:relationbetweenstructs}) we will derive recursion relations for shadow coefficients.

\subparagraph{Shadow transforming the scalar} 
\label{sec:shadow_transforming_scalar_vector_formalism}

Let us first consider the shadow transform of the scalar $\f_{\De_2}$. This is quite easy because our differential operators (which act only on points $1$ and $3$) simply commute with $\mathbf{S}$ acting on point $2$,
\be
&S(\f_{\De_1} [\f_{\De_2}] \cO_{\De_3,J}) \<\f_{\De_1} \f_{\tl \De_2} \cO_{\De_3,J}\> \nn\\
&= \<\f_{\De_1} \bS[\f_{\De_2}] \cO_{\De_3,J}\>  \nn\\
&= -2(\De_1+2-J-\De_2-\De_3)^{-1} \<(\cD^{-0} \f_{\De_1+1}) \bS[\f_{\De_2}] (\cD^{0+} \cO_{\De_3,J-1})\> \nn\\
&= -2(\De_1+2-J-\De_2-\De_3)^{-1} S(\f_{\De_1+1} [\f_{\De_2}] \cO_{\De_3,J-1}) \<(\cD^{-0} \f_{\De_1+1}) \f_{\tl \De_2} (\cD^{0+} \cO_{\De_3,J-1})\> \nn\\
&= \frac{\De_1+2-J-\tl \De_2-\De_3}{\De_1+2-J-\De_2-\De_3}
S(\f_{\De_1+1} [\f_{\De_2}] \cO_{\De_3,J-1}) \<\f_{\De_1} \f_{\tl \De_2} \cO_{\De_3,J}\>.
\ee
In other words,
\be
S(\f_{\De_1} [\f_{\De_2}] \cO_{\De_3,J}) &= \frac{\De_1+2-J-\tl \De_2-\De_3}{\De_1+2-J-\De_2-\De_3}S(\f_{\De_1+1} [\f_{\De_2}] \cO_{\De_3,J-1}).
\ee
Using (\ref{eq:scalartransform}) as a base-case, this is solved by
\be
\label{eq:scalarshadowcoeff}
S(\f_{\De_1} [\f_{\De_2}] \cO_{\De_3,J}) &= \frac{\pi^{\frac d 2}\G(\De_2 - \frac d 2)\G(\frac{\tl \De_2 + \De_1 - \De_3+J}{2}) \G(\frac {\tl \De_2 + \De_3 - \De_1+J}{2}) }{\G(d-\De_2)\G(\frac{\De_2+\De_1-\De_3+J}{2}) \G(\frac{\De_2+\De_3 - \De_1+J}{2})}.
\ee

\subparagraph{Shadow transforming the tensor}

To derive a recursion relation for the shadow transform of the tensor operator, we can follow exactly the same computation. The only new ingredient is that we must use (\ref{eq:commuteshadowandd}) to commute $\cD^{0+}$ past $\mathbf{S}$ acting on $\cO_3$. We find 
\be
S(\f_{\De_1} \f_{\De_2} [\cO_{\De_3,J}]) &= \frac{(\De_1+2-J-\De_2-\tl\De_3)(\De_3+J-2)}{(\De_1+2-J-\De_2-\De_3)(\tl\De_3+J-1)} S(\f_{\De_1+1} \f_{\De_2} [\cO_{\De_3,J-1}]).
\ee
This is solved by
\be
\label{eq:spinningshadowcoefficient}
S(\f_{\De_1} \f_{\De_2} [\cO_{\De_3,J}]) &= \frac{\pi^{\frac d 2} \G(\De_3-\frac d 2)\G(\De_3+J-1)\G(\frac{\tl\De_3+\De_1-\De_2+J}{2})\G(\frac{\tl\De_3+\De_2-\De_1+J}{2})}{\G(\De_3-1) \G(d-\De_3+J) \G(\frac{\De_3+\De_1-\De_2+J}{2})\G(\frac{\De_3+\De_2-\De_1+J}{2})}.
\ee
The results (\ref{eq:scalarshadowcoeff}) and (\ref{eq:spinningshadowcoefficient}) for shadow coefficients agree with the more laborious direct evaluation of star-triangle integrals, see e.g.\ appendix D of \cite{Liu:2018jhs}.

As a check of (\ref{eq:scalarshadowcoeff}) and (\ref{eq:spinningshadowcoefficient}), note that $\bS^2$ acts in the correct way in both cases
\be
S(\f_{\De_1}[\f_{\De_2}]\cO_{\De_3,J}) S(\f_{\De_1}[\f_{\tl \De_2}]\cO_{\De_3,J}) &= \cN(\De_2,0), \nn\\
S(\f_{\De_1}\f_{\De_2}[\cO_{\De_3,J}]) S(\f_{\De_1}\f_{\De_2}[\cO_{\tl\De_3,J}]) &= \cN(\De_3,J),
\ee
where $\cN(\De,J)$ is given in (\ref{eq:normforstts}).

\subsubsection{Example: shadow coefficients in 4d}
\label{sec:shadow_coefficients_4d}

We describe the details of integration by parts in 4d in appendix~\ref{app:details_4d_formalism}. Here, we use those results to derive several examples of 4d shadow coefficients.

\paragraph{Shadow coefficients for scalar-scalar-spin$(\ell,\ell)$}
Let us start by considering a three-point function of two scalars and a spin-$\ell$ operator, this time using the 4d embedding formalism,
\begin{align}
\label{eq:tensor_structure_scalar_scalar_spin_L}
\langle \phi_{\Delta_1}\,\phi_{\Delta_2}\,\cO^{(\ell,\ell)}_{\Delta_3}\rangle
&\equiv K_3\left[\hat J^3_{12}\right]^\ell,\\
\label{eq:invariant_J}
\hat J^k_{ij} &\equiv X_{ij}^{-1}\, (\overline S_k \mathbf{X}_i \overline{\mathbf{X}}_j S_k).
\end{align}
Here, $K_3$ is a standard product of dimensionful factors
\begin{equation}
K_3\equiv\prod_{i<j} X_{ij}^{-\frac{\kappa_i+\kappa_j-\kappa_k}{2}},
\end{equation}
where $\kappa$ is the homogeneity degree defined in~\eqref{eq:homoginety_degree}.

The structure~\eqref{eq:tensor_structure_scalar_scalar_spin_L} satisfies the following recursion relation
\begin{align}\label{eq:recursion_4D_first}
\langle \phi_{\Delta_1}\,\phi_{\Delta_2}\,\cO^{(\ell,\ell)}_{\Delta_3}\rangle
&=\mathcal{A}_1^{-1}\times
\left( D^2_{-0-}\cdot \overline D_3^{-+0}\right)
\left( D^3_{-0+}\cdot \overline D_2^{+0+}\right)
\langle \phi_{\Delta_1}\,\phi_{\Delta_2}\,\cO^{(\ell-1,\ell-1)}_{\Delta_3+1}\rangle,\\
\mathcal{A}_1(\De_1,\De_2,\De_3) &\equiv (\De_2-1)(\De_1+\De_2-\De_3+\ell-2).
\end{align}

Applying the shadow transform $\bS$ to both sides of~\eqref{eq:recursion_4D_first}, we obtain recursion relations for the corresponding shadow coefficients. Let $\bS_i$ ($i=1,2,3$) denote the shadow transform of the operator in the $i$-th position. First consider $\bS_1$. Trivially commuting $\bS_1$ with all the differential operators on the right-hand side of~\eqref{eq:recursion_4D_first}, we obtain
\begin{equation}
S([\phi_{\Delta_1}]\phi_{\Delta_2}\cO^{(\ell,\ell)}_{\Delta_3})=
\mathcal{A}^{-1}_1(\De_1,\De_2,\De_3)\mathcal{A}_1(\tl\De_1,\De_2,\De_3)\times
S([\phi_{\Delta_1}]\phi_{\Delta_2}\cO^{(\ell-1,\ell-1)}_{\Delta_3+1}).
\end{equation} 
The solution is
\begin{equation}\label{eq:basic_shadow_coefficient_1}
S([\phi_{\Delta_1}]\phi_{\Delta_2}\cO^{(\ell,\ell)}_{\Delta_3})=
\pi^2\times
\frac{\Gamma(\Delta_1-2)}{\Gamma(4-\Delta_1)}\times
\frac{
	\Gamma\left(\frac{\widetilde\Delta_1+\Delta_{23}+\ell}{2}\right)
	\Gamma\left(\frac{\widetilde\Delta_1-\Delta_{23}+\ell}{2}\right)
}{
	\Gamma\left(\frac{\Delta_1+\Delta_{23}+\ell}{2}\right)
	\Gamma\left(\frac{\Delta_1-\Delta_{23}+\ell}{2}\right)
},
\end{equation}
which agrees with~\eqref{eq:scalarshadowcoeff} after setting $d=4$ and $J=\ell$ and exchanging $\De_1\leftrightarrow\De_2$.

Next, we apply $\bS_3$ to~\eqref{eq:recursion_4D_first} and use the commutation relations~\eqref{eq:commutation1} and~\eqref{eq:commutation1b} to obtain
\begin{equation}
\nn
S(\phi_{\Delta_1}\phi_{\Delta_2}[\cO^{(\ell,\ell)}_{\Delta_3}]) =
\frac{
	(\De_3-1)
	(2 -  \Delta_{3} +  \Delta_{12}  + \ell)
	(2 -  \Delta_{3} -  \Delta_{12}  + \ell)}
{4\,(\Delta_3-2)(2 - \Delta_3 + \ell)
	(3 - \Delta_3 + \ell)}
S(\phi_{\Delta_1}\phi_{\Delta_2}[\cO^{(\ell-1,\ell-1)}_{\Delta_3+1}]).
\end{equation}
This is solved by~\eqref{eq:spinningshadowcoefficient} with $d=4$ and $J=\ell$. We provide the formula here again for convenience
\begin{equation}\label{eq:basic_shadow_coefficient}
S(\phi_{\Delta_1}\phi_{\Delta_2}[\cO^{(\ell,\ell)}_{\Delta_3}])=
\frac{\pi^2}{\Delta_3-2}\times
\frac{\Gamma(\Delta_3+\ell-1)}{\Gamma(\widetilde\Delta_3+\ell)}\times
\frac{
	\Gamma\left(\frac{\widetilde\Delta_3+\Delta_{12}+\ell}{2}\right)
	\Gamma\left(\frac{\widetilde\Delta_3-\Delta_{12}+\ell}{2}\right)
}{
	\Gamma\left(\frac{\Delta_3+\Delta_{12}+\ell}{2}\right)
	\Gamma\left(\frac{\Delta_3-\Delta_{12}+\ell}{2}\right)
}.
\end{equation}

\paragraph{Shadow coefficients for seeds structures}
Now consider a more complicated case of two different ``seed" structures in 4d~\cite{Echeverri:2015rwa}
\begin{align}\label{eq:seed_3pt_1}
\langle\phi_{\Delta_1}\,f^{(p,0)}_{\Delta_2}\,
\cO^{(\ell,\,\ell+p)}_{\Delta_3}\rangle
&\equiv K_3
\left[\hat I^{32} \right]^p\left[\hat J^{3}_{12} \right]^\ell,\\
\label{eq:seed_3pt_2}
\langle \phi_{\De_1}\,f^{(0,p)}_{\De_2}\,
\cO^{(\ell,\,\ell+p)}_{\De_3}\,\rangle
&\equiv
K_3\left[\hat{\overline K}{}^{23}_1 \right]^p\left[\hat J^{3}_{12} \right]^\ell,\\
\hat{\overline K}{}^{ij}_k &\equiv
X_{ij}^{+1/2}X_{ik}^{-1/2}X_{jk}^{-1/2}\, (\overline S_i \mathbf{X}_k \overline S_j),
\end{align}
where the tensor invariants $\hat I^{32}$ and $\hat J^{3}_{12}$ are defined in~\eqref{eq:invariant_I} and~\eqref{eq:invariant_J}. It should be understood that $\phi = f^{(0,0)}$ and thus  both~\eqref{eq:seed_3pt_1} and~\eqref{eq:seed_3pt_2} reduce to~\eqref{eq:tensor_structure_scalar_scalar_spin_L} for $p=0$. The structures~\eqref{eq:seed_3pt_1} and~\eqref{eq:seed_3pt_2} satisfy the recursion relations
\begin{align}\label{eq:recursion_4D_second}
\langle\phi_{\Delta_1}\,f^{(p,0)}_{\Delta_2}\,\cO^{(\ell,\ell+p)}_{\Delta_3}\rangle
&=
\left(D^3_{-0+}\cdot \overline D_2^{-+0}\right)
\langle\phi_{\Delta_1}\,f^{(p-1,0)}_{\Delta_2+1/2}\,\cO^{(\ell,\ell+p-1)}_{\Delta_3+1/2}\rangle,\\
\label{eq:recursion_4D_third}
\langle\phi_{\Delta_1}\,f^{(0,p)}_{\Delta_2}\, \cO^{(\ell,\ell+p)}_{\Delta_3}\rangle 
&=
-2\,\mathcal{A}_2^{-1}\times
\left( D^2_{-0+} \cdot \overline D_3^{+0+} \right)
\langle \phi_{\Delta_1}\,f^{(0,p-1)}_{\Delta_2+1/2}\,\cO^{(\ell,\ell+p-1)}_{\Delta_3-1/2}\rangle,\\
\mathcal{A}_2
&\equiv (2\De_3+p-4)(\De_1-\De_2+\De_3+\ell+p-2).
\end{align}

Applying $\mathbf{S}_1$ to~\eqref{eq:recursion_4D_second} and trivially commuting the shadow transform with differential operators, we find
\begin{equation}\label{eq:shadow_seed_1}
S([\phi_{\Delta_1}]f^{(p,0)}_{\Delta_2}\cO^{(\ell,\ell+p)}_{\Delta_3})=
S([\phi_{\Delta_1}]f^{(p-1,0)}_{\Delta_2+1/2}\cO^{(\ell,\ell+p-1)}_{\Delta_3+1/2})=
S([\phi_{\Delta_1}]\phi_{\Delta_2+p/2}\cO^{(\ell,\ell)}_{\Delta_3+p/2}).
\end{equation}
The shadow coefficient in the very last equality is simply given by~\eqref{eq:basic_shadow_coefficient_1}.
Next, applying $\mathbf{S}_2$ of~\eqref{eq:recursion_4D_second}, we find
\begin{equation}
S(\phi_{\Delta_1}[f^{(p,0)}_{\Delta_2}]\cO^{(\ell,\ell+p)}_{\Delta_3})=i\times
\frac{2-\Delta_2+\Delta_{13}+\ell+p}{6-2\Delta_2+p}\;
S(\phi_{\Delta_1}[f^{(p-1,0)}_{\Delta_2+1/2}]\cO^{(\ell,\ell+p-1)}_{\Delta_3+1/2}),
\end{equation}
with solution
\begin{equation}\label{eq:shadow_first_seed_2}
S(\phi_{\Delta_1}[f^{(p,0)}_{\Delta_2}]\cO^{(\ell,\ell+p)}_{\Delta_3})=(-i)^p\times
\frac{
	\left(\frac{\widetilde\Delta_2+\Delta_{13}+\ell-p}{2}\right)_p
}{
	\left(\Delta_2-3-p/2\right)_p
}\times
S(\phi_{\Delta_1}\,[\phi_{\Delta_2+p/2}]\,\cO^{(\ell,\ell)}_{\Delta_3+p/2}),
\end{equation}
where the shadow coefficient on the right-hand side is given by~\eqref{eq:basic_shadow_coefficient_1}.

Finally, consider the structure~\eqref{eq:seed_3pt_2}. Applying $\mathbf{S}_3$ to both sides of~\eqref{eq:recursion_4D_third}, we find
\begin{equation}
\nn
S(\phi_{\Delta_1}f^{(0,p)}_{\Delta_2}[\cO^{(\ell,\ell+p)}_{\Delta_3}])=
-2i\times
\frac{\Delta_3-3+p/2}{\Delta_3-2+p/2}
\times
\frac{\Delta_3-2+\ell+p/2}{\Delta_3-2+\Delta_{12}+\ell+p}
S(\phi_{\Delta_1}f^{(0,p-1)}_{\Delta_2+1/2}[\cO^{(\ell,\ell+p-1)}_{\Delta_3-1/2}]),
\end{equation}
which is solved by
\begin{align}
\nn
&S(\phi_{\Delta_1}f^{(0,p)}_{\Delta_2}[\cO^{(\ell,\ell+p)}_{\Delta_3}])\\
&=
(-i)^p\times
\frac{\Delta_3-2-p/2}{\Delta_3-2+p/2}\,
\frac{
	\left(\Delta_3-1+\ell-p/2\right)_p
}{
	\left(\frac{\Delta_3+\Delta_{12}+\ell-p}{2}\right)_p
}\,
S(\phi_{\Delta_1}\phi_{\Delta_2+p/2}[\cO^{(\ell,\ell)}_{\Delta_3-p/2}]),
\label{eq:shadow_seed_3}
\end{align}
where the shadow coefficient in the right-hand side is given by~\eqref{eq:basic_shadow_coefficient}.

\paragraph{Shadow coefficients for fermion-fermion-spin$(\ell,\ell)$}
We conclude by studying a more complicated example of three-point functions with two tensor structures
\begin{align}
\label{eq:fermion_3pt_1}
\langle \psi^\dag_{\Delta_1}\,\psi_{\Delta_2}\,\cO^{(\ell,\ell)}_{\Delta_3}\rangle_\Omega &=
K_3\left(
\lambda^1_{\<\psi^\dag\psi\cO\>}\hat I^{12} \hat J^3_{12}+
\lambda^2_{\<\psi^\dag\psi\cO\>}\hat I^{13} \hat I^{32}
\right)
\left[\hat J^3_{12}\right]^{\ell-1},\\
\label{eq:fermion_3pt_2}
\langle \psi_{\Delta_1}\,\psi_{\Delta_2}\,\cO^{(\ell,\ell)}_{\Delta_3}\rangle_\Omega &=
K_3\left(
\lambda^1_{\<\psi\psi\cO\>}\hat I^{31} \hat K^{23}_1+
\lambda^2_{\<\psi\psi\cO\>}\hat I^{32} \hat K^{13}_2
\right)
\left[\hat J^3_{12}\right]^{\ell-1},\\
\label{eq:fermion_3pt_3}
\langle \psi_{\Delta_1}\,\psi^\dag_{\Delta_2}\,\cO^{(\ell,\ell)}_{\Delta_3}\rangle_\Omega &=
K_3\left(
\lambda^1_{\<\psi\psi^\dag\cO\>}\hat I^{21} \hat J^3_{12}+
\lambda^2_{\<\psi\psi^\dag\cO\>}\hat I^{23} \hat I^{31}
\right)
\left[\hat J^3_{12}\right]^{\ell-1}.
\end{align}
Here $\psi$ and $\psi^\dag$ are fermions transforming in $(1,0)$ and $(0,1)$ spin representations. The subscript $\<\cdots\>_\Omega$ indicates a physical correlator, which is a linear combination of conformally-invariant structures $\<\cdots\>^{(m)}$ with OPE coefficients $\lambda^m_{\<\cdots\>}$. (Equations~(\ref{eq:fermion_3pt_1}-\ref{eq:fermion_3pt_3}) should be read as definitions of the structures $\<\cdots\>^{(m)}$.) We assume that $\ell\geq 1$.\footnote{In the special case $\ell=0$ both three-point functions have a single tensor structure.}

First, we consider the three-point function~\eqref{eq:fermion_3pt_1}. We define the following vector of weight-shifting operators
\begin{equation}
\label{eq:fermionic_dif_operators}
D^{(n)}_{12}\equiv \left(
\cD^1_{-0+}\cdot\overline \cD_2^{-+0},\quad
\overline\cD_1^{+0+}\cdot \cD^2_{++0}
\right),\quad n=1,2.
\end{equation}
We can then rewrite both tensor structures as
\begin{equation}\label{eq:fermion_bar_fermion_structures_via_differential_operators}
\langle
\psi^\dag_{\Delta_1}\,\psi_{\Delta_2}\,\cO^{(\ell,\ell)}_{\Delta_3}\rangle^{(m)}=
M^{m}_{12}{}_n\,D^{(n)}_{12}\;
\langle\phi_{\Delta_1+3/2-n}\,\phi_{\Delta_2+3/2-n}\,\cO^{(\ell,\ell)}_{\Delta_3}\rangle,
\end{equation}
where the matrix of coefficients $M$ is given by
\begin{equation}
\label{eq:fermions_matrix_differential_basis}
M^{m}_{12}{}_n \equiv
\begin{pmatrix}
1&
0\\
\frac{(\De_1+\De_2+\De_3-\ell-5)(\De_1+\De_2-\De_3+\ell-1)}{4\,\ell\,(\De_3-1)}\;\;\;\;&
-\frac{1}{4\,\ell\,(\De_3-1)\,(\De_1-3/2)(\De_2-3/2)}
\end{pmatrix}.
\end{equation}
Applying the shadow transform $\mathbf{S}_1$ to both sides of~\eqref{eq:fermion_bar_fermion_structures_via_differential_operators}, we get\footnote{Note that in this expression $n$ enters not only the indices of various objects, but also the scaling dimensions of the operators, hence we write the summation sign explicitly.}
\begin{align}
\nn
&S([\psi^\dag_{\Delta_1}]\psi_{\Delta_2}\cO^{(\ell,\ell)}_{\Delta_3})
{}^{m}{}_{n}
\<\psi_{\tl\Delta_1}\psi_{\Delta_2}O^{(\ell,\ell)}_{\Delta_3}\>^{(n)}\\
&=
\sum_{n=1}^2M^{m}_{12}{}_nD^{\prime (n)}_{12}\mathbf{S}_1\;
\<\phi_{\Delta_1+3/2-n}\phi_{\Delta_2+3/2-n}\cO^{(\ell,\ell)}_{\Delta_3}\>.
\label{eq:shadow_transformation_fermions}
\end{align}
On the left-hand side, the tensor structures are defined in~\eqref{eq:fermion_3pt_2}. On the right-hand side, we have permuted the shadow transform with the differential operators which become
\begin{equation}\label{eq:fermion_case_shadow_permuted_with_dif_operators}
D^{\prime (n)}_{12} \equiv
\left(
\mathcal{C}_1(\De_1+1/2,0,0)\times\cD^1_{++0}\cdot\overline \cD^2_{-+0},\quad
-\mathcal{C}_3(\De_1-1/2,0,0)\times\overline\cD_1^{-+0}\cdot \cD_2^{++0}
\right).
\end{equation}
The coefficient $\mathcal{C}_1$ and $\mathcal{C}_3$ are given by~\eqref{eq:commutation1} and~\eqref{eq:commutation3b} respectively. Applying the shadow transform and taking differential operators
in the right-hand side of~\eqref{eq:shadow_transformation_fermions}, we obtain
\begin{equation}\label{eq:fermions_shadow_1_rhs}
\sum_{n=1}^2
S([\phi_{\Delta_1+3/2-n}]\phi_{\Delta_2+3/2-n} \cO^{(\ell,\ell)}_{\Delta_3})\;
M^{m}_{12}{}_n
N^{n}{}_r
\<\psi_{\tl\Delta_1}\psi_{\Delta_2}\cO^{(\ell,\ell)}_{\Delta_3}\>^{(r)},
\end{equation}
where components of the new matrix $N$ are given by
\begin{equation}
N^{11}\equiv i\,\frac{\De_1 + \De_2 - \De_3 - 3 - \ell}{2 \De_1-7},\qquad
N^{12}\equiv i\,\frac{\De_1+\De_2-\De_3-3+\ell}{2 \De_1-7}
\end{equation}
and
\begin{align}
N^{21} &\equiv \frac{i}{4}\,(2 \De_1-5)(2 \De_1-3)(2 \De_2-3) (\De_1 + \De_2 + \De_3 -5-\ell),\\
N^{22} &\equiv \frac{i}{4}\,(2 \De_1-5)(2 \De_1-3)(2 \De_2-3)(\De_1 + \De_2 + \De_3-5+\ell).
\end{align}
Comparing~\eqref{eq:fermions_shadow_1_rhs} with the left-hand side of~\eqref{eq:shadow_transformation_fermions}, we finally conclude
\begin{equation}\label{eq:shadow_coefficient_1_fermions}
S([\psi^\dag_{\Delta_1}]\psi_{\Delta_2}\cO^{(\ell,\ell)}_{\Delta_3}){}^{m}{}_{r}=
\sum_{n=1}^2
S([\phi_{\Delta_1+3/2-n}]\phi_{\Delta_2+3/2-n} \cO^{(\ell,\ell)}_{\Delta_3})
M^{m}_{12}{}_n\;
N^{n}{}_r.
\end{equation}

Second, we consider the shadow coefficients for the $\bS_2$ transformation of the structures~\eqref{eq:fermion_3pt_2}. After analogous manipulations, we obtain the final result 
\begin{align}
\nn
S(\psi_{\Delta_1}[\psi_{\Delta_2}]\cO^{(\ell,\ell)}_{\Delta_3}){}^{m}{}_{n} &=
\mathcal{F}\times
\begin{pmatrix}
0 & \frac{1}{\De_{12}-\De_3+\ell+3} \\
\frac{1}{\De_{21}+\De_3+\ell-1} & \frac{\De_{12}-\De_3+\ell+1}{(\De_{12}-\De_3+\ell+3)(\De_{21}+\De_3+\ell-1)}
\end{pmatrix},\\
\mathcal{F} &\equiv -2i\pi^2\times
\frac{\Gamma(\De_2-3/2)}{\Gamma(\De_2-9/2)}
\times
\frac
{\Gamma(\frac{-\De_1+\De_{32}+\ell+4}{2})\Gamma(\frac{\De_{12}-\De_3+\ell+5}{2})}
{\Gamma(\frac{\De_1+\De_{23}+\ell}{2})\Gamma(\frac{\De_{21}+\De_3+\ell-1}{2})}.
\label{eq:shadow_coefficient_2_fermions}
\end{align}

Third, we apply $\mathbf{S}_3$ to both sides of~\eqref{eq:fermion_3pt_3}. We can use the differential operators~\eqref{eq:fermionic_dif_operators} with $1\leftrightarrow 2$, which we denote by $D_{21}^{n}$. The corresponding matrix $M_{21}$ is obtained from~\eqref{eq:fermions_matrix_differential_basis} by adding minuses in the second row. The shadow coefficient is much easier in this case since $\mathbf{S}_3$ commutes trivially with all the differential operators. We have
\begin{align}
\nn
&S(\psi_{\Delta_1}\psi^\dag_{\Delta_2}[\cO^{(\ell,\ell)}_{\Delta_3}])
{}^{m}{}_{n}
\<\psi_{\Delta_1}\psi^\dag_{\Delta_2}\cO^{(\ell,\ell)}_{\tl\Delta_3}\>^{(n)}\\
&=M_{21}^{m}{}_nD_{21}^{n}\mathbf{S}_3
\<\phi_{\Delta_1+3/2-n}\phi_{\Delta_2+3/2-n}\cO^{(\ell,\ell)}_{\Delta_3}\>.
\end{align}
Applying the shadow transform and taking derivatives we arrive at the final result
\begin{equation}\label{eq:shadow_coefficient_3_fermions}
S(\psi_{\Delta_1}\psi^\dag_{\Delta_2}[\cO^{(\ell,\ell)}_{\Delta_3}]){}^{m}{}_{r}=
\sum_{n=1}^2
S(\phi_{\Delta_1+3/2-n}\phi_{\Delta_2+3/2-n} [\cO^{(\ell,\ell)}_{\Delta_3}])
M_{21}^{m}{}_n\times
\left(M_{21}^{-1}\big|_{\De_3\rightarrow \widetilde\De_3}\right){}^{n}{}_r.
\end{equation}

\subsection{OPE coefficients in MFT}
\label{sec:mft}
Having computed the Plancherel measure, shadow coefficients, and three-point pairings, we can plug these ingredients into (\ref{eq:mftcoeffs}) to write explicit formulas for MFT OPE coefficients. Below we consider several examples.

\subsubsection{Scalar MFT}
\label{sec:scalarmft}

The partial wave coefficients for scalar MFT are given by
\be
I^\mathrm{MFT}(\De,J) &= \frac{\mu(\De,J)}{(\<\f\f\tl \cO_J\>,\<\tl \f \tl \f \cO_J\>)} S([\f_{\tl \De_1}]\f_{\tl \De_2} \cO_{\De,J}) S(\f_{\De_1}[\f_{\tl \De_2}]\cO_{\De,J}),
\ee
where the $S(\cdots)$ coefficients are given in (\ref{eq:scalarshadowcoeff}) and the three-point pairing is given in (\ref{eq:scalarscalarspinjpairing}).
The conformal block coefficients are minus the residues of
\be
&I^\mathrm{MFT}(\De,J)S(\f_{\De_1}\f_{\De_2}[\cO_{\tl \De,J}] )\nn\\
&=  \frac{2^{J-1} \G (\Delta -1) \G (\frac{d}{2}-\Delta_{1}) \G (\frac{d}{2}-\Delta_{2}) \G (\frac{d}{2}+J) \G (d+J-\Delta )
}{
\G (\Delta_{1}) \G (\Delta_{2}) \G (J+1) \G (\Delta -\frac{d}{2}) \G (\Delta+J -1)
}
\nn\\
& \qquad\x
\frac{
\G (\frac{\Delta+J +\Delta_{1}-\Delta_{2}}{2})
\G (\frac{\Delta+J -\Delta_{1}+\Delta_{2}}{2})
\G (\frac{J-\Delta +\Delta_{1}+\Delta_{2}}{2})
\G (\frac{\Delta+J-d +\Delta_{1}+\Delta_{2}}{2})
}{
\G (\frac{d-\Delta+J +\Delta_{1}-\Delta_{2}}{2})
\G (\frac{d-\Delta+J -\Delta_{1}+\Delta_{2}}{2})
\G (\frac{2 d+J-\Delta -\Delta_{1}-\Delta_{2}}{2})
\G (\frac{\Delta+J+d -\Delta_{1}-\Delta_{2}}{2})
},
\label{eq:mftthingy}
\ee
where $S(\f_{\De_1}\f_{\De_2}[\cO_{\tl \De,J}] )$ is given in (\ref{eq:spinningshadowcoefficient}).

The above expression has poles at $\De=\De_1+\De_2+J+2n$ (as expected in MFT), coming from the factor $\G (\frac{J-\Delta +\Delta_{1}+\Delta_{2}}{2})$. However, it has additional unphysical poles to the right of the principal series that deserve comment.

Firstly, the expression (\ref{eq:mftthingy}) has a set of poles at $\De=J+d-1+k$ for $k=1,2,\dots,\oo$, coming from the factor $\G(d+J-\De)$. These ``spurious" poles cancel against poles in the conformal blocks when we continue the $\De$-contour from the principal series to the real axis, as proven in \cite{Simmons-Duffin:2017nub}.

Secondly, the expression (\ref{eq:mftthingy}) has additional poles at $\De=\pm(\De_2-\De_1)-J+1-k$ for $k=1,2,\dots,\oo$. These do not contribute to the conformal block expansion if $\De_1,\De_2$ are on the principal series.  Naively, they would contribute when we allow $\De_1,\De_2$ to become real with $|\De_2-\De_1|$ sufficiently large. However, they are ultimately not present in the conformal block decomposition (as expected because MFT does not contain such operators). This can be understood as follows. Consider the partial wave expansion of MFT where $\De_1,\De_2$ start on the principal series, and then analytically continue $\De_1,\De_2$ away from the principal series. When we continue, new poles may cross into the right $\De$-half-plane, but the $\De$ contour should be deformed to avoid them. For more general correlators, we can imagine analytically continuing the completeness relation for conformal partial waves as described in \cite{Simmons-Duffin:2017nub}. Perhaps this phenomenon can alternatively be understood in terms of the appearance of new normalizable eigenstates of the Casimirs when $\De_1,\De_2$ are real, whose contributions cancel against poles in (\ref{eq:mftthingy}).

To summarize, the physical conformal block coefficients are given by
\be\label{eq:scalar_MFT_coefficients}
&P^\textrm{MFT}(\De_1,\De_2,n,J) \nn\\ 
&= -\textrm{Res}_{\De=\De_1+\De_2+2n+J}\, I^\mathrm{MFT}(\De,J)S(\f_{\De_1}\f_{\De_2}[\cO_{\tl \De,J}] ) \nn\\
&= \frac{(-1)^{n}2^{J} \Gamma (\frac{d}{2}-\Delta_{1}) \Gamma (\frac{d}{2}-\Delta_{2}) \Gamma (\frac{d}{2}+J) \Gamma (J+n+\Delta_{1}) \Gamma (J+n+\Delta_{2}) 
}{
\Gamma (\Delta_{1}) \Gamma (\Delta_{2}) \Gamma (J+1) \Gamma (n+1) \Gamma (\frac{d}{2}+J+n) \Gamma (\frac{d}{2}-n-\Delta_{1}) \Gamma (\frac{d}{2}-n-\Delta_{2})}
 \nn\\
&\qquad \x \frac{\Gamma (d-2 n-\Delta_{1}-\Delta_{2}) \Gamma (J+2 n+\Delta_{1}+\Delta_{2}-1) \Gamma (-\frac{d}{2}+J+n+\Delta_{1}+\Delta_{2})
}{
 \Gamma (d-n-\Delta_{1}-\Delta_{2}) \Gamma (2 J+2 n+\Delta_{1}+\Delta_{2}-1) \Gamma (-\frac{d}{2}+J+2 n+\Delta_{1}+\Delta_{2})
}.
\ee
In our conventions, the leading term in the OPE limit of the conformal blocks is given by (\ref{eq:blockopelimit}), which in this case becomes
\be
G_{\De,J}^{\De_i}(0,x,e,\oo) &\sim (-1)^J(e^{\mu_1}\cdots e^{\mu_J} - \textrm{traces}) R_{\mu_1\cdots\mu_J}{}^{\nu_1\cdots\nu_J}(\hat x)|x|^{\De-\De_1-\De_2} (e_{\nu_1}\cdots e_{\nu_J} - \textrm{traces}) \nn\\
&= (-1)^J(e^{\mu_1}\cdots e^{\mu_J} - \textrm{traces}) (\hat x_{\mu_1}\cdots \hat x_{\mu_J} - \textrm{traces}) |x|^{\De-\De_1-\De_2}
\nn\\
&= (-1)^J \hat C_J(e\.\hat x) |x|^{\De-\De_1-\De_2},
\ee
where $\hat C_J(x)$ is defined in (\ref{eq:chat}).
The $(-1)^J$ comes about because the two-point function 
$\<\cO^{\mu_1\cdots \mu_J}(0) \cO_{\nu_1\cdots \nu_J}(e)\>$
given in~(\ref{eq:twoptconvnentionflatspace}) is the matrix implementing a reflection in the $e$ direction. In cross-ratio space, we have\footnote{A different convention for two- and three-point structures was used in \cite{Kravchuk:2018htv} because it simplified several formulas in Lorentzian signature. In the conventions of \cite{Kravchuk:2018htv}, the leading term of the conformal blocks is $z^{\frac{\De-J}{2}} \bar z^{\frac{\De+J}{2}}$, and the shadow transform coefficients have extra $(-2)^J$'s in them.}
\be
G_{\De,J}(z,\bar z) &\sim (-1)^J (z\bar z)^{\frac {\De}{2}} \hat C_J\p{\frac{z+\bar z}{2\sqrt{z \bar z}}} + \dots & & (z\bar z \ll 1) \nn\\
&
\sim \p{-\tfrac 1 2}^J z^{\frac{\De-J}{2}} \bar z^{\frac{\De+J}{2}} + \dots & & (z\ll \bar z \ll 1).
\ee
We find that (\ref{eq:scalar_MFT_coefficients}) agrees with the result of \cite{Fitzpatrick:2011dm}, after taking into account different normalization conventions for conformal blocks,
\be
P^\textrm{MFT}_\textrm{here} &= \p{\frac{(d-2)_J}{2^J(\frac{d-2}{2})_J}}^{-1} P^\textrm{MFT}_{\mathrm{there}}.
\ee

\subsubsection{Seed correlator in 4d}

Consider the so called 4d seed correlator~\cite{Echeverri:2016dun}. In MFT it is given by
\begin{equation}
\label{eq:mft_seed_correlator}
\langle
\phi(\point_1)\,
f^{(p,0)}(\point_2)\,
\phi(\point_3)\,
f^{(0,p)}(\point_4)
\rangle=
\langle
\phi (\point_1)\,
\phi (\point_3)\,
\rangle\langle
f^{(p,0)} (\point_2)\,
f^{(0,p)} (\point_4)
\rangle.
\end{equation}
For brevity we simply denote $f^{(p,0)}$ and $f^{(0,p)}$ by $f$ and $f^\dag$ respectively.
We can decompose the left-hand side of~\eqref{eq:mft_seed_correlator} in the $s$-channel. The exchanged operators are double-twist operators of the schematic form
\begin{equation}\label{eq:seed_double_twist_operator}
\cO(x)\sim \partial^{2n}\; \partial^\ell\; \phi(x)\,f^{(0,p)}(x),
\end{equation}
together with their duals. Here both $n$ and $\ell$ are non-negative integers. The operator~\eqref{eq:seed_double_twist_operator} is in a reducible representation of the rotation group. It decomposes into $p+1$ irreducible representations as
\begin{equation}\label{eq:representation_structure}
(\ell,\ell)\otimes(0,0)\otimes(0,p)=
(\ell,\ell-p)\oplus(\ell,\ell-p+2)\oplus\ldots\oplus(\ell,\ell+p-2)\oplus
(\ell,\ell+p).
\end{equation}
From now on we focus only on the rightmost operator in~\eqref{eq:representation_structure}, which we denote by
\begin{equation}\label{eq:seed_double_twist_operator_marginal}
\cO^{(\ell,\ell+p)}_{\Delta^*},\quad
\De^*\equiv\De_\phi+\De_f+2n+\ell.
\end{equation}
Using~\eqref{eq:mftcoeffsII},~\eqref{eq:contour_integral}, and~\eqref{eq:physical spectrum_from_contour_integral} we can write the final coefficient encoding the physical spectrum
\begin{align}
\nn
C^{\textrm{MFT}}_{
\<\phi f \cO\>
\<\phi f^\dag \cO^\dag\>}
&=
\mu(\De,\ell,\ell+p)
\p{
	\<\phi_{\De_\phi} f_{\De_f} \cO_{\De}\>,\,
	\<\phi_{\tl\De_\phi} f_{\tl\De_f} \cO_{\tl\De}\>
}^{-1}\\
&\times
S([\phi_{\tl\De_\phi}]f_{\tl\De_f}\cO_{\tl\De})\;
S(\phi_{\De_\phi}[f_{\tl\De_f}]\cO_{\tl\De})\;
S(\phi_{\De_\phi} f^\dag_{\De_f} [\cO_{\tl\De}]).
\end{align}

Plugging the Plancherel measure~\eqref{eq:plancherel_measure_4d}, the three-point pairing~\eqref{eq:seed_pairing_4d}, the shadow coefficients~\eqref{eq:shadow_seed_1}, \eqref{eq:shadow_first_seed_2} and~\eqref{eq:shadow_seed_3} and taking into account~\eqref{eq:relation_volumes_SO}, we get
\begin{align}\nn
C^{\textrm{MFT}}_{\<\phi f \cO\>\<\phi f^\dag \cO^\dag\>}
&=
\frac{1}{2}\,
(-1)^{p+1}(\ell+1)\left(\Delta +p/2-2\right)
\\\nn
&\times
\frac{\Gamma (2-\De_\phi)}{\Gamma (\De_\phi)}
\frac{\Gamma \left(1-\De_f-p/2\right)}{ \Gamma\left(\De_f-1-p/2\right)}
\frac{
	\Gamma \left(4-\Delta+\ell+p/2\right)
}{
	\Gamma \left(\Delta-1+\ell+p/2\right)
}\\\nn
&\times
\frac{
	\Gamma \left(\frac{\De_\phi+\De_f-\Delta+\ell}{2}\right)
	\Gamma \left(\frac{\De_\phi-\De_f+\Delta+\ell}{2}\right)
}{
	\Gamma\left(\frac{4-\De_\phi-\De_f+\Delta+\ell}{2}\right) \Gamma \left(\frac{4-\De_\phi+\De_f-\Delta+\ell}{2}\right)
}\\
\label{eq:C_coefficients}
&\times
\frac{
	\Gamma \left(\frac{-\De_\phi + \De_f + \Delta + \ell + p}{2}\right)
	\Gamma \left(\frac{-4 + \De_\phi + \De_f + \Delta + \ell + p}{2}\right)
}{
	\Gamma\left(\frac{4 + \De_\phi - \De_f - \Delta + \ell  + p}{2}\right) \Gamma \left(\frac{8 - \De_\phi - \De_f - \Delta + \ell + p}{2}\right)
}.
\end{align}
The poles $\De^*$ reproducing the spectrum of our operator~\eqref{eq:seed_double_twist_operator_marginal} come from the factor
\begin{equation}
	\Gamma \left(\frac{\De_\phi+\De_f-\Delta+\ell}{2}\right).
\end{equation}
The OPE coefficients are
\begin{align}\nn
\lambda_{\<\phi f \cO\>}\lambda_{\<\phi f^\dag \cO^\dag\>}
&= -\Res_{\De^*} C^{\textrm{MFT}}_{\<\phi f \cO\>\<\phi f^\dag \cO^\dag\>}
\nn\\
\nn
&=\frac{(-1)^{n+p+1}(\ell+1)(\De_\phi+\De_f+2n+\ell-2+p/2)
}{
\Gamma(n+1)\Gamma(n+\ell+2)}\\\nn
&\times
\frac{\Gamma(2-\De_\phi)}{\Gamma(\De_\phi)}\frac{\Gamma(\De_\phi+n+\ell)}{\Gamma(2-\De_\phi-n)}
\frac{\Gamma(\De_f+n+\ell+p/2)}{\Gamma(2 - \De_f - n + p/2)}
\frac{\Gamma(1-\De_f-p/2)}{\Gamma(\De_f-1-p/2)}\\
\label{eq:seed_MFT_coefficients}
&\times
\frac{\Gamma(4 - \De_\phi - \De_f - 2 n + p/2)}{\Gamma(4 - \De_\phi - \De_f - n + p/2)}
\frac{\Gamma(-2 + \De_\phi + \De_f + n + \ell + p/2)}{\Gamma(-1 + \De_\phi + \De_f + 2 n + 2 \ell + p/2)}.
\end{align}
For $p=0$ the expression~\eqref{eq:seed_MFT_coefficients} is equal to the scalar one~\eqref{eq:scalar_MFT_coefficients} for $d=4$ and $J=\ell$ times an overall  $2^{-\ell}$ factor which is a matter of normalization in two different formalisms.\footnote{More precisely the difference comes from the three-point pairing. See~\eqref{eq:scalar_three_point_pairing_4d} and the discussion below.}
For $p=1$~\eqref{eq:seed_MFT_coefficients} matches the first formula in (3.56) in~\cite{Elkhidir:2017iov}.\footnote{To match the result one needs to compare definitions of tensor structures here and there. For $p=1$ one finds that
$
\lambda^{\text{here}}_{\<\phi f \cO\>}\lambda^{\text{here}}_{\<\phi f^\dag \cO^\dag\>}=
-\lambda^{\text{there}}_{\<\phi f \cO\>}\lambda^{\text{there}}_{\<\cO^\dag \phi f^\dag\>} =
- P_{\overline{\mathcal{X}}\mathcal{X}}$.
}

\subsubsection{Fermionic correlator in 4d}
Finally, consider a four-fermion correlator
\begin{align}
\nn
&\<
\psi^\dag(\point_1)\psi(\point_2)\psi(\point_3)\psi^\dag(\point_4)
\>\\
&=
\<\psi^\dag(\point_1)\psi(\point_2)\>
\<\psi(\point_3)\psi^\dag(\point_4)\>
-
\<\psi^\dag(\point_1)\psi(\point_3)\>
\<\psi(\point_2)\psi^\dag(\point_4)\>.
\label{eq:fermionic_correlator}
\end{align}
As before, $\psi$ transforms in the representation $(1,0)$ and $\psi^\dag$ in the representation $(0,1)$. We for simplicity consider only one species of $\psi$.
In~\eqref{eq:fermionic_correlator} the minus sign comes from permuting two fermionic operators.
We decompose~\eqref{eq:fermionic_correlator} in the s-channel. The exchanged operators are the identity, which reproduces the first term in the right-hand side of~\eqref{eq:fermionic_correlator}, and double-twist bosonic operators of the schematic form
\begin{equation}\label{eq:fermionic_double_twist_operator}
\cO(x)\sim \partial^{2n}\; \partial^\ell\; 
\psi^\dag(x)\,\psi(x),\qquad n,\ell \geq 0
\end{equation}
which reproduce the second term in the right-hand side of~\eqref{eq:fermionic_correlator}. 
The set of operators~\eqref{eq:fermionic_double_twist_operator} is in a reducible representation of the rotation group. They decompose into irrducibles as follows
\begin{equation}\label{eq:representation_structure_fermions}
(\ell,\ell)\otimes(0,1)\otimes(1,0)=
(\ell-1,\ell-1)\oplus
(\ell-1,\ell+1)\oplus
(\ell+1,\ell-1)\oplus
(\ell+1,\ell+1).
\end{equation}
Here, we focus on the traceless symmetric operators only (the leftmost and the rightmost operators in~\eqref{eq:representation_structure_fermions}). Let us shift their spins for convenience and denote them as follows
\begin{align}
\cO_1\equiv \cO_{\De_1}^{(\ell,\ell)},\quad\De_1 &=2\De_\psi+2n+\ell+1,\quad \ell\geq 0,\\
\cO_2\equiv \cO_{\De_2}^{(\ell,\ell)},\quad\De_2 &=2\De_\psi+2n+\ell-1,\quad \ell\geq 1.
\end{align}
It is important to note that these operators have a degenerate spectrum since
\begin{equation}\label{eq:fermionic_degeneracy}
\De_1(\ell,\,n)=\De_2(\ell,\,n+1),\quad n\geq 0,\quad \ell\geq 1.
\end{equation}
We thus cannot generically distinguish their contributions to the four-point function (apart from boundary cases: $\cO_1$ with $\ell=0,\;\forall n$ and $\cO_2$ with $\forall\ell\geq 1,\;n=0$). In other words, we compute the total contribution to the conformal block expansion from operators with scaling dimension $\De^*$ given by
\begin{equation}\label{eq:spectrum_fermions}
\De^*=2\De_\psi+2n+\ell-1,\quad \ell\geq 1.
\end{equation}
This equation defines the meaning of $n$ in all what follows. For simplicity we will ignore the case $\ell=0$.

According to~\eqref{eq:mftcoeffsII},~\eqref{eq:contour_integral}, and~\eqref{eq:physical spectrum_from_contour_integral} we can write the final coefficient encoding the physical spectrum
\begin{align}
\nn
C^{\mathrm{MFT}}_{ae\,\<\psi^\dag\psi\cO\>\<\psi\psi^\dag\cO^\dag\>}
&=-\mu(\De,\ell,\ell)
\p{
\<\psi^\dag_{\De_\psi}\psi_{\De_\psi}\cO_{\De}\>^{(a)},
\<\psi^\dag_{\tl\De_\psi}\psi_{\tl\De_\psi}\cO_{\tl\De}\>^{(b)}}^{-1}\\
&\times
S([\psi^\dag_{\tl\De_\psi}]\psi_{\tl\De_\psi}\cO_{\tl\De})^b{}_c\,
S(\psi_{\De_\psi}[\psi_{\tl\De_\psi}]\cO_{\tl\De})^c{}_d
S(\psi_{\De_\psi}\psi^\dag_{\De_\psi}[\cO_{\tl\De}])^d{}_e.
\label{eq:I_fermions_4d}
\end{align}
Note the extra minus in the formula above compared to~\eqref{eq:mftcoeffsII}. This minus comes from~\eqref{eq:fermionic_correlator} since it is not present in~\eqref{eq:basicgff}. We get the final expression for the coefficient $C^{\mathrm{MFT}}$ by plugging the three-point pairing~\eqref{eq:three_point_pairing_fermions} and the shadow coefficients~\eqref{eq:shadow_coefficient_1_fermions}, \eqref{eq:shadow_coefficient_2_fermions} and~\eqref{eq:shadow_coefficient_3_fermions} and using~\eqref{eq:relation_volumes_SO}.

The final explicit form of~\eqref{eq:I_fermions_4d} has poles corresponding to the physical spectrum~\eqref{eq:spectrum_fermions} and can be used to extract the products of OPE coefficients as minus residues  of these poles
\begin{equation}
\label{eq:product_ope_coefficients_4d}
\lambda^a_{\<\psi^\dag\psi O\>}
\lambda^b_{\<\psi\psi^\dag\cO\>}=
-\Res_{\De^*}
C^{\mathrm{MFT}}_{ab\,\<\psi^\dag\psi\cO\>\<\psi\psi^\dag\cO^\dag\>}
= -2\Res_{n}
C^{\mathrm{MFT}}_{ab\,\<\psi^\dag\psi\cO\>\<\psi\psi^\dag\cO^\dag\>}.
\end{equation}
Let us make a simple rotation of the coefficients~\eqref{eq:product_ope_coefficients_4d} and define the following object
\begin{equation}
P^{ac}\equiv
\lambda^a_{\<\psi^\dag\psi O\>}
\lambda^b_{\<\psi\psi^\dag\cO\>}
R^{bc},\qquad
R\equiv
\begin{pmatrix}
+1 & 0\\
0  & -1
\end{pmatrix}.
\end{equation}
We will now write the final answer for the coefficients $P^{ac}$.\footnote{This is done for an easy comparison of the results obtained here with the ones in~\cite{Karateev:2017}. Looking at their definition of structures one has $P^{ac}=
(-1)^{\ell+1}\times
\left(\lambda^a_{\<\psi^\dag\psi\cO\>}
\lambda^c_{\<\cO\psi^\dag\psi\>}\right)_{\text{there}}$. }

In the boundary case of $n=0$ only the $C^{22}$ component develops a pole corresponding to~\eqref{eq:spectrum_fermions}. The associated product of OPE coefficients is
\begin{equation}
P^{22}=8\sqrt\pi\;\frac{4^{-\De_\psi-\ell}}{(\ell-1)!}
\frac{\Gamma\left(\De_\psi-1/2+\ell\right)\Gamma\left(2\De_\psi-1+\ell\right)}
{\Gamma\left(\De_\psi+1/2\right)\Gamma\left(\De_\psi-1+\ell\right)}.
\end{equation}
In the generic case of $n\geq 1$, all four components of $C^{mn}$ develop poles associated to~\eqref{eq:spectrum_fermions}. The related products of OPE coefficients are
\begin{align}
P^{11} &=S\times\left((\ell+n+1)-
\frac{\ell+1}{\ell}\;(2\De_\psi-3+n)(2\De_\psi-4+\ell+2n)
\right),\\
P^{12}&=P^{21}=S\times(2\De_\psi-5+n)(2\De_\psi-2+\ell+2n),\\
P^{22} &=-S\times\frac{1}{n}\;(2\De_\psi-2+\ell+2n)
\left((2\De_\psi-4+\ell)(2+\ell+2n)+2n(n+1)\right),
\end{align}
where we have defined $S$ to be
\begin{align}\nn
S &\equiv 128\pi^2\;\frac{(-1)^{n+1}\, 4^{-2\De_\psi-\ell-2n}}{(n-1)!(\ell+n+1)!}\;
\frac{\ell(\ell+1)}{\ell+2}\;
\frac{2\De_\psi-3+\ell+2n}{2\De_\psi-4+\ell+2n}\;
\frac{\sec\left(\pi\,\De_\psi\right)^2}{\Gamma\left(\De_\psi-3/2\right)^2\Gamma\left(\De_\psi+1/2\right)^2}\\
&\times
\frac{\Gamma\left(\De_\psi-1/2+\ell+n\right)\Gamma\left(2\De_\psi-3+\ell+n\right)}
{\Gamma\left(\De_\psi-1+\ell+n\right)}\;
\frac{\Gamma\left(3-\De_\psi-n\right)}
{\Gamma\left(5-2\De_\psi-n\right)\Gamma\left(5/2-\De_\psi-n\right)}.
\end{align}

The coefficients $P^{ab}$ have been guessed in~\cite{Karateev:2017} by using explicit expressions for conformal blocks. Our result looks apparently different but is in fact in a perfect agreement with the one of~\cite{Karateev:2017}.

\section{Shadow coefficients from Fourier transforms}
\label{eq:shadowfourier}

In this section, we discuss a general approach to the shadow transform based on going through momentum space. The advantage of this approach is that it gives a straightforward algorithm for computing arbitrary shadow coefficients from representation-theoretic data, which will allow us to compute MFT OPE coefficients for rather complicated correlators, such as $\<JJJJ\>$ and $\<TTTT\>$ in 3d.

The shadow transform of a correlation function $\<\cO^a(x)\cdots\>$ is defined as
\be
\int d^d y \<\widetilde\cO^a(x) \widetilde\cO^\dagger_b(y)\>\<\cO^b(y)\cdots\>.
\ee
Due to translation invariance we have
\be\label{eq:KPositionDefinition}
	\<\widetilde\cO^a(x) \widetilde\cO^\dagger_b(y)\>=K^a{}_b(x-y),
\ee
and thus the shadow transform is a convolution with the kernel $K$. The basic idea now is to recall that a convolution in position space is equivalent to multiplication in momentum space. In particular, if we define
\be
	\cO^b(p)&\equiv\int d^dx\, e^{-ipx}\cO^b(x),\\
	K^a{}_b(p)&\equiv \int d^dx\, e^{-ipx} K^a{}_b(x),\label{eq:Kshadowfourier}
\ee
then the shadow-transformed correlator in momentum space is given by
\be
	K^a{}_b(p)\<\cO^b(p)\cdots\>.
\ee
We can then perform the shadow transform by first doing a Fourier transform, then multiplying by the Fourier transform of the two-point function, and finally Fourier-transforming back. 

This observation is only useful if the Fourier transform is easier to perform than the shadow transform. In this work, we are mainly interested in shadow transforms of three-point functions. Three-point functions in momentum space have been studied in~\cite{Bzowski:2013sza,Bzowski:2015pba,Bzowski:2017poo,Bzowski:2018fql} and are in general quite complicated if all three-operators are taken to be in momentum space. However, for our purposes it is sufficient to Fourier-transform only one of the operators, and we are free to choose the kinematics for the other two. We will see shortly that in a particular choice of kinematics both the direct and the inverse Fourier transforms are straightforward to perform on general three-point functions. This leads to an efficient algorithm for computation of  shadow coefficients.

\subsection{An algorithm for computing shadow transform}
\label{sec:shadowfourieralgorithm}

\subsubsection{Two-point function in momentum space}

Our first goal is to understand the kernel $K^a{}_b(p)$, which is the Fourier-transform of the two point function
\be
	K^a{}_b(p)=\int d^d x e^{-ipx}\<\tl\cO^{a}(x)\tl\cO_b^\dagger(0)\>.
\ee

Let us first understand the constraints imposed on $K^a{}_b(p)$ by $\SO(d)$ invariance. Recall that in $K^a{}_b(p)$, the index $b$ transforms in $\rho^*$, while the index $a$ transforms in $\rho^R$, where $\rho$ is the $\SO(d)$ representation of $\cO$. Therefore, for each $p$ the kernel $K$ is a map $\rho\to \rho^R$. This map has to be invariant under the $\SO(d-1)$ which stabilizes $p$. 

Such maps are easy to classify. Indeed, any irreducible $\rho$ decomposes into a direct sum of $\SO(d-1)$ representations $\lambda_{d-1}$ without multiplicities~\cite{RepresentationsAndSpecialFunctions},
\be
	\rho\simeq\bigoplus_{\lambda_{d-1}\in \rho}\,\lambda_{d-1}\qquad (\text{as a }\SO(d-1)\text{ representation}),
\ee
and the same holds for $\rho^R$. Moreover, the $\SO(d-1)$ decompositions of $\rho^R$ and $\rho$ coincide, since one can take the reflection direction to be the one preserved by $\SO(d-1)$,\footnote{In other words, the outer automorphism of $\SO(d)$ induced by the reflection can be taken to act trivially on $\SO(d-1)$.}
\be
\rho^R\simeq\bigoplus_{\lambda_{d-1}\in \rho}\,\lambda_{d-1}\qquad (\text{as a }\SO(d-1)\text{ representation}).
\ee
By Schur's lemma for $\SO(d-1)$ irreps and scaling invariance, we must have
\be
	K(p)=\sum_{\lambda_{d-1}\in \rho} \cA_{\lambda_{d-1}} \Pi_{\lambda_{d-1}} (p) p^{2\tl\De-d},
\ee
where $\Pi_{\lambda_{d-1}} (p)$ is the unique\footnote{Unique up to a multiplicative factor. One can implement $\rho^R$ so that $\rho$ and $\rho^R$ are realized in the same vector space and coincide as representations of $\SO(d-1)$ which preserves some standard momentum $q$. In this case one can canonically normalize $\Pi_{\lambda_{d-1}}(q)$ to be a projector, and then extend this to all $p$ by $\SO(d)$ invariance, see below.} $\SO(d-1)$-invariant map which maps $\lambda_{d-1}$ in $\rho$ to $\lambda_{d-1}$ in $\rho^R$ and everything else to $0$, $\cA_{\lambda_{d-1}}$ are some yet undetermined coefficients. Here the argument $p$ in $\Pi_{\lambda_{d-1}} (p)$ means that it is invariant under the $\SO(d-1)$ which preserves $p$. We have
\be
	\Pi_{\lambda_{d-1}} (p) = R(p) \Pi_{\lambda_{d-1}} (q) R^{-1}(p),
\ee
where $q$ is some standard momentum and $R(p)$ is any $\SO(d)$ rotation which takes $p$ to $q$. Note that $R$ is represented in $\rho$ on the right and $\rho^R$ on the left.

The problem of computing $K(p)$ is now reduced to the problem of determining the numbers $\cA_{\lambda_{d-1}}$. Up to an overall coefficient these numbers are fixed by requiring invariance under special conformal transformations. In this way, $\cA_{\lambda_{d-1}}$ have been determined for general $\rho$ in 4d CFTs in~\cite{Mack:1975je}. In appendix~\ref{app:twoptfourier}, we compute $\cA_{\lambda_{d-1}}$ in 3d CFTs. Here, we merely state the 3d result.

The $\SO(3)$ representations are labeled by non-negative (half-)integer spin $j$, and $\SO(2)$ representations are labeled by (half-)integer spin $m$. We have $j^R=j$ and
\be
	j=\bigoplus_{m=-j}^j m.
\ee
The corresponding decomposition of $K$ is then
\be
	K(p) = \sum_{m=-j}^j \cA_{j,m} \Pi_m(p)p^{2\tl\De-3},
\ee
where
\be\label{eq:AjmMain}
	\cA_{j,m}=i^{-2m}\frac{4\pi\sin\pi(\tl\De+j)\Gamma(2-2\tl\De)(\tl\De-1)}{\tl\De+j-1}\frac{\Gamma(\tl\De+m-1)\Gamma(\tl\De-m-1)}{\Gamma(\tl\De+j-1)\Gamma(\tl\De-j-1)}.
\ee
Here $\tl\De$ and $j$ are the scaling dimension and spin of $\tl\cO$.

\subsubsection{Three-point functions in momentum space}

The next step is to understand the Fourier transform of three-point functions. Let us first consider a simple example, the three-point function 
\be
	\<\phi_1(x_1) \cO_2^{\mu_1\ldots \mu_J}(x_2) \phi_3(x_3)\>.
\ee
Here $\phi_i$ are scalars and $\cO_2$ is a traceless-symmetric spin-$J$ tensor. This three-point function has only one conformally-invariant tensor structure, and to write it down it is convenient to set $x_1=0$ and $x_3=\infty$. We then find
\be
	\<\phi_1(0) \cO_2^{\mu_1\ldots \mu_J}(x) \phi_3(\infty)\>=(x^{\mu_1}\cdots x^{\mu_J}-\text{traces})x^{-\De_1-\De_2+\De_3-J}.
\ee
This form is fixed completely by $\SO(d)$ and scale invariance. It is obvious that the Fourier transform of this tensor structure is of the same form,
\be
	\<\phi_1(0) \cO_2^{\mu_1\ldots \mu_J}(p) \phi_3(\infty)\>=\cF_{\De_1+\De_2-\De_3,J}(p^{\mu_1}\cdots p^{\mu_J}-\text{traces})p^{\De_1+\De_2-\De_3-J-d},
\ee
where the constant $\cF_{\lambda,J}$ is defined by
\be\label{eq:Fcoefficient}
	\cF_{\l,J}(p^{\mu_1}\cdots p^{\mu_J}-\text{traces})p^{\l-J-d}
	=\int d^d x e^{-ipx}(x^{\mu_1}\cdots x^{\mu_J}-\text{traces})x^{-\l-J}.
\ee
We can compute $\cF_{\l,J}$ by setting $p=(1,0,\ldots)$ and contracting all indices on both sides with a null vector $z=(1,i,0,\ldots)$. The result is
\be\label{eq:fouriercoefficient}
	\cF_{\l,J}
	&=i^{-J}2^{d-\l}\pi^{\frac{d}{2}}\frac{  \Gamma\p{\frac{d+J-\l}{2}}}{
		\Gamma\p{\frac{\l+J}{2}} }.
\ee
This in fact turns out to be the only Fourier transform we ever need to compute.

Indeed, consider the most general three-point function
\be\label{eq:general3ptCF}
	\<\cO_1^a(0)\cO_2^b(x)\cO_3^c(\infty)\>
\ee
with operators $\cO_i$ in $\SO(d)$ representations $\rho_i$. The allowed tensor structures are in one-to-one correspondence~\cite{Mack:1976pa,Kravchuk:2016qvl} with $\SO(d-1)$-invariants in\footnote{We write $\rho_3^R$ because this operator has been put at infinity. This is irrelevant from the point of view of $\SO(d-1)$, but will become relevant from the point of view of $\SO(d)$ below.}
\be\label{eq:rhoproduct}
	\rho_1\otimes\rho_2\otimes\rho_3^R.
\ee
This follows simply from $\SO(d)$ invariance, the $\SO(d-1)$ subgroup being the subgroup which preserves $x$. The only $\SO(d)$ representations which contain $\SO(d-1)$ invariants are the traceless-symmetric tensors; moreover, each traceless-symmetric representation contains precisely one $\SO(d-1)$ invariant. This means that the tensor structures are also in one-to-one correspondence with traceless-symmetric $\SO(d)$ representations in~\eqref{eq:rhoproduct}.

This latter point of view allows one to write down the general form of~\eqref{eq:general3ptCF}. Indeed, it follows that for each tensor structure, there is a number $J$ and an $\SO(d)$-invariant tensor $\Q^{abc}_{\mu_1\ldots \mu_J}$, traceless and symmetric in indices $\mu_i$. We can then write\footnote{Here we mix the index of $\rho^R_3$ with $\rho_3$. To do so we assume that the representation $\rho^R$ was defined according to~\eqref{eq:reflecteddefinition} with reflection $R$ being along the axis along which $\cO_3$ was taken to $\infty$. In other words, here it is essential to use a separate realization for $\rho^R$ even if $\rho^R\simeq \rho$.}
\be\label{eq:3ptfromQ}
	\<\cO_1^a(0)\cO_2^b(x)\cO_3^c(\infty)\>=\Q^{abc}_{\mu_1\ldots \mu_J} (x^{\mu_1}\ldots x^{\mu_J}-\text{traces})x^{-\De_1-\De_2+\De_3-J}.
\ee
Clearly, the Fourier transform is just
\be
	\<\cO_1^a(0)\cO_2^b(p)\cO_3^c(\infty)\>=\cF_{\De_1+\De_2-\De_3,J}\Q^{abc}_{\mu_1\ldots \mu_J} (p^{\mu_1}\ldots p^{\mu_J}-\text{traces})p^{\De_1+\De_2-\De_3-J-d}.
\ee
We thus see that in this basis of tensor structures, the Fourier transform is essentially diagonal. We can label the three-point tensor structures by pairs $(\rho_{13},J)$, where $\rho_{13}\in \rho_1\otimes\rho_3^R$ and $\rho_{13}\otimes \rho_2$ contains a traceless-symmetric spin-$J$ tensor. The same classification applies also in momentum space, and we have
\be
	(\rho_{13},J) &\xleftrightarrow[]{Fourier} \cF_{\De_1+\De_2-\De_3,J}\times (\rho_{13},J).	
\ee
We will call this basis of three-point tensor structures the $\SO(d)$-basis.

\subsubsection{The final algorithm}
\label{sec:fourieralgo}

With Fourier transforms understood, it only remains to act on the Fourier-transformed three-point function by $K(p)$. The only problem with this is that $K(p)$ acts diagonally on $\SO(d-1)$ components of $\rho_2$, while in $\SO(d)$-basis structures introduced above these components are mixed when we combine $\rho_{13}$ and $\rho_2$ into an $\SO(d)$ traceless-symmetric tensor. It is therefore convenient at this stage to use a different basis of tensor structures, in which the action of $K(p)$ will be diagonal.

It is the easiest to define this basis for a standard momentum $p=q$. In this standard configuration we can build $\SO(d-1)$ invariants by taking dual pairs of $\SO(d-1)$ representations from $\rho_{13}$ and $\rho_2$. Since in any $\SO(d)$ irrep a given $\SO(d-1)$ irrep appears at most once, we can label these structures by pairs $(\rho_{13},\lambda_{d-1})$, where the $\SO(d-1)$ irrep $\lambda_{d-1}\in \rho_2$ and $\lambda^*_{d-1}\in \rho_{13}$. We then have
\be
	K(q)\.(\rho_{13},\lambda_{d-1})=\cA_{\lambda_{d-1}}\times (\rho_{13},\lambda_{d-1}).
\ee
We will call this basis of tensor structures the $\SO(d-1)$-basis. 

To sum up, we have introduced two bases of tensor structures for a generic three-point function, the $\SO(d)$- and the $\SO(d-1)$-basis. The Fourier transform is diagonal, in an appropriate sense, in the $\SO(d)$-basis, while momentum-space shadow transform is diagonal in the $\SO(d-1)$-basis. The algorithm for computing the shadow transform of a given three-point function is then as follows:
\begin{enumerate}
	\item Start in the $\SO(d)$-basis. Fourier-transform to momentum space. (Diagonal operation.)
	\item Convert to the $\SO(d-1)$-basis. (Representation-theoretic problem.)
	\item Apply shadow transform in momentum space. (Diagonal operation.)
	\item Convert to the $\SO(d)$-basis. (Representation-theoretic problem.)
	\item Fourier-transform to position space. (Diagonal operation.)
\end{enumerate}

It will not be important in our examples, but let us note that the representation-theoretic data which goes into the conversion between $\SO(d)$- and $\SO(d-1)$-bases is given by $\SO(d):\SO(d-1)$ isoscalar factors,\footnote{Also known as reduced Clebsch-Gordan coefficients or reduction factors (see also~\cite{Kravchuk:2017dzd} for a recent discussion in the context of CFT).} rather then the complete Clebsch-Gordan coefficients.

\subsubsection[A note on simplifying the expressions]{A note on simplifying the expressions\protect{\footnote{This subsection can be safely skipped on the first reading.}}}
\label{sec:simplifyingfourier}

The algorithm described above involves several non-diagonal operations (two changes of basis), which means that the Fourier-transform coefficients $\cF_{\De,J}$ and $\cA_{\lambda_{d-1}}$ for different $J$ and $\lambda_{d-1}$ will mix. Since these coefficients are given by products of $\Gamma$-functions, the intermediate expressions can get quite complicated. It is the purpose of this section to point out how one can easily factor out all $\Gamma$-factors and work only with rational functions.

The coefficients $\cA_{\lambda_{d-1}}$ are the easiest to deal with since, as can be seen for example from the explicit expressions in 3d~\eqref{eq:AjmMain} and in 4d~\cite{Mack:1975je}, the ratios $\cA_{\lambda_{d-1}}/\cA_{\lambda'_{d-1}}$ for two $\SO(d-1)$ irreps are rational. Thus $\cA_{j,0}$ or $\cA_{j,\half}$ can be factored out when applying the shadow transform in momentum space. The more problematic factor is the Fourier transform coefficient $\cF_{\De,J}$ since it is $\frac{J}{2}$ which enters the arguments of the $\Gamma$-functions, cf.~\eqref{eq:Fcoefficient}, and at this point we have not constrained the parity of $J$.

Fortunately, the parity of $J$ is constrained automatically in even dimensions. Indeed, $\SO(2n)$ has a non-trivial central element,
\be
	\cC=e^{\pi M_{12}+\pi M_{34} +\ldots + \pi M_{d-1,d}}
\ee
which rotates by $\pi$ in all planes, thereby sending any vector $x$ to $-x$. Since it commutes with all elements of $\SO(2n)$, it acts by a $c$-number in any irrep of $\SO(2n)$. In particular, in traceless-symmetric representations it acts by $(-1)^J$. Thus, if it acts by $\cC_i$ in representations $\rho_i$, the product $\cC_1\cC_2\cC_3=\pm 1$ determines the parity of $J$ of traceless-symmetric irreps appearing in~\eqref{eq:rhoproduct}.\footnote{For bosonic representations this can also be explained by noting that the invariant tensors $\delta_{\mu\nu}$ and $\e_{\mu_1\ldots \mu_{2n}}$ of $\SO(2n)$ both have an even number of indices. This means that $\Q$ in~\eqref{eq:3ptfromQ} necessarily has an even number of vector indices (including those which go into labels $a,b,c$).\label{footnote:SO2nJparity}}

In odd dimensions we can still define $\cC$ as $\cC=\cP$, the parity transformation which sends $x\to -x$. However, now it is not part of $\SO(d)$. Nevertheless, it belongs to the center of $\mathrm{O}(d)$, and since there are no chiral irreps in odd dimensions, we can always assume that our representations are also representations of $\mathrm{O}(d)$. Then again $\cC$ will act by a $c$-number in irreducible representations, and the action on~\eqref{eq:rhoproduct} will be by $\cC_1\cC_2\cC_3$. The only difference is that the three-point structures now can be parity-even, in which case we have $(-1)^J=\cC_1\cC_2\cC_3$ or parity-odd, in which case $(-1)^J=-\cC_1\cC_2\cC_3$.\footnote{Similarly to footnote~\ref{footnote:SO2nJparity}, for bosonic irreps we can observe that $\delta_{\mu\nu}$ is parity-even and has an even number of indices, while $\e_{\mu_1\ldots \mu_d}$ is parity-odd and has an odd number of indices.}

The conclusion is that in the algorithm of section~\ref{sec:fourieralgo} all $\Gamma$-functions can be factored out from all expressions in even dimensions, and separately for parity-even and parity-odd structures in odd dimensions, leaving only rational functions of spins and scaling dimensions. We will see this at work in 3d examples below.

\subsection{Implementation in 3d}
\label{sec:3dfourier}

In this section we implement the algorithm of section~\ref{sec:fourieralgo} in 3d. Our conventions are described in appendix~\ref{app:3dconventions}. 

\subsubsection{Plancherel measure}
\label{sec:plancherel_3d}
As a warm up, let us compute the square of shadow transform and the Plancherel measure. As computed in appendix~\ref{app:twoptfourier}, the Fourier-transformed shadow kernel~\eqref{eq:Kshadowfourier} for $p=(0,0,1)$ is
\be
	K_\De(p)=\sum_{m=-j}^j \cA_{j,m}(\tilde \De)|j,m\>\<j,m|,
\ee
with $\cA_{j,m}(\De)$ given by~\eqref{eq:Ajm}. The square of the shadow transform is
\be
	K_{\tilde\De}(p)K_\De(p)=\sum_{m=-j}^j \cA_{j,m}(\De)\cA_{j,m}(\tilde \De)|j,m\>\<j,m|.
\ee
Using the explicit expressions we can easily verify that for $j\in \Z$
\be
	\cN(\De,j)\equiv\cA_{j,m}(\De)\cA_{j,m}(\tilde \De)=\frac{2\pi^3 \tan\pi\De}{(2\De-3)(\De+j-1)(2+j-\De)},
\ee
and for $j\in \Z+\half$
\be
	\cN(\De,j)\equiv\cA_{j,m}(\De)\cA_{j,m}(\tilde \De)=\frac{2\pi^3 \cot\pi\De}{(2\De-3)(\De+j-1)(2+j-\De)},
\ee
independently of $m$.
In either case we have
\be
	K_{\tilde\De}(p)K_\De(p)=\sum_{m=-j}^j \cN(\De,j)|j,m\>\<j,m|=\cN(\De,j)\mathbf{1}.
\ee
This is then valid for all $p$ by $\SO(3)$ and scale invariance. According to~\eqref{eq:relationbetweenshadowsqandplancherel} we have
\be
	\mu(\De,j)=\frac{1}{\cN(\De,j)}\frac{\mathrm{tr} [K_{\tl\De}(-\infty)K_\De(\infty)]}{2^3 8\pi^2},
\ee
where $K$ is in position space. Here we have $K_\De(\infty) = \lim L^{2\tl\De} K_\De(L\hat e_2)$. From~\eqref{eq:3dtwopt} we find $K_{\tl\De}(-\infty)=(-1)^{2j}K_{\De}(\infty)=(-1)^{2j}K_{\De}(\hat e_2)$ and then using~\eqref{eq:Kstandard} we conclude
\be
	\mathrm{tr} [K_{\tl\De}(-\infty)K_\De(\infty)]=(-1)^{2j}(2j+1).
\ee
Therefore,
\be\label{eq:3dplancherel}
	\mu(\De,j)=\frac{1}{128\pi^5}(2j+1)(2\De-3)(\De+j-1)(2+j-\De)\cot \pi (\De-j).
\ee
This is in precise agreement with~\eqref{eq:generalplanchereloddd}. As an extra consistency check, we note that by using spinor weight-shifting operators from~\cite{Karateev:2017jgd} and following the same reasoning as in section~\ref{sec:plancherelrecursions} we can also derive the following recursion relation for $\mu(\De,j)$,
\be
	\mu(\De,j)=\frac{(2j+1)(2\De-3)(\De+j-1)}{2j(2\De-4)(\De+j-2)}\mu\p{\De-\half,j-\half},
\ee
which is obviously consistent with~\eqref{eq:3dplancherel}.

\subsubsection{Example: scalar-vector-spin-$j$}

Let us now see how the algorithm of section~\ref{sec:fourieralgo} can be used to compute the shadow transform of 
\be
	\<\phi(x_1)v^\mu(x_2)\cO^{\nu_1\ldots \nu_j}(x_3)\>
\ee
with respect to $x_2$. First, let us note that there exist two parity-even structures tensor structures, in the notation of~\cite{Costa:2011mg} given by
\be
	\mathbb{T}^+_1&=V_2 V_3^j,\\
	\mathbb{T}^+_2&=H_{23} V_3^{j-1},
\ee
and one parity-odd tensor structure, 
\be
	\mathbb{T}^-_1=\e_{23} V_3^{j-1}.
\ee
These structures in what we will call the $q$-basis of~\cite{Kravchuk:2016qvl} are given by
\be
	\mathbb{T}^+_1&=(-i)^{j+1}2^{\frac{j-1}{2}}[000]^+,\\
	\mathbb{T}^+_2&=(-i)^{j-1}2^{\frac{j-1}{2}}([000]^++[0\bar 1 1]^+),\\
	\mathbb{T}^-_1&=-(-i)^j2^{\frac{j-1}{2}}[0\bar 1 1]^-,
\ee
where for compactness we used notation $\bar 1 \equiv -1$. As explained in appendix~\ref{app:3dconventions:tensorstructures}, the $q$-basis structures can then be directly translated into the notation
\be
	|j_1,m_1\>\otimes |j_2,m_2\>\otimes |j_3,m_3\>
\ee
where $j_1=0, j_2=1, j_3=j$. In particular, we have
\be
	[000]^+&=2i^{j-1}|0,0\>\otimes|1,0\>\otimes|j,0\>,\\
	[0\bar 1 1]^\pm&=i^{j+1}\sqrt{\frac{2(j+1)}{j}}(
		|0,0\>\otimes|1,-1\>\otimes|j,+1\>\pm
		|0,0\>\otimes|1,+1\>\otimes|j,-1\>
	)
\ee

To implement the algorithm, we now need to define the $\SO(d)=\SO(3)$ and $\SO(d-1)=\SO(2)$ bases. Let us do this in full generality for three arbitrary spins $j_1,j_2,j_3$ and then specialize to our example. We first define $\SO(3)$ basis using the standard $\SO(3)$ Clebsch-Gordan coefficients as
\be
	\mathbb{T}^{j_{13},J}_{\SO(3)}\equiv \sum_{m_1+m_2+m_3=0}& \<j_2,m_2;j_{13},m_1+m_3|J,0\>\<j_1,m_1;j_3,m_3|j_{13},m_1+m_3\>\times\nn\\
	&\quad\times|j_1,m_1\>\otimes |j_2,m_2\>\otimes |j_3,m_3\>.
\ee
Namely, we first take the tensor product $j_{13}\in j_1\otimes j_3$ and then take the tensor product $J\in j_{13}\otimes j_2$. Then we define the $\SO(2)$ basis as
\be
	\mathbb{T}^{j_{13},m_2}_{\SO(2)}\equiv \sum_{m_1+m_3=-m_2}& \<j_1,m_1;j_3,m_3|j_{13},m_1+m_3\>\,|j_1,m_1\>\otimes |j_2,m_2\>\otimes |j_3,m_3\>,
\ee
which makes it intermediate between $q$-basis and $\SO(3)$ basis. We have
\be
	\mathbb{T}^{j_{13},J}_{\SO(3)}&=\sum_{m_2}
	\<j_2,m_2;j_{13},-m_2|J,0\>\mathbb{T}^{j_{13},m_2}_{\SO(2)},\\
	\mathbb{T}^{j_{13},m_2}_{\SO(2)}&=\sum_{J\in j_{13}\otimes j_2}
	\<J,0|j_2,m_2;j_{13},-m_2\>\mathbb{T}^{j_{13},J}_{\SO(3)}.
\ee

In our example the transition from $q$-basis to $\SO(2)$-basis is essentially trivial, since we have $j_1=0$. Translating then to $\SO(3)$ basis we find
\be
	\mathbb{T}_1^+&=(-1)^{1+j}2^{\frac{1+j}{2}}\sqrt{\frac{j}{2j+1}}\mathbb{T}_{\SO(3)}^{j,j-1}+(-1)^{j}2^{\frac{1+j}{2}}\sqrt{\frac{j+1}{2j+1}}\mathbb{T}_{\SO(3)}^{j,j+1},\\
	\mathbb{T}_2^+&=(-1)^{j}2^{\frac{1+j}{2}}\sqrt{\frac{2j+1}{j}}\mathbb{T}_{\SO(3)}^{j,j-1},\\
	\mathbb{T}_1^-&=i(-1)^j2^{\frac{j+1}{2}}\sqrt{\frac{j+1}{j}}\mathbb{T}^{j,j}_{\SO(3)}.
\ee
As promised in section~\ref{sec:simplifyingfourier}, parity of $J$ is determined by the parity of the structures. In particular, for parity-even structures we see that $J=j\pm 1$, which has the same parity as $j_1+j_2+j_3=j+1$, while for parity odd structures we have $J=j$, which has the opposite parity. We will now compute the shadow transform in $\SO(3)$ basis and then reinterpret the result in terms of $\mathbb{T}^\pm_i$ structures.

\paragraph{Shadow transform of $\SO(3)$ structures}
Let us first apply the shadow transform to $\mathbb{T}_{\SO(3)}^{j,j-1}$. We first Fourier transform 
\be
	\mathbb{T}_{\SO(3)}^{j,j-1}\to \cF_{\De_1+\De_2-\De_3,j-1}\mathbb{T}_{\SO(3)}^{j,j-1}.
\ee
We will keep this factor of $\cF_{\De_1+\De_2-\De_3,j-1}$ in mind, but not write it explicitly until the end. Then we convert to $\SO(2)$ basis, obtaining
\be
	\mathbb{T}_{\SO(3)}^{j,j-1}=\sqrt{\frac{j+1}{2(2j+1)}}\mathbb{T}_{\SO(2)}^{j,-1}+\sqrt{\frac{j+1}{2(2j+1)}}\mathbb{T}_{\SO(2)}^{j,+1}-\sqrt{\frac{j}{2j+1}}\mathbb{T}_{\SO(2)}^{j,0}.
\ee
Now we apply shadow transform in momentum space to obtain,
\be
	&\cA_{1,-1}(\tl\De)\sqrt{\frac{j+1}{2(2j+1)}}\mathbb{T}_{\SO(2)}^{j,-1}+\cA_{1,+1}(\tl\De)\sqrt{\frac{j+1}{2(2j+1)}}\mathbb{T}_{\SO(2)}^{j,+1}-\cA_{1,0}(\tl\De)\sqrt{\frac{j}{2j+1}}\mathbb{T}_{\SO(2)}^{j,0}=\nn\\
	&=\cA_{1,0}(\tl\De)\p{-\frac{\De_2-1}{\De_2-2}\sqrt{\frac{j+1}{2(2j+1)}}(\mathbb{T}_{\SO(2)}^{j,-1}+\mathbb{T}_{\SO(2)}^{j,+1})-\sqrt{\frac{j}{2j+1}}\mathbb{T}_{\SO(2)}^{j,0}}.
\ee
We again temporarily remove the factor $\cA_{1,0}(\tl\De)$ to keep the expressions simpler. We now transform this expression back to $\SO(3)$ basis to obtain
\be
	\frac{2+j-\De_2}{(2j+1)(\De_2-1)}\mathbb{T}_{\SO(3)}^{j,j-1}+
	\frac{\sqrt{j(j+1)}(3-2\De_2)}{(2j+1)(\De_2-1)}\mathbb{T}_{\SO(3)}^{j,j+1}.
\ee
Now we need to Fourier-transform back to position space, which is easy in $\SO(3)$ basis,
\be
	\frac{\cF_{-\De_1+\De_2+\De_3,j-1}^*}{(2\pi)^3}\frac{2+j-\De_2}{(2j+1)(\De_2-1)}\mathbb{T}_{\SO(3)}^{j,j-1}+
	\frac{\cF_{-\De_1+\De_2+\De_3,j+1}^*}{(2\pi)^3}\frac{\sqrt{j(j+1)}(3-2\De_2)}{(2j+1)(\De_2-1)}\mathbb{T}_{\SO(3)}^{j,j+1}.
\ee
Here $-\De_1+\De_2+\De_3=-(\De_1+\tl\De_2-\De_3)+3$ is the power of $p$ in the shadow transform of momentum space three-point function. At this point, we use the fact that only $J$ of the same parity as $j-1$ enter this expression, and write
\be
	\cF_{-\De_1+\De_2+\De_3,j-1}^*&=(-1)^{j-1}\cF_{-\De_1+\De_2+\De_3,j-1},\\
	\cF_{-\De_1+\De_2+\De_3,j+1}^*&=(-1)^{j}\frac{\De_1-\De_2-\De_3+j+2}{\De_3+\De_2-\De_1+j-1}\cF_{-\De_1+\De_2+\De_3,j-1}.
\ee
This allows us to factor out $\cF_{-\De_1+\De_2+\De_3,j-1}$. Collecting everything together we now have the shadow transform
\be\label{eq:shadowexamplefactored}
	\mathbb{T}_{\SO(3)}^{j,j-1}\longrightarrow&
	C_+\left[\frac{(-1)^{j-1}}{(2\pi)^3}\frac{2+j-\De_2}{(2j+1)(\De_2-1)}\mathbb{T}_{\SO(3)}^{j,j-1}\right.\nn\\&\left.+
	\frac{(-1)^j}{(2\pi)^3}\frac{\De_1-\De_2-\De_3+j+2}{\De_3+\De_2-\De_1+j-1}\frac{\sqrt{j(j+1)}(3-2\De_2)}{(2j+1)(\De_2-1)}\mathbb{T}_{\SO(3)}^{j,j+1}\right],
\ee
where $C_+$ is given by
\be
	C_+\equiv \cF_{\De_1+\De_2-\De_3,j-1}\cA_{1,0}(\tl\De_2)\cF_{-\De_1+\De_2+\De_3,j-1}
\ee
and contains all some non-trivial products of $\Gamma$-functions. Note that by restricting to definite parity structures we were able to restrict parity of $J$ in $\cF_{\l,J}$, and thus factor all $\cF$-coefficients out, leaving only rational functions of scaling dimensions inside the square brackets of~\eqref{eq:shadowexamplefactored}.
We can now perform the same calculation for shadow transform of $\mathbb{T}_{\SO(3)}^{j,j+1}$, factoring out the same $C_+$, to obtain
\be
	\mathbb{T}_{\SO(3)}^{j,j+1}\longrightarrow C_+\frac{\De_3-\De_1-\De_2+j+2}{\De_1+\De_2-\De_3+j-1}\left[
		\frac{(-1)^{j+1}}{(2\pi)^3}\frac{\sqrt{j(1+j)}(2\De_2-3)}{(2j+1)(\De_2-1)}\mathbb{T}_{\SO(3)}^{j,j-1}
	\right.\nn\\
	\left.
		+\frac{(-1)^j}{(2\pi)^3}\frac{\De_1-\De_2-\De_3+j+2}{\De_3+\De_2-\De_1+j-1}\frac{\De_2+j-1}{(2j+1)(\De_2-1)}\mathbb{T}_{\SO(3)}^{j,j+1}
	\right].
\ee
A much more straightforward calculation implies for the parity-odd structure
\be
	\mathbb{T}_{\SO(3)}^{j,j}\longrightarrow C_-\frac{(-1)^{j+1}}{(2\pi)^3}\frac{\De_2-2}{\De_2-1}\mathbb{T}_{\SO(3)}^{j,j},
\ee
where 
\be
	C_-\equiv \cF_{\De_1+\De_2-\De_3,j}\cA_{1,0}(\tl\De_2)\cF_{-\De_1+\De_2+\De_3,j}.
\ee

\paragraph{Shadow transform of $\mathbb{T}^\pm_i$ structures}

We can now translate these results into results for $\mathbb{T}^\pm_i$ structures by a straightforward change of basis and find
\be
	C_+^{-1}\mathbb{T}_1^+\longrightarrow& \frac{(-1)^j j(2\De_2-3)(\De_1-\De_3)}{8\pi^3(\De_2-1)(\De_1+\De_2-\De_3+j-1)(\De_2+\De_3-\De_1+j-1)}\mathbb{T}_2^+\nn\\
	&+\frac{(-1)^j \left(-\Delta _1+\Delta _2+\Delta _3-j-2\right) \left(\Delta _2^2+\Delta _1 \Delta _2-\Delta _3 \Delta _2-3 \Delta _2-\Delta _1+\Delta _3+\Delta _2 j-2 j+2\right)}{8 \pi ^3 \left(\Delta _2-1\right) \left(\Delta _1+\Delta _2-\Delta _3+j-1\right) \left(-\Delta _1+\Delta _2+\Delta _3+j-1\right)}\mathbb{T}_1^+,\\
	C_+^{-1}\mathbb{T}_2^+\longrightarrow& 
	\frac{(-1)^j\left(2 \Delta _2-3\right)  \left(-\Delta _1+\Delta _2+\Delta _3-j-2\right)}{8 \pi ^3 \left(\Delta _2-1\right) \left(-\Delta _1+\Delta _2+\Delta _3+j-1\right)}\mathbb{T}_1^+\nn\\
	&+\frac{(-1)^j \left(\Delta _2^2-\Delta _1 \Delta _2+\Delta _3 \Delta _2-3 \Delta _2+2 \Delta _1-2 \Delta _3-\Delta _2 j+j+2\right)}{8 \pi ^3 \left(\Delta _2-1\right) \left(-\Delta _1+\Delta _2+\Delta _3+j-1\right)}\mathbb{T}_2^+,\\
	C_-^{-1}\mathbb{T}^-_1\longrightarrow&
	\frac{(-1)^{j+1}\left(\Delta _2-2\right) }{8 \pi ^3 \left(\Delta _2-1\right)}\mathbb{T}^-_1.
\ee

\paragraph{Comparing with weight-shifting operators}

Let us compare this result with the one which may be obtained using weight-shifting operators for the parity-even structures. We are only going to check a very particular linear combination of structures, namely the one which is obtained by acting with a specific weight-shifting operator on the scalar-scalar-spin-$j$ structure,
\be
	\mathbb{T}_\cD\equiv(\cD^{0+}_2\.\cD_1^{-0})\<\f_{\De_1+1}\f_{\De_2}\cO_{\De_3,j}\>=\cD_{21}\<\f_{\De_1+1}\f_{\De_2}\cO_{\De_3,j}\>=\frac{j}{2}\mathbb{T}^+_2+\frac{\De_1-\De_2-\De_3+j+1}{2}\mathbb{T}^+_1,
\ee
where $\cD_{21}$ is in notation of~\cite{Costa:2011dw}. By using our results above, we find that under the shadow transform
\be
	\mathbb{T}_\cD\longrightarrow&
	C_+\frac{(-1)^{j+1}   \left(\Delta _1-\Delta _2-\Delta _3+j+2\right)}{8 \pi ^3 \left(\Delta _1+\Delta _2-\Delta _3+j-1\right)}\left(\frac{j}{2}\mathbb{T}^+_2+\frac{\Delta _1+\Delta _2-\Delta _3+j-2}{2}\mathbb{T}^+_1\right).
\ee
On the other hand, we have from equation~\eqref{eq:commuteshadowandd}
\be
	&\mathbf{S}_2(\cD^{0+}_2\.\cD_1^{-0})\<\f_{\De_1+1}\f_{\De_2}\cO_{\De_3,j}\>=\frac{\De_2-1}{3-\De_2}(\cD^{0+}_2\.\cD_1^{-0})\<\f_{\De_1+1}\mathbf{S}[\f_{\De_2}]\cO_{\De_3,j}\>\nn\\
	&=\frac{\De_2-1}{3-\De_2}S(\f_{\De_1+1}[\f_{\De_2}]\cO_{\De_3,j})(\cD^{0+}_2\.\cD_1^{-0})\<\f_{\De_1+1}\f_{3-\De_2}\cO_{\De_3,j}\>,
\ee
and thus under shadow transform
\be
	\mathbb{T}_\cD\longrightarrow\frac{\De_2-1}{3-\De_2}S(\f_{\De_1+1}[\f_{\De_2}]\cO_{\De_3,j})\left(\frac{j}{2}\mathbb{T}^+_2+\frac{\Delta _1+\Delta _2-\Delta _3+j-2}{2}\mathbb{T}^+_1\right).
\ee
The two calculations are thus consistent if
\be
	C_+\frac{(-1)^{j+1}   \left(\Delta _1-\Delta _2-\Delta _3+j+2\right)}{8 \pi ^3 \left(\Delta _1+\Delta _2-\Delta _3+j-1\right)}=\frac{\De_2-1}{3-\De_2}S(\f_{\De_1+1}[\f_{\De_2}]\cO_{\De_3,j}).
\ee
This identity indeed holds. 

It may appear that the above derivation of shadow transform from weight-shifting operators is simpler than using the Fourier-space algorithm. While this is true in some cases, in general the former approach requires computing the action of differential operators on tensor structures, which can be quite non-trivial for more complicated correlators. It also requires one to choose a basis of differential operators to generate all tensor structure, which is in principle a non-canonical procedure. At the same time, the approach based on Fourier space only involves handling rational\footnote{Modulo some square roots of functions of $j$, which come from CG coefficients and can perhaps be avoided by a suitable rescaling of basis tensor structures.} functions built out of closed-form expressions such as $\cA_{j,m}(\De)/\cA_{j,0}(\De)$ and $\cF_{\l,J+2n}/\cF_{\l,J}$. Together with the fact that it works on bases of tensor structures which are defined uniformly for all choices of external representations, this allows one to efficiently automate the calculation.

\subsubsection{Three-point pairings}

The last non-trivial element is the computation of three-point pairings. It is again convenient to work in $q$-basis of~\cite{Kravchuk:2016qvl}. From~\eqref{eq:threeptpairinggeneral} and the discussion of appendix~\ref{app:3dconventions:tensorstructures} it is easy to see that the only non-zero pairings are
\be
([m_1,m_2,m_3],[-m_1,-m_2,-m_3])=\frac{(-1)^{j_1+j_2+j_3}}{2^3 2\pi}
\p{\binom{2j_1}{j_1+m_1}
	\binom{2j_2}{j_2+m_2}
	\binom{2j_3}{j_3+m_3}}^{-1}.
\ee
This makes computing the inverse pairing in~\eqref{eq:jjjjpairinginversedefinition} below especially easy.

\subsection{OPE coefficients of 3d current MFT}
\label{sec:opecurrent3d}

As an application, we use the results of~\ref{sec:conformal_partial_waves_and_MFT} to compute the conformal block expansion of the MFT current four-point function in $3d$,
\be
	&\<J(x_1)J(x_2)J(x_3)J(x_4)\>=\nn\\
	&\quad=\<J(x_1)J(x_2)\>\<J(x_3)J(x_4)\>+\<J(x_1)J(x_3)\>\<J(x_2)J(x_4)\>+\<J(x_1)J(x_4)\>\<J(x_2)J(x_3)\>,
\ee
where we kept the vector indices implicit. The calculation is rather involved, so we do not reproduce here the intermediate results, such as shadow transform coefficients. Instead, we will describe the key points of the computation and state the final results. Same comments apply to the analysis of the four-point function of stress-energy tensors in the next section.

We agree to normalize two-point functions of our operators as in appendix~\ref{app:3dconventions:twopt}. We also need to fix a basis of three-point structures for $\<JJ\cO_j\>$ for $\cO$ a spin-$j$ operator. These three-point structures have been studied in~\cite{Dymarsky:2017xzb}. The conclusion of their analysis is that after taking into account the conservation of $J$ and the permutation symmetry between the two $J$'s,
\begin{itemize}
	\item for $j=0$ there exist one parity-even and one parity-odd structure,
	\item for $j=1$ there exist no structures,
	\item for even $j\geq 2$ there exist two parity-even and one parity-odd structure,
	\item for odd $j\geq 3$ there exists one parity-odd structure.
\end{itemize}
We will not completely reproduce these tensor structures here, but only give the minimal information sufficient to define the OPE coefficients. In particular, we write
\be
	\<JJ\cO_0\> &= \lambda^{(1)}_{JJ\cO^+}[000]^++\lambda^{(1)}_{JJ\cO^-}[1\bar 1 0]^-+\ldots,\\
	\<JJ\cO_j\> &= \lambda^{(1)}_{JJ\cO^+}[000]^+ + \lambda^{(2)}_{JJ\cO^+}[01\bar 1]^+ +\lambda^{(1)}_{JJ\cO^-}[1\bar 1 0]^-+\ldots,\quad(j\geq 2\text{ even})\\
	\<JJ\cO_j\> &= \lambda^{(1)}_{JJ\cO^-}[01\bar 1]^-+\ldots,\quad(j\geq 3\text{ odd}).
\ee
Here we are using the $q$-basis as defined in~\cite{Kravchuk:2016qvl}, and $\bar 1=-1$. Furthermore, the dots in these equations represent terms which contain other $q$-basis structures, which are needed to restore permutation symmetry and conservation of $J$. Finally, we note that MFT preserves parity, so parity-odd operators have only parity-odd OPE coefficients $\lambda^{(i)}_{JJ\cO^-}$ and parity-even operators have only parity-even OPE coefficients $\lambda^{(i)}_{JJ\cO^+}$.

\subsubsection{Exchanged operators}
The primary operators exchanged in the $\<JJJJ\>$ four-point function are normal-ordered products 
\be\label{eq:JJfamily}
	&:\!J^{\mu_1}(x)\ptl^{\nu_1}\cdots \ptl^{\nu_k}\ptl^{2n}J^{\mu_2}(x)\!:+\text{desc.}-\text{traces}(\nu_i)
\ee
We have computed the degeneracies of these operators by taking the symmetric square of the character of the Verma module corresponding to $J$, and decomposing into irreducible characters. We find that there exist the following operators,\footnote{Here $n$ is not necessarily the same as $n$ in~\eqref{eq:JJfamily}.}
\begin{itemize}
	\item parity-even operators $\cO_j^+$ with $\De^+_{n,j}=2n+j+2$, for all even $j\geq 0$, $n\geq 0$, and $\De^+_{n,j}\geq 4$. The degeneracies of the operators are the same as the number of three-point tensor structures, except that for $n=0$ there is at most one operator.
	\item parity-odd operators $\cO_j^-$ with $\De^-_{n,j}=2n+j+3$, for all $j\geq 0$, $n\geq 0$, and $\De^-_{n,j}\geq 5$, except $j=1$. All degeneracies are equal to $1$.
\end{itemize}

\subsubsection{Shadow transforms and pairings with conserved operators}
\label{sec:conservationconditions}

Our goal now is to compute $P_{ab}(\De,j,\pm)$ which is defined as
\be
	P_{ab}(\De,j,\pm)=\sum_{\cO^\pm} \lambda^{(a)}_{JJ\cO^\pm}\lambda^{(b)}_{JJ\cO^\pm},
\ee
where the sum is over operators with given parity, dimension and spin. It should generically be non-zero for the operators described above, and its rank should generically be equal to the number of degenerate operators (but no more than the dimension of the matrix, of course). According to \eqref{eq:physical spectrum_from_contour_integral} we have
\be
	P_{ac}(\De_*,j,\pm)=-\mathrm{Res}_{\De\to \De_*}\p{I_{ab}(\De,j,\pm) S(J_3J_4[\cO_{\tl\De,j,\pm}])^b{}_{c}},
\ee
where we have labeled the $J$ operators by their positions in the four-point function. In turn, the matrix $I_{ab}$ is given by~\eqref{eq:mftcoeffs},
\be\label{eq:JJJJ_MFT_formula_pre}
	I_{ab}(\De,j,\pm)=\mu(\De,j)S([\tl J_1]\tl J_2 \cO_{\De,j,\pm})^d{}_eS(J_1[\tl J_2]\cO_{\De,j\pm})^e{}_a\p{\<J_1J_2\tl\cO\>^{(b)},\<\tl J_1\tl J_2\cO\>^{(d)}}^{-1},
\ee
where $\p{\<J_1J_2\tl\cO\>^{(b)},\<\tl J_1\tl J_2\cO\>^{(d)}}^{-1}$ is defined by 
\be
	\label{eq:jjjjpairinginversedefinition}
	\p{\<J_1J_2\tl\cO\>^{(b)},\<\tl J_1\tl J_2\cO\>^{(d)}}^{-1}\p{\<J_1J_2\tl\cO\>^{(a)},\<\tl J_1\tl J_2\cO\>^{(d)}}=\delta^a_b
\ee
To be precise, this only gives the contribution of $\<J(x_1)J(x_3)\>\<J(x_2)J(x_4)\>$. The contribution of $\<J(x_1)J(x_4)\>\<J(x_2)J(x_3)\>$ can be then obtained by applying a permutation of $J$'s either to the right or to the left three-point structure of $I_{ab}$.

Here we encounter a slight difficulty, since we need to invert the three-point pairing between structures for $\<JJ\cO\>$ and $\<\tl J\tl J\tl \cO\>$. Now, while our structures $\<JJ\cO\>$ are taken to be conserved, the structures $\<\tl J\tl J\tl \cO\>$ do not have the right scaling dimensions in order for us to impose conservation constraints. In particular, $\tl J$ has scaling dimension $1$, which makes it a one-form. If we do not impose any constraints on $\tl J$, the three-point pairing matrix is not going to be a square matrix, so we have to be careful with what is meant by the inverse. 

Specifically, we need to satisfy~\eqref{eq:jjjjpairinginversedefinition}. In this expression, the sum is over index $d$ for which there are more choices, as discussed above, than for $b$ or $a$. This implies that the definition~\eqref{eq:jjjjpairinginversedefinition} of the inverse pairing is ambiguous. In particular, we can always replace
\be
	\p{\<J_1J_2\tl\cO\>^{(b)},\<\tl J_1\tl J_2\cO\>^{(d)}}^{-1}\longrightarrow \p{\<J_1J_2\tl\cO\>_{(b)},\<\tl J_1\tl J_2\cO\>_{(d)}}^{-1}+ v_d y_b
\ee
as long as
\be\label{eq:vdefinition}
	v_d\p{\<J_1J_2\tl\cO\>^{(a)},\<\tl J_1\tl J_2\cO\>^{(d)}}=0.
\ee
In order for~\eqref{eq:JJJJ_MFT_formula_pre} to be consistent, we must have
\be
	\label{eq:vconsistency}
	v_dS([\tl J_1]\tl J_2 \cO_{\De,j,\pm})^d{}_e=0
\ee
for all such $v_d$.

Vectors $v_d$ subject to~\eqref{eq:vdefinition} have to exist simply due to a mismatch of the number of structures $\<JJ\tl\cO\>$ and $\<\tl J\tl J\cO\>$, but it is also possible to understand their existence more conceptually. Equation~\eqref{eq:vdefinition} simply means that the structure $v_d\<\tl J_1\tl J_2\cO\>^{(d)}$ has vanishing pairing with all $\<JJ\tl\cO\>$ structures. Such structures are easy to find. Indeed, let us set
\be
	\label{eq:vconstruction}
	v_d\<\tl J(x_1)\tl J(x_2)\tl \cO(x_3)\>^{(d)}=\<d\phi(x_1) \tl J(x_2) \tl \cO(x_3)\>,
\ee
where $\phi$ is a scalar of dimension $0$ (i.e. a 0-form), $d$ is the exterior derivative, and the structure on the right is arbitrary. Pairings of this structure with $\<JJ\tl\cO\>$ then vanish by integration by parts and conservation of $J$. 

This means that from the point of view of three-point pairings we should be looking at the gauge equivalence classes of $\tl J$ under the gauge equivalence
\be
	\tl J\sim \tl J+d\f.
\ee
It is easy to check that the counting of $\<\tl J\tl J\cO\>$ structures, along the lines of~\cite{Kravchuk:2016qvl}, gives the same number of gauge equivalence classes as the number of $\<JJ\cO\>$ structures. Thus all vectors $v_d$ can be obtained from~\eqref{eq:vconstruction}.\footnote{We have verified this claim by an explicit calculation in several examples.}

This notion of gauge equivalence extends to other types of conserved currents. For example, for the shadow of stress-tensor $\tl T$, the role of $\f$ is played by vectors $\xi$ of dimension $-1$ and the exterior derivative is replaced by the operator which participates in the conformal Killing equation,
\be
	\tl T^{\mu\nu}\sim \tl T^{\mu\nu}+\ptl^{(\mu} \xi^{\nu)}-\text{trace}.
\ee

This point of view makes it clear why~\eqref{eq:vconsistency} must hold: interpreting this equation as taking the shadow transform of a pure gauge structure, we can integrate by parts and use the conservation of the $\<JJ\>$ two-point function to show that the shadow transform is 0.\footnote{This is a bit subtle due to the need to regularize the shadow transform. We always regularize by analytic continuation in scaling dimension, and one way to do this is to perform shadow transform in Fourier space, as described earlier in this section. In even dimensions Fourier transform of current two-point function diverges due to existence of possible contact terms, but in odd dimensions, which is what we care about here, it shows that regularized two-point function is conserved.}

In practice, however, we will simply start with the MFT four-point function for more general operators $J$ with $\De_J\neq d-1$ and continue analytically to $\De_J=d-1$. In this way we find a larger matrix
\be
	\hat I_{\hat a\hat b}(\De,j,\pm),
\ee
where $\hat a$ and $\hat b$ run through non-conserved structures for $\<JJ\cO\>$. At $\De_J=d-1$ we find that
\be
	\hat I_{\hat a\hat b}(\De,j,\pm) = T^{a}_{\hat a}T^{b}_{\hat b}I_{ a b}(\De,j,\pm),
\ee
where $T^{a}_{\hat a}$ are the coordinates of conserved structure $a$ in the non-conserved basis $\hat a$. This equation then determines $I_{ab}$.\footnote{The fact that $\hat I$ takes this form provides a non-trivial consistency check on the whole calculation.} This analytic continuation also helps to distinguish the physical poles in $I_{ab}(\De,j,+)$ at $\De=2\De_J+2n+j-2$ (or $\De=2\De_j+2n+j-1$ for $I_{ab}(\De,j,-)$) from various unphysical poles similarly to the case of scalar operators in section~\ref{sec:scalarmft}.

\subsubsection{The OPE coefficients}
It is now straightforward to use the Fourier algorithm to compute the shadow coefficients and combine them into OPE coefficients. As mentioned above, care should be taken to first compute the residues at $\De=2\De_J+\ldots$ and then take the limit $\De_J\to 2$. The resulting coefficients are as follows.

\paragraph{Parity-even operators}

For parity-even operators we first define
\be
	P_{ab}(\De_{n,j}^+,j,+)=\frac{\pi ^{3/2} 2^{-2 (n+1)} \Gamma \left(j+\frac{3}{2}\right) \Gamma (j+n+1) \Gamma (j+2 n-1)}{\Gamma (j+2) \Gamma \left(\frac{1}{2}-n\right)^2 \Gamma (2 n+1) \Gamma \left(j+n+\frac{3}{2}\right) \Gamma \left(j+2 n+\frac{1}{2}\right)}p_{ab}(n,j,+),
\ee
where $\De_{n,j}^+=2n+j+2$. Note that only even $j$ is allowed. We then have for $n>0$,
\be
	p_{11}(n,j,+)=&(j+1) \left(2 (j-1) j \left(j \left(2 j^2+j-7\right)-5\right) n+(j-1)^2 j^2 (j+1) (j+2)+\right.\nn\\
	&\qquad\qquad\left.+4 \left(j^4-4 j^3+11 j+4\right) n^2-16 ((j-3) j-3) n^3+32 n^4\right),\\
	p_{12}(n,j,+)=&j (j+1) (j+2 n-1) (j+2 n+1) \left(j \left(j^2+j-2\right)+2 ((j-1) j-3) n-4 n^2\right),\\
	p_{22}(n,j,+)=&j (j+2 n-1) (j+2 n+1) \left(4 \left(j^2+j-1\right) n^2+2 (2 j+1) \left(j^2+j-1\right) n+\right.\nn\\
	&\qquad\qquad\qquad\qquad\qquad\qquad\left.+(j-1) j (j+1) (j+2)\right),
\ee
and $p_{21}=p_{12}$. For $n=0$ we have
\be
	p_{11}(0,j,+)=p_{12}(0,j,+)=p_{22}(0,j,+)=\half(j-1)^2 j^2 (j+1)^2 (j+2).
\ee
These values are half those obtained from analytic continuation in $n$.\footnote{An analogue for this situation is $\lim_{\e\to 0}\mathrm{res}_{x=n}\frac{x+\e}{(x-n)(x+2\e)}$, where $x$ plays the role of $\De$ and $\e$ the role of $\De_J-2$. This shows the importance of taking the residues before the limit $\De_J\to 2$.} Note that for $n=0$ there are no operators with $j=0$. For $n>0$ and $j=0$ only $p_{11}$ is non-vanishing, consistently with our definition of the basis of three-point structures. 

Note that according to~\eqref{eq:3dconjugationrules} even-spin operators in spinor convention are hermitian, and thus their OPE coefficients in the $q$-basis are real. Thus, the matrices $P_{ab}(\De_{n,j}^+,j,+)$ should be positive-definite. We have explicitly verified this for $n,j\leq 20$.

\paragraph{Parity-odd operators}
For parity-odd operators we define, similarly to the parity-even case,
\be
	P(\De_{n,j}^-,j,-)=\frac{\pi ^{3/2} j 2^{-2 (n+1)} \Gamma \left(j+\frac{3}{2}\right) \Gamma (j+n+2) \Gamma (j+2 n+2)}{\Gamma (j) \Gamma \left(-n-\frac{1}{2}\right)^2 \Gamma (2n+2) \Gamma \left(j+n+\frac{3}{2}\right) \Gamma \left(j+2 n+\frac{3}{2}\right)}p(n,j,-),
\ee
where we don't use the structure indices $a,b$ since there is at most one parity-odd structure for any spin and $\De^-_{n,j}=2n+j+3$. Here both even and odd $j$ are allowed, and we have for $n\geq 0$ (only $n\geq 1$ is allowed for $j=0$)
\be
	p(n,j,-)&=\begin{cases}
		\frac{(j+2 n) (j+2 n+2) (j+2 n+3)}{j^2 (j+2 n+1)},&\text{even }j,\\
		-\frac{(j-1) (j+2)}{j (j+1)},&\text{odd }j.
	\end{cases}
\ee
Note that according to~\eqref{eq:3dconjugationrules}, odd-spin operators in our spinor convention are anti-hermitian, and thus their OPE coefficients in the $q$-basis are pure imaginary, which leads to the negative values of $p(n,j-)$ for odd $j$.

\paragraph{Comparison with numerical results}

We can compare our results with the table (A.24) in appedix A.6 of~\cite{Dymarsky:2017xzb}. We define, with the right hand side in their notation
\be
	q_{11}=``(\tl\lambda_{JJ\cO^+}^{(1)})^2",\quad
	q_{12}=q_{21}=``\tl\lambda_{JJ\cO^+}^{(1)}\tl\lambda_{JJ\cO^+}^{(2)}",\quad
	q_{11}=``(\tl\lambda_{JJ\cO^+}^{(2)})^2",
\ee
where we use quotes because these coefficients are in general sums corresponding to several operators. We then have
\be
	p=MqM^T,
\ee
where 
\be
	M=2^{-\De}\frac{i^{-j}2^{-j/2}}{(j+2)(\De-j)}\begin{pmatrix}
		1 & -1 \\
		j(\De-2) & 2\De-j^2
	\end{pmatrix}.
\ee
For parity odd structures we have (for $j>0$)
\be
	M=2^{-\De+2}\begin{cases}
		i^{j}2^{\frac{j}{2}-3}j,& \text{even }j,\\
		i^{j}2^{\frac{j}{2}-3}(j+2),& \text{odd }j.
	\end{cases}
\ee
This transformation rule can be obtained by tracing relation of $\tl\lambda$ of~\cite{Dymarsky:2017xzb} to $V_i,H_{ij}$ structures, and then the relation of $H_{ij}$ structures to the $q$-basis.\footnote{We also had to add the $2^{-\De}$ factor in parity-even case and $i2^{-\De+2}$ factor in parity-odd case to match the results of~\cite{Dymarsky:2017xzb}. This is possibly due to our misinterpretation of their conventions. We have checked that our normalization is consistent with our conventions by explicitly computing the contribution of the operators $J_\mu J^\mu$ and $\epsilon^{\mu\nu\lambda}J_\mu \ptl_\nu J_\lambda$.} With this identification, our formula matches all the data presented in~\cite{Dymarsky:2017xzb}.

\subsection{OPE coefficients of 3d stress-tensor MFT}
\label{sec:mftopecoeffs}

In this section, we state the results for the OPE coefficients of the MFT stress-tensor four-point function
\be\label{eq:TTTTMFT}
&\<T(x_1)T(x_2)T(x_3)T(x_4)\>=\nn\\
&\quad=\<T(x_1)T(x_2)\>\<T(x_3)T(x_4)\>+\<T(x_1)T(x_3)\>\<T(x_2)T(x_4)\>+\<T(x_1)T(x_4)\>\<T(x_2)T(x_3)\>,
\ee
which can be computed in exactly the same way as for $\<JJJJ\>$. Normalization of the $\<TT\>$ two-point function we use is the same as for all 3d operators,
\be
	\<T(x_1,z_1)T(x_2,z_2)\>=\frac{(z_1\.I(x_{12})\.z_2)^2}{x_{12}^6}.
\ee
The three-point tensor structures for $\<TT\cO\>$ follow the same pattern as for $\<JJ\cO\>$, except for low spins. In particular we have the following possibilities for $\cO$ of spin $j$ and dimension $\De$,
\begin{itemize}
	\item for $j=0$ (all $\De\geq \half$) and $j=2$ ($\De>3$) there exists one parity-even and one parity-odd structure,
	\item for $j=2$ and $\De=3$ there exist two parity-even structures and one parity-odd structure,
	\item for even $j\geq 4$ there exist two parity-even structures and one parity-odd structure,
	\item for odd $j\geq 5$ there exist one parity-odd structure.
\end{itemize}

\paragraph{Parity-even structures}
We are going to use the same basis of tensor structures as in~\cite{Dymarsky:2017yzx}. For even $j\geq 4$ the parity-even structures can be identified as
\be
	\<TT\cO_j^+\>^{(1)}&=(-i)^{j} 2^{\frac{j}{2}-3} (j-3)_4 (j+4) (\Delta -1) (4 \Delta (\Delta -2) -3 j (j+1)+6)[12\bar 3]^++\nn\\
	&\quad+
	(-i)^j2^{\frac{j}{2}-5}(j-3)_8[\bar 2 20]+
	\ldots,\\
	\<TT\cO_j^+\>^{(2)}&=-(-i)^j 2^{\frac{j}{2}-4}(j-2)_3  (\Delta -1) \left(\Delta(\De+1) -3 j(j+1)+36\right)[12\bar 3]^++\nn\\
	&\quad+
	(-i)^j 2^{\frac{j}{2}-6} (j-3)_4 \left(\Delta  (7 \Delta-13 )-j(j+1)+12\right)[22\bar 4]+
	\ldots,
\ee
where as before we use $\bar n=-n$ inside $q$-basis structures. For $j=0$ we define
\be\label{eq:TTOevenj0}
	\<TT\cO_0^+\>^{(1)}=\frac{1}{(\De-3)(\De+1)}\lim_{j\to 0}\<TT\cO_j^+\>^{(2)},
\ee
and for $j=2$ we define
\be\label{eq:TTOevenj2}
	\<TT\cO_2^+\>^{(1)}=-8\lim_{j\to 2}\<TT\cO_j^+\>^{(2)}.
\ee

\paragraph{Parity-odd structures}
For even $j\geq 4$ the parity-odd structures can be identified as
\be
	\<TT\cO_j^-\>^{(1)}=-(\Delta -1)_4 (j-2)_3 (-i)^{j+1} 2^{\frac{j}{2}+1} [12\bar 3]^-+\ldots,
\ee
while for odd $j\geq 5$ we have
\be
	\<TT\cO_j^-\>^{(1)}=-(-i)^{j+1}2^{j/2} (\Delta -2) (\Delta -1) (j-3)_4 (j+4) \left(4  \Delta (\Delta -2)-j(j+1)+6\right)[12\bar 3]^-+\ldots.
\ee
For $j=0$ we write
\be\label{eq:TTOoddj0}
	\<TT\cO_0^-\>^{(1)} = \frac{1}{4(\De-5)(\De-2)(\De-1)\De(\De+2)}\lim_{j\to 0}\<TT\cO_j^-\>^{(1)},
\ee
and for $j=2$ we write
\be\label{eq:TTOoddj2}
	\<TT\cO_2^-\>^{(1)} = \frac{1}{4(\De-5)(\De+2)}\lim_{j\to 2}\<TT\cO_j^-\>^{(1)},
\ee
where we take the limits of even-spin structures.

\subsubsection{Exchanged operators}

Similarly to current MFT, the exchanged operators are given by normal-ordered products
\be\label{eq:TTfamily}
	&T^{\mu_1\nu_1}(x)\ptl^{\s_1}\cdots \ptl^{\s_k}\ptl^{2n}T^{\mu_2\nu_2}(x)+\text{desc.}-\text{traces}(\s_i)
\ee
We have computed the degeneracies of these operators by taking the symmetric square of the character of the Verma module corresponding to $T$, and decomposing into irreducible characters. We find that there exist the following operators,\footnote{Here $n$ is not necessarily the same as $n$ in~\eqref{eq:TTfamily}.}
\begin{itemize}
	\item parity-even operators $\cO_j^+$ with $\De^+_{n,j}=2n+j+2$, for all even $j\geq 0$, $n\geq 0$, and $\De^+_{n,j}\geq 6$. The degeneracies of the operators are the same as the number of three-point tensor structures, except that for $n=0$ there is at most one operator.
	\item parity-odd operators $\cO_j^-$ with $\De^-_{n,j}=2n+j+3$, for all $j\geq 0$, $n\geq 0$, and $\De^-_{n,j}\geq 7$, except $j=1$ and $j=3$. All degeneracies are equal to $1$.
\end{itemize}

\subsubsection{The OPE coefficients}

\paragraph{Parity-even operators}

For parity-even operators of even spin $j\geq 4$ and $n\geq 1$ we have
\be\label{eq:firstTTTT}
	P_{ab}(\De_{n,j}^+,j,+)=
	\frac{\pi ^{3/2} 2^{-j-2 n-1} \Gamma \left(j+\frac{3}{2}\right) \Gamma (j+n+1) \Gamma (j+2 n-3)}{9 j! (j+2 n+1)_4 \Gamma \left(\frac{1}{2}-n\right)^2 \Gamma (2 n) \Gamma \left(j+n+\frac{1}{2}\right) \Gamma \left(j+2 n+\frac{1}{2}\right)}
	p_{ab}(n,j,+),
\ee
where $\De_{n,j}^+=2n+j+2$ and
\be
	p_{ab}(n,j,+)=\begin{pmatrix}
		\frac{1+\frac{(2n+j-3)_8}{(j-3)_8}}{4n(1+2j+2n)} & -1 \\
		-1 & 4n(1+2j+2n)
	\end{pmatrix}.
\ee
Note that in these expressions, we used the rising Pochhammer symbols $(a)_k=a(a+1)\cdots (a+k-1)$ with $k=4$ and $k=8$. For even $j\geq 4$ and $n=0$ we have
\be
	P_{ab}(\De_{0,j}^+,j,+)=\frac{\sqrt{\pi } 2^{-j-4} (2 j+1) \Gamma (j+1)^2}{9 (j-3) (j-2) (j-1) j \Gamma \left(j+\frac{3}{2}\right) \Gamma (j+5)}\begin{pmatrix}
		1 & 0 \\ 0 &0
	\end{pmatrix}.
\ee
As in the case of four currents, this is $\half$ of the $n\to 0$ limit of the previous expression.

\noindent For $j=2$ we have $n\geq 1$ and 
\be
	P(\De_{n,2}^+,2,+)=\frac{5 \sqrt{\pi } 16^{-n-3} n \Gamma (2 n+3)}{3 (n+3) (2 n-1) (2 n+1)^2 (2 n+3)^2 \Gamma \left(2 n+\frac{5}{2}\right)}.
\ee
For $j=0$ we have $n\geq 2$ and
\be
	P(\De_{n,0}^+,0,+)=\frac{(n-1)^2 n \Gamma (2 n-3) \Gamma (2 n+4)}{36 (n+1)^2 (n+2) (2 n+1) \Gamma (4 n)}.
\ee

The expressions for $j=0$ and $j=2$ are just analytic continuations of $P_{22}$ from $j\geq 4$ case, times an extra factor needed to account for definitions~\eqref{eq:TTOevenj0} and~\eqref{eq:TTOevenj2} of tensor structures at $j=0$ and $j=2$. For $j=0$ this extra factor is $(\De_{n,0}^+-4)^{2}(\De_{n,0}^++1)^{2}$, while for $j=2$ it is $\frac{1}{64}$.

\paragraph{Parity-odd operators}

For parity-odd operators with spin $j\geq 4$ and $n\geq 0$ we have
\be
	P(\De_{n,j}^-,j,-)=-\frac{\pi  2^{j+2 n-6} \Gamma \left(j+\frac{3}{2}\right) \Gamma (j+n+2) \Gamma (j+2 n+1)^2}{9 j! (j+2 n+2)^2 \Gamma \left(-n-\frac{1}{2}\right)^2 \Gamma (2 n+2) \Gamma \left(j+n+\frac{3}{2}\right) \Gamma (2 j+4 n+3)}p(n,j,-),
\ee
where $\De^-_{n,j}=2n+j+3$ and
\be
	p(n,j,-)&=\begin{cases}
	\frac{(2n+j+1)^2(2n+j+2)^2}{4(2n+j-2)_8},&\text{even }j,\\
	\frac{1}{(j-3)_8},&\text{odd }j.
	\end{cases}
\ee
For $j=2$ we have $n\geq 1$ and
\be
	P(\De_{n,2}^-,2,-)=-\frac{5 n (2 n+7) \Gamma (2 n+3)^2}{384 (2 n+1) (2 n+5)^2 \Gamma (4 n+7)}.
\ee
For $j=0$ we have $n\geq 2$ and
\be\label{eq:lastTTTT}
	P(\De_{n,0}^-,0,-)=-\frac{(n-1) n (2 n+3) (2 n+5) \Gamma (2 n+3)^2}{576 (n+2) (2 n-1) (2 n+1) \Gamma (4 n+3)}.
\ee
Note that the coefficients $P$ are negative because unitarity requires the OPE coefficients of our parity-odd structures to be pure imaginary.

Again, the expressions for $j=0$ and $j=2$ are analytic continuation of $P$ from $j\geq 4$ case, times an extra factor needed to account for definitions~\eqref{eq:TTOoddj0} and~\eqref{eq:TTOoddj2} of tensor structures for $j=0$ and $j=2$. For $j=0$ this extra factor is $16(\De_{n,0}^--5)^{2}(\De_{n,0}^--3)^{2}(\De_{n,0}^--1)^{2}(\De_{n,0}^-)^{2}(\De_{n,0}^-+2)^{2}$, while for $j=2$ it is $16(\De_{n,2}^--5)^{2}(\De_{n,2}^-+2)^{2}$.

\paragraph{Comparison with numerical calculation}

We have also numerically determined the OPE coefficients for intermediate operators with dimension $\De\leq 10$ by series expanding the four-point function~\eqref{eq:TTTTMFT} and matching the coefficients to explicit $\<TTTT\>$ conformal blocks computed in~\cite{Dymarsky:2017yzx}. The resulting numerical values are listed in appendix~\ref{app:TTTTMFTcoefficients}. The analytical expressions~\eqref{eq:firstTTTT}-\eqref{eq:lastTTTT} perfectly match these results.

\section*{Acknowledgements}

We thank Murat Kolo\u{g}lu, Eric Perlmutter and Matt Walters for  discussions. DSD and PK are supported by Simons Foundation grant 488657 (Simons Collaboration on the Nonperturbative Bootstrap), a Sloan Research Fellowship, and a DOE Early Career Award under
grant No.\ DE-SC0019085. PK is supported by DOE grant No.\ DE-SC0009988.
DK is supported by Simons Foundation grant 488649 (Simons Collaboration on the Nonperturbative Bootstrap) and by the National Centre of Competence in
Research SwissMAP funded by the Swiss National Science Foundation.

\newpage

\appendix

\section{Alternative derivation of inner product between partial waves}
\label{eq:explicitinnerproduct}

The inner product between scalar partial waves was computed in \cite{Caron-Huot:2017vep} by examining the OPE limit of conformal blocks. Here, we follow the same approach to verify formula (\ref{eq:innerproductcpw}) for the inner product between partial waves in general representations.

An inner product between two partial waves is given by
\be
\label{eq:innerproductwerecomputing}
\p{\Psi^{\cO_i(pq)}_\cO, \Psi_{\tl \cO'^\dag}^{\tl \cO_i^\dag (p'q')}}
&=
\int \frac{d^d x_1 \cdots d^d x_4}{\vol\SO(d+1,1)} \Psi^{\cO_i(pq)}_\cO(x_1,x_2,x_3,x_4) \Psi_{\tl \cO'^\dag}^{\tl \cO_i^\dag (p'q')}(x_1,x_2,x_3,x_4) \nn\\
&= \frac{1}{2^d \vol\SO(d-1)} \int d^d x \Psi^{\cO_i(pq)}_\cO(0,x,e,\oo) \Psi_{\tl \cO'^\dag}^{\tl \cO_i^\dag(p'q')}(0,x,e,\oo).
\ee
In the second line, we gauge-fixed $x_1=0,x_3=e,x_4=\oo$, where $e$  is a unit vector, and set $x_2=x$. The factor $\vol\SO(d-1)$ is the volume of the stabilizer group of the points $0,e,\oo$. The factor $1/2^d$ is a Fadeev-Popov determinant. (We could additionally gauge-fix $x$ to lie inside a 2-plane, resulting in an integral over cross-ratios, but that will not be useful in this computation.) Let us assume that $\cO,\cO'$ have dimensions $\De=\frac d 2 + is, \De'=\frac d 2 + is'$ with $s,s'>0$. (The case where $s$ or $s'$ is negative can be obtained by using a symmetry transformation of the partial waves under $s\to -s$.)

Recall that a partial wave has the decomposition into conformal blocks
\be
\Psi^{\cO_i(pq)}_\cO(x_1,x_2,x_3,x_4) &= S(\cO_3\cO_4[\tl \cO^\dag])^{q}{}_{c} G^{(pc)}_{\cO}(x_1,x_2,x_3,x_4) + S(\cO_1\cO_2[\cO])^p{}_c G^{(cq)}_{\tl \cO}(x_1,x_2,x_3,x_4).
\ee
Thus, the inner product contains a term
\be
\p{\Psi^{(pq)}_\cO,\Psi_{\tl \cO'^\dag}^{(p'q')}}
&\supset
\frac{S(\cO_3\cO_4[\tl\cO^\dag])^q{}_c S(\tl\cO_3^\dag \tl \cO_4^\dag[\cO'])^{q'}{}_{c'}}{2^d \vol\SO(d-1)} \int d^d x G^{(pc)}_\cO(0,x,e,\oo) G_{\tl \cO'}^{(p'c')}(0,x,e,\oo),
\label{eq:containsaterm}
\ee
together with three other terms.

The inner product is expected to be proportional to $\de(s-s')$, where $\De=\frac d 2 + is, \De'=\frac d 2 + is'$. Such a $\de$-function singularity can only come from the OPE limit $x\sim 0$, where the radial integral is of Mellin type. Thus, let us examine the OPE singularity in more detail. Let us introduce the shorthand notation
\be
p^{a_1 a_2 a} &\equiv \<\cO_1^{a_1}(0) \cO_2^{a_2}(e) \cO^a(\oo)\>^{(p)},\nn\\
 q^{a_3 a_4 \bar a} & \equiv \<\cO_3^{a_1}(0) \cO_4^{a_2}(e) \cO^{\dag \bar a}(\oo)\>^{(q)},\nn\\
 g^{b \bar c} &\equiv \<\cO^b(0) \cO^{\dag \bar c}(\oo)\> = \lim_{L\to \oo} L^{2\De}\<\cO^b(0) \cO^{\dag \bar c}(L e)\> ,
\ee
where $e$ is a unit vector.
Here, the $a_i$ are indices for the Lorentz representations $\rho_i$ of the operators $\cO_i$, $a$ is an index for the representation $\rho$ of $\cO$, and $\bar a$ is an index for the dual reflected representation $\rho^\dag = (\rho^R)^*$.

The leading term in the OPE limit of a block formed from $p$ and $q$ is
\be
\label{eq:blockopelimit}
G^{(pq)a_1a_2a_3a_4}_{\cO}(0,x,e,\oo) &\sim p^{a_1 a_2 a} R_a{}^b(\hat x)|x|^{\De-\De_1-\De_2} (g^{-1})_{b\bar b} q^{a_3 a_4 \bar b}.
\ee
Here $R(\hat x)$ is any rotation that takes the unit vector $e$ to $\hat x=x/|x|$. (The result is invariant under different choices of $\R(\hat x)$ because the three-point structures are $\SO(d-1)$-invariant.)

In the OPE limit, the integral in (\ref{eq:containsaterm}) becomes
\be
\int d^d x G^{(pc)}_\cO G^{(p'c')}_{\tl \cO'} &\sim \int \frac{d^d x}{|x|^d} |x|^{\De-\De'} p^{a_1 a_2 a} R_a{}^b(\hat x) (g^{-1})_{b\bar b} c^{a_3 a_4 \bar b}  p'_{a_1 a_2 e} R^{e}{}_{f}(\hat x) (g'^{-1})^{f\bar f} c'_{a_3 a_4 \bar f}.
\ee
Here, $g'_{f \bar f} = \< \tl \cO^\dag_{f}(0)\tl \cO_{\bar f}(\oo)\>$.
The angular integral can be performed using the Schur orthogonality relation,
\be
\int_{S^{d-1}} d\hat x R_a{}^b(\hat x) R^{e}{}_{f}(\hat x) &= \frac{\vol S^{d-1}}{\dim \rho} \de_a^e \de^b_f.
\ee
(To derive the coefficient out front, take the trace with respect to the indices $a,e$ and use that $R_a{}^b(\hat x)$ and $R^a{}_b(\hat x)$ are inverse-transposes.)
Furthermore, we have
\be
(g^{-1})_{b\bar b} (g'^{-1})^{b\bar f} &= \frac{\dim\rho}{g_{b\bar b} g'^{b \bar b}}\de_{\bar b}^{\bar f},
\ee
since the product $g_{b\bar b} g'^{b \bar f}$ is proportional to the identity matrix.
Thus, we get
\be
&\sim \frac{\vol S^{d-1}}{g_{b\bar b} g'^{b \bar b}} (p^{a_1 a_2 a} p'_{a_1 a_2 a})(c^{a_3 a_4 \bar b}c_{a_3 a_4 \bar b}) \int \frac{dr}{r} r^{\De-\De'} \nn\\
&= \frac{\vol S^{d-1}(2^d \vol \SO(d-1))^2}{\<\cO^b(0) \cO^{\dag \bar b}(\oo)\>\<\tl \cO_{\bar b}(\oo) \tl \cO^\dag_{b}(0)\>} 
 \p{\<\cO_1 \cO_2 \cO\>^{(p)},\<\tl \cO_1^\dag \tl \cO_2^\dag \tl \cO^\dag\>^{(p')}}\p{\<\cO_3 \cO_4 \cO^\dag \>^{(c)},\<\tl \cO_3^\dag \tl \cO_4^\dag \tl \cO\>^{(c')}} \nn\\
&\qquad \x
\pi \de(s-s')  + \dots
\ee
Here, we have computed the $\de$-function term coming from the region near $r=0$. Larger $r$ contributes non-singular terms that must cancel in the final answer, since the inner product is proportional to a $\de$-function. We have also re-introduced the three-point pairings using (\ref{eq:threeptpairinggeneral}).

Once we have a $\de$-function restricting $\De=\De'$, we can use
\be
&S(\cO_3\cO_4[\tl\cO^\dag])^q{}_c S(\tl\cO_3^\dag \tl \cO_4^\dag[\cO])^{q'}{}_{c'} \p{\<\cO_3 \cO_4 \cO^\dag \>^{(c)},\<\tl \cO_3^\dag \tl \cO_4^\dag \tl \cO\>^{(c')}} \nn\\
&= \p{\<\cO_3 \cO_4 \bS[\tl \cO^\dag] \>^{(q)},\<\tl \cO_3^\dag \tl \cO_4^\dag \bS[\cO]\>^{(q')}} \nn\\
&= \p{\<\cO_3 \cO_4 \tl \cO^\dag \>^{(q)},\<\tl \cO_3^\dag \tl \cO_4^\dag \bS^2[\cO]\>^{(q')}} \nn\\
&= \cN(\De,\rho) \p{\<\cO_3 \cO_4 \tl \cO^\dag \>^{(q)},\<\tl \cO_3^\dag \tl \cO_4^\dag \cO\>^{(q')}}.
\ee
Putting everything together, we find that the term (\ref{eq:containsaterm}) contributes
\be
\p{\Psi^{(pq)}_\cO,\Psi_{\tl \cO'^\dag}^{(p'q')}}&\supset 
\frac{2^d \vol S^{d-1}\vol \SO(d-1) \cN(\De,\rho)}{\<\cO^b(0) \cO^{\dag \bar b}(\oo)\>\<\tl \cO_{\bar b}(\oo) \tl \cO^\dag_{b}(0)\>}  \nn\\
&\quad \x \p{\<\cO_1 \cO_2 \cO\>^{(p)},\<\tl \cO_1^\dag \tl \cO_2^\dag \tl \cO^\dag\>^{(p')}}
 \p{\<\cO_3 \cO_4 \tl \cO^\dag \>^{(q)},\<\tl \cO_3^\dag \tl \cO_4^\dag \cO\>^{(q')}} \pi \de(s-s') \nn\\
&= \frac{\p{\<\cO_1 \cO_2 \cO\>^{(p)},\<\tl \cO_1^\dag \tl \cO_2^\dag \tl \cO^\dag\>^{(p')}}
 \p{\<\cO_3 \cO_4 \tl \cO^\dag \>^{(q)},\<\tl \cO_3^\dag \tl \cO_4^\dag \cO\>^{(q')}}}{\mu(\De,\rho)} \pi \de(s-s'),
 \label{eq:halftheanswer}
\ee
where in the last line we used (\ref{eq:relationbetweenshadowsqandplancherel}) to rewrite the answer in terms of the Plancherel measure.

The inner product (\ref{eq:innerproductwerecomputing}) contains three other terms involving integrals of $G_{\tl \cO} G_{\tl \cO'}$, $G_{\cO}G_{\cO'}$, and $G_{\tl \cO} G_{\cO'}$. The terms $G_{\tl \cO} G_{\tl \cO'}$, $G_{\cO}G_{\cO'}$ give radial integrals of the form $\int dr r^{\tl \De+\tl \De'-d-1}$ and $\int dr r^{\De+\De'-d-1}$, neither of which can contribute a $\de$-function near $r=0$ because we have restricted $s,s'>0$. Meanwhile the term $G_{\tl \cO} G_{\cO'}$ can contribute a $\de$-function and does so with the same coefficient as in (\ref{eq:halftheanswer}). Thus, the final answer for the inner product is twice (\ref{eq:halftheanswer}), which agrees with  (\ref{eq:innerproductcpw}).

\section{Plancherel measure for spinor representation}
\label{app:spinorplancherel}

In this appendix we compute $\mu(\De,S)$. For simplicity, first assume that $d$ is odd. The spinor representation is then the representation on which we have the action of Dirac gamma matrices
\be
	\{\gamma^\mu,\gamma^\nu\}=2\delta^{\mu\nu}.
\ee
The kernel which implements the shadow transform is then given by
\be
	K_\De(x)=\frac{\gamma^\mu x_\mu}{x^{2\De+1}},
\ee
and its Fourier transform is given by
\be
	K_\De(p)=\cF_{2\De,1}\gamma^\mu p_\mu p^{2\De-1-d},
\ee
where $\cF_{\De,J}$ was defined in~\eqref{eq:fouriercoefficient}. Square of the shadow transform is then
\be
	\cN(\De,S)\mathbf{1}=K_{\tl \De}(p)K_\De(p)=\cF_{2\De,1}\cF_{2\tl\De,1}\gamma^\mu\gamma^\nu p_\mu p_\nu p^{-2}=\cF_{2\De,1}\cF_{2\tl\De,1}\mathbf{1}=\frac{(-1)^n\pi^d\cot\pi\De}{(\De+\half-d)_d}\mathbf{1},
\ee
where we used that $d=2n+1$. According to~\eqref{eq:relationbetweenshadowsqandplancherel} we have
\be\label{eq:spinorKtomu}
	\mu(\De,S) = \frac{1}{\cN(\De,S)}\frac{\mathrm{tr}[K_{\tl\De}(-\infty)K_{\tl\De}(\infty)]}{2^d\vol\SO(d)},
\ee
where $K$ is position space. We have
\be
	\mathrm{tr}[K_{\tl\De}(-\infty)K_{\tl\De}(\infty)]=-\mathrm{tr}[\gamma^1\gamma^1]=-\dim S,
\ee
and thus
\be\label{eq:plancherelspinorodddim}
	\mu(\De,S)=(-1)^{n+1}(\De+\half-d)_d\dim S\frac{\tan\pi\De}{(2\pi)^d\vol\SO(d)}.
\ee
For $d=3$ and $n=1$ this agrees with $j=\half$ case of~\eqref{eq:3dplancherel}. Also, one can check that~\eqref{eq:plancherelspinorodddim} is positive on the principal series $\De=\frac{d}{2}+is$.

The above discussion gives the answer in the case of odd dimensions. In the case of even dimension we have a slight complication since the Dirac representation is the direct sum $S\oplus \bar S$, where $\bar S$ is the reflected representation of $S$, with $m_{d,n}=-\half$. The shadow kernel exchanges the two representations, and thus we now have
\be
	\cN(\De,S)\mathbf{1}\oplus \mathbf{1}=K_{\tl \De}(p)K_\De(p)=\cF_{2\De,1}\cF_{2\tl\De,1}\gamma^\mu\gamma^\nu p_\mu p_\nu p^{-2}=\cF_{2\De,1}\cF_{2\tl\De,1}\mathbf{1}\oplus \mathbf{1}.
\ee
Simplifying with $d=2n$ we obtain in this case
\be
	\cN(\De,S)=\frac{(-1)^{n+1}\pi^{2n}}{(\De+\half-d)_d}.
\ee
Equation~\eqref{eq:spinorKtomu} should be replaced with
\be
	\mu(\De,S) = \frac{1}{\cN(\De,S)}\frac{\mathrm{tr}_S[K_{\tl\De}(-\infty)K_{\tl\De}(\infty)]}{2^d\vol\SO(d)},
\ee
where we trace only over the irreducible component $S$. Using this we obtain
\be\label{eq:plancherelspinorevendim}
	\mu(\De,S)=(-1)^{n}(\De+\half-d)_d\dim S\frac{1}{(2\pi)^d\vol\SO(d)}.
\ee
This is again positive on the principal series and in $d=4$ agrees with $(\ell,\bar\ell)=(1,0)$ case of~\eqref{eq:plancherel_measure_4d}.

\section{Details of the 4d formalism}
\label{app:details_4d_formalism}

In this appendix we provide all the necessary details of the 4d formalism. In secttion~\ref{app:conventions_4d} we start by setting up the notation and conventions. In section~\ref{app:conformally_invariant_pairing} we define the main object used in harmonic analysis, namely the conformally invariant pairing. We proceed in section~\ref{app:properties_weight_shifting_operators} by carefully discussing properties of the 4d weight-shifting operators under integration by parts, their bubble coefficients and the two-point 6j-symbols. We provide the definition of the shadow transform, compute its commutation properties with the 4d weight-shifting operators and derive the shadow square transformation in section~\ref{app:shadow_transform_4d}.
We conclude in section~\ref{sec:two_and_three_point_pairings} by computing conformally invariant pairing for the two-point structure and for several examples of three-point structures. 

We point out right away that in this appendix we work in the Minkowski metric. This is in contrast to section~\ref{sec:harmonic_analysis_review} where it is more natural to use the Euclidean metric to introduce harmonic analysis. All our results obtained here however remain unchanged also in the Euclidean metric.

\subsection{Conventions}
\label{app:conventions_4d}

In the Minkowski metric the conformal group is $\SO(4,2)$. We denote the representation of local operators as $V_{\De,\rho}$, where $\rho$ is the representation of the spin sub-group $\SO(1,3)$. We will work with its covering group $\SL(2,C)$. All its representations are labeled by a pair of non-negative integers $(\ell,\bar\ell)$ which also provide the number of undotted and dotted indices respectively in the two-component formalism.\footnote{\label{foot:spin_irreps_4d}
There are in fact four different representations $(\ell,\bar\ell)$, $(\ell^*,\bar\ell)$, $(\ell,\bar\ell{}^*)$ and $(\ell^*,\bar\ell{}^*)$ which are however equivalent to each other. The superscript $*$ stays for dual. The equivalence is established by products of $\epsilon$-symbols $\epsilon_{\alpha\beta}$, $\epsilon^{\alpha\beta}$, $\epsilon_{\dot\alpha\dot\beta}$ and $\epsilon^{\dot\alpha\dot\beta}$. The representation $(\ell,\bar\ell)$ is defined as an object with $\ell$ lower undotted and $\bar\ell$ lower dotted indices. The dual version in each index is represented by rising relevant indices (undotted, dotted or both). In index-full notation we never distinguish between equivalent representation since it is obvious what representation we deal with from the position of indices.}{}$\,$\footnote{In the language of section~\ref{sec:shadow_representations} the reflected representation is $(\ell,\bar\ell)^R = (\bar\ell, \ell)$, the dual representation is  $(\ell,\bar\ell)^* =(\ell{}^*,\bar\ell{}^*)\sim(\ell,\bar\ell)$ and the conjugate representation $(\ell,\bar\ell)^\dag=(\ell,\bar\ell)^{*R}=(\bar\ell{}^*,\ell^*)\sim(\bar\ell, \ell)$. Here $\sim$ stands for equivalent to.} In what follows we use the 4d conventions of appendix A in~\cite{Cuomo:2017wme}.

We define a local operator transforming in the representation $V_{\De,\ell,\bar\ell}$ in the index-free notation as
\begin{equation}
\label{eq:index_free_4d}
\cO^{(\ell,\bar\ell)}_\Delta(x,s,\bar s) \equiv 
\cO_{\alpha_1\ldots\alpha_\ell}^{\dot\beta_1\ldots\dot\beta_{\bar\ell}}\times
\left(s^{\alpha_1}\cdots s^{\alpha_\ell}\right)
\left(\bar s_{\dot\beta_1}\cdots\bar s_{\dot\beta_{\bar\ell}}\right),
\end{equation}
where $s$ and $\bar s$ are  spinor polarizations.\footnote{In the language of the footnote~\ref{foot:spin_irreps_4d} we actually encode an operator transforming in $(\ell,\bar\ell{}^*)$ spin representation.}
To go from index-free to index-full notation one takes derivatives with respect to spinor polarizations. For convenience we define the following short-hand notation
\begin{equation}\label{eq:derivatives_polarizations_definitions}
[\partial_{s}]^\alpha\equiv\frac{\partial}{\partial s_\alpha},\quad
[\partial_{\bar s}]^{\dot\alpha}\equiv\frac{\partial}{\partial \bar s_{\dot\alpha}},\quad
[\partial_{s}]_\alpha\equiv-\frac{\partial}{\partial s^\alpha},\quad
[\partial_{\bar s}]_{\dot\alpha}\equiv-\frac{\partial}{\partial \bar s^{\dot\alpha}}.
\end{equation}
The minus signs are introduced to consistently rise and lower indices in a standard way
\begin{equation}
[\partial_s]_\alpha=\epsilon_{\alpha\beta}[\partial_s]^\beta,\qquad
[\partial_{\bar s}]_{\dot\alpha}=\epsilon_{\dot\alpha\dot\beta}[\partial_{\bar s}]^{\dot\beta}.
\end{equation}
We have then
\begin{equation}
\label{eq:index_free_to_index_full}
\cO_{\alpha_1\ldots\alpha_\ell}^{\dot\beta_1\ldots\dot\beta_{\bar\ell}}=\frac{(-1)^\ell}{\ell!\,\bar\ell!}\;
[\partial_s]_{\alpha_1}\ldots[\partial_s]_{\alpha_\ell}\;
[\partial_{\bar s}]^{\dot\alpha_1}\ldots[\partial_{\bar s}]^{\dot\alpha_{\bar\ell}}\;
\cO^{(\ell,\bar\ell)}_\De(x,s,\bar s).
\end{equation}

We will do several computations in the conformal frame. For that we will also need the decomposition of auxiliary spinors into their components
\begin{equation}
s_\alpha \equiv \begin{pmatrix}
\xi\\
\eta
\end{pmatrix},\quad
s^\alpha = \epsilon^{\alpha\beta} s_\beta =
\begin{pmatrix}
\eta\\
-\xi
\end{pmatrix},\quad
\bar s_{\dot\alpha} \equiv \begin{pmatrix}
\bar\xi\\
\bar\eta
\end{pmatrix},\quad
\bar s^{\dot\alpha} = \epsilon^{\dot\alpha\dot\beta} s_{\dot\beta} =
\begin{pmatrix}
\bar\eta\\
-\bar\xi
\end{pmatrix},
\end{equation} 
where the $\epsilon$-symbol has the following components
\begin{equation}
\epsilon^{12} = - \epsilon^{21} = 
- \epsilon_{12} = \epsilon_{21} = +1.
\end{equation}
The convention for $\sigma^\mu_{\alpha\dot\beta}$ and $\bar\sigma^{\mu\,\dot\alpha\beta}$ matrices is as follows $\bar\sigma^0=+\sigma^0$, $\bar\sigma^1=-\sigma^1$, $\bar\sigma^2=-\sigma^2$ and $\bar\sigma^3=-\sigma^3$, 
where
\begin{equation}
\sigma^{0}=\begin{pmatrix}
-1 &  0\\
0  & -1
\end{pmatrix},\quad
\sigma^{1}=\begin{pmatrix}
0 &  +1\\
+1  & 0
\end{pmatrix},\quad
\sigma^{2}=\begin{pmatrix}
0 &  -i\\
+i  & 0
\end{pmatrix},\quad
\sigma^{3}=\begin{pmatrix}
+1 &  0\\
0  & -1
\end{pmatrix}.
\end{equation}

\subsection{Conformally invariant pairing}
\label{app:conformally_invariant_pairing}

Let us now define a full contraction of spin indices for a pair of local operators $\cO$ in the spin representation $(\ell, \bar\ell)$ and $\cO'$ in the dual representation $(\ell, \bar\ell)^*$
\begin{equation}
\label{eq:contraction_indices_4d}
\cO_{\alpha_1\ldots\alpha_\ell}^{\dot\beta_1\ldots\dot\beta_{\bar\ell}}
\;
\cO'{}_{\dot\beta_1\ldots\dot\beta_{\bar\ell}}^{\alpha_1\ldots\alpha_\ell}=
\epsilon^{\alpha_1\beta_1}\ldots\epsilon^{\alpha_\ell\beta_\ell}
\epsilon_{\dot\beta_1\dot\alpha_1}\ldots\epsilon_{\dot\beta_{\bar\ell}\dot\alpha_{\bar\ell}}
\cO{}_{\alpha_1\ldots\alpha_\ell}^{\dot\beta_1\ldots\dot\beta_{\bar\ell}}\;
\cO'{}_{\beta_1\ldots\beta_\ell}^{\dot\alpha_1\ldots\dot\alpha_{\bar\ell}}.
\end{equation}
In index-free formalism the contraction of indices~\eqref{eq:contraction_indices_4d}
is written as follows
\begin{equation}\label{eq:inde-free_pairing}
\cO^{(\ell,\bar\ell)}_\De(x,s,\bar s)
\;\overleftrightarrow{G}_{s,\bar s}^{(\ell,\bar\ell)}\;
\cO{}^{(\ell,\bar\ell)}_{\De'}(x,s,\bar s),
\end{equation}
where $\overleftrightarrow{G}_{s,\bar s}$ is the ``gluing'' differential operator defined using~\eqref{eq:index_free_to_index_full} as
\begin{equation}\label{eq:glueing_differential_operator}
\overleftrightarrow{G}_{s,\bar s}^{(\ell,\bar\ell)}\equiv
\frac{1}{\ell!^2}\,
\prod_{i=1}^\ell
[\overleftarrow\partial_s]_{\alpha_i}
\,\epsilon^{\alpha_i\beta_i}\,
[\overrightarrow\partial_s]_{\beta_i}
\;\times\;
\frac{1}{\bar{\ell}!^2}\,
\prod_{j=1}^{\bar{\ell}}
[\overleftarrow\partial_{\bar s}]^{\dot\beta_j}
\,\epsilon_{\dot\beta_j\dot\alpha_j}\,
[\overrightarrow\partial_{\bar s}]^{\dot\alpha_j}.
\end{equation}

We are finally in the position to define a conformally invariant pairing
\begin{equation}
\label{eq:pairing_4d}
\p{
\cO^{(\ell,\bar\ell)}_\De,
\cO{}^{(\ell,\bar\ell)}_{\tl\De}
}\equiv
\int d^4 x\,
\cO^{(\ell,\bar\ell)}_\De(x,s,\bar s)
\;\overleftrightarrow{G}_{s,\bar s}^{(\ell,\bar\ell)}\;
\cO{}^{(\ell,\bar\ell)}_{\tl\De}(x,s,\bar s),\quad
\tl\De\equiv 4-\De.
\end{equation}
We stress that the expression~\eqref{eq:pairing_4d} is formal in a sense that the operators 
here are merely an indication of the representations used. Thus, despite the seeming spin-statistics these operators must always commute. Due to the definitions~\eqref{eq:inde-free_pairing} and~\eqref{eq:glueing_differential_operator} the following property holds
\begin{equation}
\label{eq:property_4d}
\p{
	\cO^{(\ell,\bar\ell)}_\De,
	\cO{}^{(\ell,\bar\ell)}_{\tl\De}
}=
(-1)^{\ell+\bar\ell}\times
\p{
	\cO{}^{(\ell,\bar\ell)}_{\tl\De},
	\cO^{(\ell,\bar\ell)}_\De
}.
\end{equation}

\subsection{Properties of weight-shifting operators}
\label{app:properties_weight_shifting_operators}

We study here in details the properties of the 4d weight-shifting operators~\eqref{eq:4Doperators_4} and~\eqref{eq:4Doperators_dual_4}. They are naturally defined in 6d embedding formalism. We discuss their 4d form in section~\ref{app:6d_4d_projection}. We proceed in section~\ref{app:integration_by_parts} by studying their integration by parts properties with respect to conformally invariant pairings. We summarize some other properties of the 4d weight-shifting in section~\ref{app:weight-shifting-formulas-4d} such as the two-point 6j-symbols and the bubble coefficients.

\subsubsection{6d to 4d projection}
\label{app:6d_4d_projection}
The 6d embedding space in the light-cone coordinates is defined as follows
\begin{equation}
X^M=(X^\mu,X^+,X^-),\quad \mu=1,\ldots,4,
\end{equation}
with the metric
\begin{equation}
\eta_{\mu\nu}=\{-,+,+,+\},\quad
\eta_{+-}=\eta_{-+}=\frac{1}{2}.
\end{equation}

In what follows we use indices $a,b$ for the fundamental and anti-fundamental representations of $SU(2,2)$. The isomorphism between $SO(2,4)$ and $SU(2,2)$ conformal algebras is established via the matrices
\begin{align}
\Sigma_{ab}^\mu &=\begin{pmatrix}
0 & -(\sigma^\mu\epsilon)_\alpha^{\;\;\dot\beta} \\
(\bar\sigma^\mu\epsilon)^{\dot\alpha}_{\;\;\beta} & 0
\end{pmatrix},\quad 
&
\Sigma_{ab}^+&=\begin{pmatrix}
0 & 0 \\
0 & 2\,\epsilon^{\dot\alpha\dot\beta}
\end{pmatrix},\quad
&
\Sigma_{ab}^-&=\begin{pmatrix}
-2\,\epsilon_{\alpha\beta} & 0 \\
0 & 0
\end{pmatrix},\\
\overline\Sigma^{\mu\,ab}&=\begin{pmatrix}
0 & -(\epsilon\sigma^\mu)^\alpha_{\;\;\dot\beta} \\
(\epsilon\bar\sigma^\mu)_{\dot\alpha}^{\;\;\beta} & 0
\end{pmatrix},\quad
&
\overline\Sigma^{+\,ab}&=\begin{pmatrix}
-2\,\epsilon^{\alpha\beta} & 0 \\
0 & 0
\end{pmatrix},\quad
&
\overline\Sigma^{-\,ab}&=\begin{pmatrix}
0 & 0 \\
0 & 2\,\epsilon_{\dot\alpha\dot\beta}
\end{pmatrix}.
\end{align}

We choose the following simple uplift of the 4d operator $\cO$ into the 6d operator $O$
\begin{equation}\label{eq:uplift}
O(X,S,\overline S)=\left(X^+\right)^{-\kappa}\cO\left(\frac{X^\mu}{X^+},S_\alpha,\overline S_{\dot\alpha}\right),\quad\kappa\equiv\Delta+\frac{\ell+\bar{\ell}}{2}.
\end{equation}
We always set $X^+=1$ where there is no place for confusion.
For 6d coordinates we have
\begin{equation}
X^M\Big|_{proj}=(x^\mu,\;1,\;-x^2)
\end{equation}
or equivalently
\begin{equation}
X_{ab}\Big|_{proj}=
\begin{pmatrix}
-\epsilon_{\alpha\beta}  &\;\; -x_{\alpha}{}^{\dot\beta}\\
x^{\dot\alpha}{}_{\beta}  &\;\; -\epsilon^{\dot\alpha\dot\beta}x^2
\end{pmatrix},\qquad
\overline X^{ab}\Big|_{proj}=
\begin{pmatrix}
\epsilon^{\alpha\beta}x^2  &\;\; -x^{\alpha}{}_{\dot\beta}\\
x_{\dot\alpha}{}^{\beta}  &\;\; \epsilon_{\dot\alpha\dot\beta}
\end{pmatrix},
\end{equation}
where we have defined
\begin{equation}\label{eq:definitions_x_1}
x_{\alpha\dot\beta}\equiv x_\mu\sigma^\mu_{\alpha\dot\beta},\qquad
x^{\dot\alpha\beta}\equiv x_\mu\bar\sigma^\mu{}^{\dot\alpha\beta},
\end{equation}
\begin{equation}\label{eq:definitions_x_2}
x_{\alpha}{}^{\dot\beta}\equiv x_\mu (\sigma^\mu\epsilon)_\alpha{}^{\dot\beta},\quad
x^{\dot\alpha}{}_{\beta}\equiv x_\mu (\bar\sigma^\mu\epsilon)^{\dot\alpha}{}_{\beta},\quad
x^{\alpha}{}_{\dot\beta}\equiv x_\mu (\epsilon\sigma^\mu)^\alpha{}_{\dot\beta},\quad
x_{\dot\alpha}{}^{\beta}\equiv x_\mu (\epsilon\bar\sigma^\mu)_{\dot\alpha}{}^{\beta}.
\end{equation}
For 6d polarizations we have
\begin{equation}
S_a\Big|_{proj}=(s_\alpha,\;-x^{\dot\alpha\lambda}s_{\lambda}),\qquad
\overline S^a\Big|_{proj}=(\bar s_{\dot\lambda} x^{\dot\lambda\alpha},\;\bar s_{\dot\alpha}).
\end{equation}
Using the uplift~\eqref{eq:uplift} one can also deduce the projection of the 6d derivatives
\begin{equation}
\label{eq:derivatives_projection}
\frac{\partial}{\partial X^M}\Bigg|_{proj}=(\partial_{\mu},\;-\kappa-x\cdot\partial,\;0),\quad
\frac{\partial}{\partial S_b}\Bigg|_{proj}=(\partial_{s}^{\beta},0),\quad
\frac{\partial}{\partial\overline S^b}\Bigg|_{proj}=(0,\;\partial_{\bar s}^{\dot\beta}).
\end{equation}
See definition~\eqref{eq:derivatives_polarizations_definitions}. Extra care should be taken when dealing with derivatives of second order and higher in coordinates, we will not need them however in this work. The first entry in~\eqref{eq:derivatives_projection} can be written equivalently as
\begin{equation}
\partial_{ab}\Big|_{proj}=
\begin{pmatrix}
0  &\;\; -\partial_{\alpha}{}^{\dot\beta}\\
\partial^{\dot\alpha}{}_{\beta}  &\;\; -2\,(\kappa+x\cdot\partial)\,\epsilon^{\dot\alpha\dot\beta}
\end{pmatrix},\qquad
\overline \partial^{ab}\Big|_{proj}=
\begin{pmatrix}
2\,(\kappa+x\cdot\partial)\,\epsilon^{\alpha\beta}  &\;\; -\partial^{\alpha}{}_{\dot\beta}\\
\partial_{\dot\alpha}{}^{\beta}  &\;\; 0
\end{pmatrix},
\end{equation}
where the 4d derivatives are defined in an exact manner as the 4d coordinates in~\eqref{eq:definitions_x_2} with $x_\mu$ replaced by $\partial_\mu$.

Let us project some elementary pieces entering the weight-shifting operators~\eqref{eq:4Doperators_4}
\begin{align}
T_1^a &\equiv S_b\overline\partial^{ab},
&
T_1^a\Big|_{proj}&= (
2\,s^\alpha\,(\kappa+x\cdot\partial)+
(sx\partial)^{\alpha},\;
-(s\partial)_{\dot\alpha}
),\\
T_2 &\equiv \left(S\overline\partial\partial_{\overline S}\right)=-T_1^a\frac{\partial}{\partial\overline S^a},
&
T_2\Big|_{proj}&=(s\partial\partial_{\bar s}),\\
T_3 &\equiv(\partial_{\overline S}\partial_S),
&
T_3\Big|_{proj}&=0,\\
T_4^a&\equiv X_{bc}\overline\partial^{ab}\partial_{S}^c,
&
T_4^a\Big|_{proj}&=(
-2\,(\kappa+x\cdot\partial)\partial_s^\alpha
-(\partial_s x \partial)^{\alpha},\;
(\partial_s\partial)_{\dot\alpha}
),\\
T_5&\equiv X_{bc}\overline\partial^{ba}\partial_{S}^c\partial_{\overline S,\,a}=-T_4^a\,\partial_{\overline S,\,a},
&
T_5\Big|_{proj}&=-(\partial_s\partial\partial_{\bar s}).
\end{align}
Combining them we get the projection of the weight-shifting operators from the 6d embedding space to the 4d physical space
\begin{align}
\label{eq:D1_projection}
\cD^a_{-0+}\Big|_{proj}
&=
\left(\bar s_{\dot\alpha}\bar x^{\dot\alpha\alpha},\;\bar s_{\dot\alpha}\right),\\
\label{eq:D2_projection}
\cD^a_{-0-}\Big|_{proj}
&=
\left(\partial_{\bar s\,\dot\alpha}\bar x^{\dot\alpha\alpha},\;\partial_{\bar s\,\dot\alpha}\right),\\
\label{eq:D3_projection}
\cD^a_{++0}\Big|_{proj}
&=
\big(
\ldots,\;
-\bar a\times(s\partial)_{\dot{\alpha}}+\bar s_{\dot\alpha}(s\partial\partial_{\bar s})
\big),\\
\label{eq:D4_projection}
\cD^a_{+-0}\Big|_{proj}
&=
\big(
\ldots,\;
(c+\bar\ell)\times(\partial_s\partial)_{\dot\alpha}+
(\partial_s\partial\;\bar s)\partial_{\bar s\,\dot\alpha}
\big).
\end{align}
In the very last expression we have used the relation
\begin{equation}\label{eq:useful_relation}
\bar s_{\dot\alpha}(\partial_s\partial\partial_{\bar s})=
(\bar s\partial_{\bar s})(\partial_s\partial)_{\dot\alpha}+(\partial_s\partial\;\bar s)\partial_{\bar s\,\dot\alpha},
\end{equation}
which follows from the two identities
\begin{equation}
\epsilon_{\dot\alpha\dot\lambda}\epsilon^{\dot\beta\dot\tau}=
-\delta_{\dot\alpha}^{\dot\beta}\delta_{\dot\lambda}^{\dot\tau}
+\delta_{\dot\alpha}^{\dot\tau}\delta_{\dot\lambda}^{\dot\beta}
\;\Rightarrow\;
\bar s_{\dot\alpha}\partial_{\bar s}^{\dot\beta}=
(\bar s\partial_{\bar s})\,\delta_{\dot\alpha}^{\dot\beta}+
\bar s^{\dot\beta}\partial_{\bar s\,\dot\alpha},
\quad
(\bar s\partial_{\bar s})\,\cO^{(\ell,\bar\ell)}=\bar\ell\,\cO^{(\ell,\bar\ell)}.
\end{equation}
We omit some parts of the formulas~\eqref{eq:D1_projection}-\eqref{eq:D4_projection}, namely the $\alpha$ component for the last two operators. For understanding the integration by parts properties (addressed in the next section) it is enough to look only at the $\dot\alpha$ component. Analogous expressions can be written for the weight-shifting operators~\eqref{eq:4Doperators_dual_4}.
\begin{align}
\overline\cD_a^{-+0}\Big|_{proj}
&=
+(s_\alpha,\; -x^{\dot\alpha\beta}s_\beta),\\
\overline\cD_a^{--0}\Big|_{proj}
&=
-(
\partial_{s\,\alpha},\;
-x^{\dot\alpha\beta}\partial_{s\,\beta}
),\\
\overline\cD_a^{+0+}\Big|_{proj}
&=
\big(
-a\times{}_\alpha(\partial\bar s) +s_\alpha (\bar s\partial \partial_s),
\;\ldots
\big),\\
\overline\cD_a^{+0-}\Big|_{proj}
&=
\big(
-(c+\ell)\times{}_\alpha(\partial\partial_{\bar s})-(s\partial\partial_{\bar s})\partial_{s\,\alpha},
\;\ldots
\big).
\end{align}

\subsubsection{Integration by parts}
\label{app:integration_by_parts}
We discuss here the conformally covariant pairing~\eqref{eq:inde-free_pairing} in a presence of a weight-shifting operator on one side. We will see how in this situation the weight-shifting operator can be moved to another side of the pairing. We call this an integration by parts procedure. We start by considering spinor polarizations and then move to discussing the 4d weight-shifting operators~\eqref{eq:4Doperators_4} and~\eqref{eq:4Doperators_dual_4}.

\paragraph{Pairing of covariant objects}
We start by considering the following pairing
\begin{align}
\nn
\p{
	s^\alpha\,\cO^{(\ell,\,\bar\ell)}_\De,
	\cO^{(\ell+1,\,\bar\ell)}_{\tl\De}
}
&=
\int d^4 x\,
s^\alpha\,\cO^{(\ell,\,\bar\ell)}_\De(x,s,\bar s)
\;\overleftrightarrow{G}_{s,\bar s}^{(\ell+1,\bar\ell)}\;
\cO^{(\ell+1,\,\bar\ell)}_{\tl\De}(x,s,\bar s)\\
\nn
&=
\frac{(-1)^{\ell+1}}{(\ell+1)!\,\bar\ell!}\,
\int d^4 x\,
\cO^{(\ell,\,\bar\ell)}_\De(x,\partial_s,\partial_{\bar s})\partial_s^\alpha\;
\cO^{(\ell+1,\,\bar\ell)}_{\tl\De}(x,s,\bar s)\\
\nn
&=
-\frac{1}{\ell+1}
\int d^4 x\,
\cO^{(\ell,\,\bar\ell)}_\De(x,s,\bar s)
\;\overleftrightarrow{G}_{s,\bar s}^{(\ell,\bar\ell)}\;
\partial_s^\alpha\;
\cO^{(\ell+1,\,\bar\ell)}_{\tl\De}(x,s,\bar s)\\
&=
-\frac{1}{\ell+1}
\p{
	\cO^{(\ell,\,\bar\ell)}_\De,
	\partial_s^\alpha\cO^{(\ell+1,\,\bar\ell)}_{\tl\De}
}.
\label{eq:property1_4d}
\end{align}
By using~\eqref{eq:property_4d} and making the replacement $\ell\rightarrow\ell-1$ and $\De \leftrightarrow \tl\De$ the result~\eqref{eq:property1_4d} can be brought into an equivalent form which reads as
\begin{equation}\label{eq:rule_1}
\p{
	\partial_s^\alpha\cO^{(\ell,\,\bar\ell)}_\De,
	\cO^{(\ell-1,\,\bar\ell)}_{\tl\De}
}
=+\ell\;
\p{
	\cO^{(\ell,\,\bar\ell)}_\De,
	s^\alpha\,\cO^{(\ell-1,\,\bar\ell)}_{\tl\De}
}.
\end{equation}
In an analogous way one can write the following expressions
\begin{align}\label{eq:sbar}
\p{
	\bar s_{\dot\alpha}\cO^{(\ell,\,\bar\ell)}_\De,
	\cO^{(\ell,\,\bar\ell+1)}_{\tl\De}
}
&=+
\frac{1}{\bar\ell+1}\;\,
\p{
	\cO^{(\ell,\,\bar\ell)}_\De,
	\partial_{\bar s\,\dot\alpha}\cO^{(\ell,\,\bar\ell+1)}_{\tl\De}
},\\
\label{eq:rule_2}
\p{
	\partial_{\bar s\,\dot\alpha}\cO^{(\ell,\,\bar\ell)}_\De,
	\cO^{(\ell,\,\bar\ell-1)}_{\tl\De}
}
&=\;\;\;\;-\bar\ell\;\;\;\;
\p{
	\cO^{(\ell,\,\bar\ell)}_\De,
	\bar s_{\dot\alpha}\cO^{(\ell,\,\bar\ell-1)}_{\tl\De}
}.
\end{align}
Finally let us use~\eqref{eq:sbar} to write
\begin{equation}\label{eq:result_1}
\p{
	\bar s_{\dot\alpha}\partial_{\bar s}^{\dot\beta}\,
	\cO^{(\ell,\,\bar\ell)}_\De,
	\cO^{(\ell,\,\bar\ell)}_{\tl\De}
}=
+\frac{1}{\overline\ell}
\p{
	\partial_{\bar s}^{\dot\beta}\,
	\cO^{(\ell,\,\bar\ell)}_\De,
	\partial_{\bar s\,\dot\alpha}\cO^{(\ell,\,\bar\ell)}_{\tl\De}
}.
\end{equation}
We can also rewrite this relation using~\eqref{eq:property_4d}, replacing $\De \leftrightarrow \tl\De$ and renaming the spinor indices as $\alpha\leftrightarrow\beta$. We get
\begin{equation}\label{eq:result_2}
\p{
	\cO^{(\ell,\,\bar\ell)}_\De,
	\bar s_{\dot\beta}\partial_{\bar s}^{\dot\alpha}\,
	\cO^{(\ell,\,\bar\ell)}_{\tl\De}
}=
-\frac{1}{\overline\ell}
\p{
	\partial_{\bar s\,\dot\beta}\,
	\cO^{(\ell,\,\bar\ell)}_\De,
	\partial_{\bar s}^{\dot\alpha}\cO^{(\ell,\,\bar\ell)}_{\tl\De}
}.
\end{equation}
Comparing the expressions~\eqref{eq:result_1} and~\eqref{eq:result_2} we finally conclude that
\begin{equation}\label{eq:rule_3}
\p{
	\bar s_{\dot\alpha}\partial_{\bar s}^{\dot\beta}\, 
	\cO^{(\ell,\,\bar\ell)}_\De,
	\cO^{(\ell,\,\bar\ell)}_{\tl\De}
}=-
\p{
	\cO^{(\ell,\,\bar\ell)}_\De,
	\bar s^{\dot\beta}\partial_{\bar s\,\dot\alpha}\,
	\cO^{(\ell,\,\bar\ell)}_{\tl\De}
}.
\end{equation}
Analogously the following holds
\begin{equation}\label{eq:rule_4}
\p{
	s^{\alpha}\partial_{s\,\beta}\, 
	\cO^{(\ell,\,\bar\ell)}_\De,
	\cO^{(\ell,\,\bar\ell)}_{\tl\De}
}=-
\p{
	\cO^{(\ell,\,\bar\ell)}_\De,
	s_{\beta}\partial_s^\alpha\,
	\cO^{(\ell,\,\bar\ell)}_{\tl\De}
}.
\end{equation}

\paragraph{Integration by parts procedure}
Consider now the conformally invariant pairing defined in~\eqref{eq:pairing_4d}.
Let us study it in the presence of weight-shifting operators. As discussed in section~\ref{eq:weightshiftingshadowcoeffs} the following relation holds
\begin{equation}
\p{
	\cD^a\cO^{(\ell,\bar\ell)}_\De,
	\cO{}^{(\ell{}',\bar\ell{}')}_{\tl\De{}'}
}
=
\p{
	\cO^{(\ell,\bar\ell)}_\De,
	(\cD^a)^*\cO{}^{(\ell{}',\bar\ell{}')}_{\tl\De{}'}
},
\end{equation}
where $(\cD^a)^*$ is the adjoint weight-shifting operator to $\cD^a$. To determine its explicit expression one needs to know first how each element of the weight-shifting operator transforms when moved from one side of the pairing to another: $\cD^a\rightarrow (\cD^a)^*$. For coordinates and space-time derivatives we have
\begin{equation}
x^\mu\rightarrow x^\mu,\quad
\partial_\mu \rightarrow -\partial_\mu.
\end{equation}
The second entry follows from an ordinary integration by parts procedure with dropped boundary terms. Polarizations and polarization derivatives  transform according to the previous paragraph as
\begin{equation}\label{eq:transformation_rules}
s^\alpha\rightarrow-\frac{1}{\ell+1}\partial_s^\alpha,
\quad
\partial_s^\alpha \rightarrow +\ell s^\alpha,
\quad
\bar s_{\dot\alpha}\rightarrow +\frac{1}{\bar\ell+1} \partial_{\bar s\,\dot\alpha}
\quad
\partial_{\bar s\,\dot\alpha} \rightarrow -\bar\ell \bar s_{\dot\alpha},
\end{equation}
together with
\begin{equation}
\bar s_{\dot\alpha}\partial_{\bar s}^{\dot\beta} \rightarrow -
\bar s^{\dot\beta}\partial_{\bar s\,\dot\alpha},\qquad
s^{\alpha}\partial_{s\,\beta} \rightarrow -
s_{\beta}\partial_s^\alpha.
\end{equation}
Here we assume that they act on the operator in the left-hand side in a representation $[\De,\ell,\bar\ell]$.
Using these rules, the relation~\eqref{eq:useful_relation} and the definitions~\eqref{eq:definitions_parameters_weight_shifting_operators_4d} we see that under the ``integration by parts'' we have for instance
\begin{equation}
\cD^a_{-0+}\Big|_{proj} \longleftrightarrow\; \cD^a_{-0-}\Big|_{proj},\quad
\cD^a_{++0}\Big|_{proj} \longleftrightarrow\; \cD^a_{+-0}\Big|_{proj}.
\end{equation}
Performing a straightforward computation we can restore also all the proportionality coefficients. We provide here the final result
\begin{align}
\label{eq:integration_by_parts_1}
\left(
\cD^a_{-0+}
O^{(\ell,\bar{\ell})}_\Delta,\;
O^{(\ell,\bar\ell+1)}_{\tl\Delta+1/2}\right) &= +\frac{1}{\bar\ell+1}
\left(
O^{(\ell,\bar{\ell})}_\Delta,\;
\cD^{a}_{-0-}\;
O^{(\ell,\bar\ell+1)}_{\tl\Delta+1/2}\right),\\
\label{eq:integration_by_parts_2}
\left(
\cD^a_{-0-}
O^{(\ell,\bar{\ell})}_\Delta,\;
O^{(\ell,\bar\ell-1)}_{\tl\Delta+1/2}\right) &= \;\;\;-\bar\ell\;\;\;\,
\left(
O^{(\ell,\bar{\ell})}_\Delta,\;
\cD^{a}_{-0+}\;
O^{(\ell,\bar\ell-1)}_{\tl\Delta+1/2}\right),\\
\label{eq:integration_by_parts_3}
\left(
\cD^a_{++0}
O^{(\ell,\bar{\ell})}_\Delta,\;
O^{(\ell+1,\bar\ell)}_{\tl\Delta-1/2}\right) &= -\frac{1}{\ell+1}
\left(
O^{(\ell,\bar{\ell})}_\Delta,\;
\cD^{a}_{+-0}\;
O^{(\ell+1,\bar\ell)}_{\tl\Delta-1/2}\right),\\
\label{eq:integration_by_parts_4}
\left(
\cD^a_{+-0}
O^{(\ell,\bar{\ell})}_\Delta,\;
O^{(\ell-1,\bar\ell)}_{\tl\Delta-1/2}\right) &= \;\;\;+\ell\;\;\;\,
\left(
O^{(\ell,\bar{\ell})}_\Delta,\;
\cD^{a}_{++0}\;
O^{(\ell-1,\bar\ell)}_{\tl\Delta-1/2}\right).
\end{align}
Analogously one can also obtain 
\begin{align}
\label{eq:integration_by_parts_1_bar}
\left(
\overline\cD_a^{-+0}
O^{(\ell,\bar{\ell})}_\Delta,\;
O^{(\ell+1,\bar\ell)}_{\tl\Delta+1/2}\right) &= +\frac{1}{\ell+1}
\left(
O^{(\ell,\bar{\ell})}_\Delta,\;
\overline\cD_{a}^{--0}\;
O^{(\ell+1,\bar\ell)}_{\tl\Delta+1/2}\right),\\
\label{eq:integration_by_parts_2_bar}
\left(
\overline\cD_a^{--0}
O^{(\ell,\bar{\ell})}_\Delta,\;
O^{(\ell-1,\bar\ell)}_{\tl\Delta+1/2}\right) &= \;\;\;-\ell\;\;\;\,
\left(
O^{(\ell,\bar{\ell})}_\Delta,\;
\overline\cD_{a}^{-+0}\;
O^{(\ell-1,\bar\ell)}_{\tl\Delta+1/2}\right),\\
\label{eq:integration_by_parts_3_bar}
\left(
\overline\cD_a^{+0+}
O^{(\ell,\bar{\ell})}_\Delta,\;
O^{(\ell,\bar\ell+1)}_{\tl\Delta-1/2}\right) &= -\frac{1}{\bar\ell+1}
\left(
O^{(\ell,\bar{\ell})}_\Delta,\;
\overline\cD_{a}^{+0-}\;
O^{(\ell,\bar\ell+1)}_{\tl\Delta-1/2}\right),\\
\label{eq:integration_by_parts_4_bar}
\left(
\overline\cD_a^{+0-}
O^{(\ell,\bar{\ell})}_\Delta,\;
O^{(\ell,\bar\ell-1)}_{\tl\Delta-1/2}\right) &= \;\;\;+\bar\ell\;\;\;\,
\left(
O^{(\ell,\bar{\ell})}_\Delta,\;
\overline\cD_{a}^{+0+}\;
O^{(\ell,\bar\ell-1)}_{\tl\Delta-1/2}\right).
\end{align}
From these expressions the form of $(\cD^a)^*$ is obvious.

\subsubsection{Bubble coefficients and the 6j-symbols}
\label{app:weight-shifting-formulas-4d}

One of the most important ingredients is the two-point tensor structure. We make the following choice in this work
\begin{align}\label{eq:2PointFunction}
\<
\cO_{\Delta}^{(\ell,\bar\ell)}(x_1, s_1, \bar s_1)
\cO_{\Delta}^{(\bar\ell,\ell)}(x_2, s_2, \bar s_2)
\>
&=
i^{\bar\ell-\ell}\,
X_{12}^{-\kappa}
\big[ \hat I^{12}\big]^{\ell}\big[\hat I^{21}\big]^{\bar\ell}\Big|_{\text{proj}},\\
\label{eq:invariant_I}
\hat I^{ij} &\equiv \overline S_i^a\; S_{j\,a},\\
X_{ij} &\equiv - 2\,(X_i\cdot X_j).
\end{align}
It is naturally defined in the 6d formalism and then projected ot 4d, see section~\ref{app:6d_4d_projection} for details. The form of~\eqref{eq:2PointFunction} is chosen to be identical to (2.15) in~\cite{Cuomo:2017wme}. Notice that no bar (representing hermitian conjugation in that work) is present here. This is because we do not deal with a full correlator but rather with its tensor structure.
For the scalar case $\ell=\bar\ell=0$ the expression~\eqref{eq:2PointFunction} is identical to~\eqref{eq:two_point_structure_vector_formalism}.

\paragraph{Bubble coefficients} Let us consider a conformally invariant pair of weight-shifting operators (meaning that their conformal indices are contracted) acting on the same point. Only the pairs effectively not changing the spin and scaling dimensions give a non zero result. We call such conformally invariant operators the bubble operators, their action is trivial and just amounts to multiplication by a constant which we call a bubble coefficient, see~\eqref{eq:definition_bubble_constant}.
We compute these coefficients by acting with the bubble differential operators on the two-point structure~\eqref{eq:2PointFunction}. They remain valid however also for three-point functions.

Starting from the first set of weight-shifting operators~\eqref{eq:4Doperators_4} one can form four such pairs
\begin{align}\label{eq:bubble_1}
\left(\cD_{-0+}\cdot\overline{\cD}^{+0-}\right)\big|_{[\De,\ell,\bar\ell]} =
\left(\overline{\cD}^{-+0}\cdot   \cD_{+-0}\right)\big|_{[\De,\bar\ell,\ell]} =
\mathcal{B}_1(\Delta,\ell,\bar\ell),\\
\label{eq:bubble_2}
\left(\cD_{-0-}\cdot\overline{\cD}^{+0+}\right)\big|_{[\De,\ell,\bar\ell]} =
\left(\overline{\cD}^{--0}\cdot   \cD_{++0}\right)\big|_{[\De,\bar\ell,\ell]} =
\mathcal{B}_2(\Delta,\ell,\bar\ell),\\
\label{eq:bubble_3}
\left(\cD_{++0}\cdot\overline{\cD}^{--0}\right)\big|_{[\De,\ell,\bar\ell]} =
\left(\overline{\cD}^{+0+}\cdot   \cD_{-0-}\right)\big|_{[\De,\bar\ell,\ell]} =
\mathcal{B}_3(\Delta,\ell,\bar\ell),\\
\label{eq:bubble_4}
\left(\cD_{+-0}\cdot\overline{\cD}^{-+0}\right)\big|_{[\De,\ell,\bar\ell]} =
\left(\overline{\cD}^{+0-}\cdot   \cD_{-0+}\right)\big|_{[\De,\bar\ell,\ell]} =
\mathcal{B}_4(\Delta,\ell,\bar\ell).
\end{align}
The coefficients read as
\begin{align}
\mathcal{B}_1(\Delta,\ell,\bar\ell)
&\equiv
-\frac{1}{2}\,\bar\ell\,(2-\ell+\bar\ell-2\Delta)(4+\ell+\bar\ell-2\Delta),\\
\mathcal{B}_2(\Delta,\ell,\bar\ell)
&\equiv
+\frac{1}{2}\,(2+\bar\ell)\,(2+\ell-\bar\ell-2\Delta)(\ell+\bar\ell+2\Delta),\\
\mathcal{B}_3(\Delta,\ell,\bar\ell)
&\equiv
-\frac{1}{2}\,\ell\,(-6+\ell-\bar\ell+2\Delta)(-4+\ell+\bar\ell+2\Delta),\\
\mathcal{B}_4(\Delta,\ell,\bar\ell)
&\equiv
-\frac{1}{2}\,(2+\ell)\,(6+\ell-\bar\ell-2\Delta)(8+\ell+\bar\ell-2\Delta).
\end{align}

\paragraph{The 6j-symbols}
One of the most striking properties of weight-shifting operators is that they satisfy a sort of crossing~\cite{Karateev:2017jgd} when acting on two- or three-point tensor structures. Since there is a unique two-point structure defined in~\eqref{eq:2PointFunction} it is straightforward to compute all the 6j-symbols associated to it. For the first set of operators~\eqref{eq:4Doperators_4} we have
\begin{align}
\label{eq:6j_2pt_1}
\cD^{1\,a}_{-0+}
\langle O^{(\ell,\bar\ell)}_{\Delta}O^{(\bar\ell,\ell)}_{\Delta}\rangle&=
+\mathcal{J}_1(\De, \ell, \bar\ell)\times
\cD^{2\,a}_{+-0}
\langle O^{(\ell,\bar\ell+1)}_{\Delta-1/2}
O^{(\bar\ell+1,\ell)}_{\Delta-1/2}\rangle,\\
\label{eq:6j_2pt_2}
\cD^{1\,a}_{-0-}
\langle O^{(\ell,\bar\ell)}_{\Delta} O^{(\bar\ell,\ell)}_{\Delta}\rangle&=
+\mathcal{J}_2(\De, \ell, \bar\ell)\times
\cD^{2\,a}_{++0}
\langle O^{(\ell,\bar\ell-1)}_{\Delta-1/2}
 O^{(\bar\ell-1,\ell)}_{\Delta-1/2}\rangle,\\
\label{eq:6j_2pt_3}
\cD^{1\,a}_{++0}
\langle O^{(\ell,\bar\ell)}_{\Delta} O^{(\bar\ell,\ell)}_{\Delta}\rangle&=
+\mathcal{J}_3(\De, \ell, \bar\ell)\times
\cD^{2\,a}_{-0-}
\langle O^{(\ell+1,\bar\ell)}_{\Delta+1/2}
 O^{(\bar\ell,\ell+1)}_{\Delta+1/2}\rangle,\\
\label{eq:6j_2pt_4}
\cD^{1\,a}_{+-0}
\langle O^{(\ell,\bar\ell)}_{\Delta} O^{(\bar\ell,\ell)}_{\Delta}\rangle&=
+\mathcal{J}_4(\De, \ell, \bar\ell)\times
\cD^{2\,a}_{-0+}
\langle O^{(\ell-1,\bar\ell)}_{\Delta+1/2}
 O^{(\bar\ell,\ell-1)}_{\Delta+1/2}\rangle.
\end{align}
For the second set of operators~\eqref{eq:4Doperators_dual_4} we have
\begin{align}
\label{eq:6j_2pt_1b}
\overline\cD_{1\,a}^{-+0}
\langle O^{(\ell,\bar\ell)}_{\Delta} O^{(\bar\ell,\ell)}_{\Delta}\rangle&=
-\mathcal{J}_1(\De,\bar\ell, \ell)\times
\overline\cD_{2\,a}^{+0-}
\langle O^{(\ell+1,\bar\ell)}_{\Delta-1/2}
 O^{(\bar\ell,\ell+1)}_{\Delta-1/2}\rangle,\\
\label{eq:6j_2pt_2b}
\overline\cD_{1\,a}^{--0}
\langle O^{(\ell,\bar\ell)}_{\Delta} O^{(\bar\ell,\ell)}_{\Delta}\rangle&=
-\mathcal{J}_2(\De,\bar\ell, \ell)\times
\overline\cD_{2\,a}^{+0+}
\langle O^{(\ell-1,\bar\ell)}_{\Delta-1/2}
 O^{(\bar\ell,\ell-1)}_{\Delta-1/2}\rangle,\\
\label{eq:6j_2pt_3b}
\overline\cD_{1\,a}^{+0+}
\langle O^{(\ell,\bar\ell)}_{\Delta} O^{(\bar\ell,\ell)}_{\Delta}\rangle&=
-\mathcal{J}_3(\De,\bar\ell, \ell)\times
\overline\cD_{2\,a}^{--0}
\langle O^{(\ell,\bar\ell+1)}_{\Delta+1/2}
 O^{(\bar\ell+1,\ell)}_{\Delta+1/2}\rangle,\\
\label{eq:6j_2pt_4b}
\overline\cD_{1\,a}^{+0-}
\langle O^{(\ell,\bar\ell)}_{\Delta} O^{(\bar\ell,\ell)}_{\Delta}\rangle&=
-\mathcal{J}_4(\De,\bar\ell, \ell)\times
\overline\cD_{2\,a}^{-+0}
\langle O^{(\ell,\bar\ell-1)}_{\Delta+1/2}
 O^{(\bar\ell-1,\ell)}_{\Delta+1/2}\rangle.
\end{align}
The indices $1$ and $2$ indicate the coordinates the weight-shifting operators act on.
The coefficients read as follows
\begin{align}
\mathcal{J}_1(\De, \ell, \bar\ell)
&\equiv
\frac{2i}{(1+\bar\ell)(4-\ell+\bar\ell-2\Delta)(6+\ell+\bar\ell-2\Delta)},\\
\mathcal{J}_2(\De, \ell, \bar\ell)
&\equiv
\frac{2i\bar\ell}{(4+\ell-\bar\ell-2\Delta)(-2+\ell+\bar\ell+2\Delta)},\\
\mathcal{J}_3(\De, \ell, \bar\ell)
&\equiv
\frac{i(2-\ell+\bar\ell-2\Delta)(\ell+\bar\ell+\Delta)}{2(1+\ell)},\\
\mathcal{J}_4(\De, \ell, \bar\ell)
&\equiv
\frac{i\ell}{2}\,(2+\ell-\bar\ell-2\Delta)(4+\ell+\bar\ell-2\Delta).
\end{align}

\subsection{Shadow transform and weight-shifting operators}
\label{app:shadow_transform_4d}

We define the shadow transform in the 4d index-free through a conformally invariant pairing with the two-point structure~\eqref{eq:2PointFunction} as
\begin{align}
\nn
\bS[\cO^{(\ell,\bar\ell)}_\De](x,s,\bar s)
&\equiv
\int d^4 y\,
\<
\cO^{(\bar\ell,\ell)}_{\tl\De}(x,s,\bar s)
\cO{}^{(\ell,\bar\ell)}_{\tl\De}(y,t,\bar t)
\>
\;\overleftrightarrow{G}_{t,\bar t}^{(\ell,\bar\ell)}\;
\cO^{(\ell,\bar\ell)}_\De(y,t,\bar t)\\
\label{eq:shadow_transform_4d}
 &=
\int d^4 y\,
\cO^{(\ell,\bar\ell)}_\De(y,t,\bar t)
\;\overleftrightarrow{G}_{t,\bar t}^{(\ell,\bar\ell)}\;
\<
\cO{}^{(\ell,\bar\ell)}_{\tl\De}(y,t,\bar t)
\cO^{(\bar\ell,\ell)}_{\tl\De}(x,s,\bar s)
\>.
\end{align}
In the last equality we have used the property~\eqref{eq:property_4d} and the symmetry of the two-point structure~\eqref{eq:2PointFunction}. Notice that all the alternating signs cancel out.

Using the integration by parts properties~\eqref{eq:integration_by_parts_1}-\eqref{eq:integration_by_parts_4_bar} and the 6j-symbols~\eqref{eq:6j_2pt_1}-\eqref{eq:6j_2pt_4b} it is straightforward to find the commutation relations of weight-shifting operators with the shadow transform. They read
\begin{align}\label{eq:commutation1}
\mathbf{S}\,\cD^a_{-0+}\,\big|_{[\De,\ell,\bar\ell]} &=
+\mathcal{C}_1(\De,\ell,\bar\ell)\times
\cD^a_{++0}\,\mathbf{S}\,\big|_{[\De,\ell,\bar\ell]},
\\
\label{eq:commutation2}
\mathbf{S}\,\cD^a_{-0-}\,\big|_{[\De,\ell,\bar\ell]} &=
+\mathcal{C}_2(\De,\ell,\bar\ell)\times
\cD^a_{+-0}\,\mathbf{S}\,\big|_{[\De,\ell,\bar\ell]},\\
\label{eq:commutation3}
\mathbf{S}\,\cD^a_{++0}\,\big|_{[\De,\ell,\bar\ell]} &=
+\mathcal{C}_3(\De,\ell,\bar\ell)\times
\cD^a_{-0+}\,\mathbf{S}\,\big|_{[\De,\ell,\bar\ell]},\\
\label{eq:commutation4}
\mathbf{S}\,\cD^a_{+-0}\,\big|_{[\De,\ell,\bar\ell]} &=
+\mathcal{C}_4(\De,\ell,\bar\ell)\times
\cD^a_{-0-}\,\mathbf{S}\,\big|_{[\De,\ell,\bar\ell]}
\end{align}
together with
\begin{align}\label{eq:commutation1b}
\mathbf{S}\,\overline\cD_a^{-+0}\,\big|_{[\De,\ell,\bar\ell]} &=
-\mathcal{C}_1(\De,\bar\ell,\ell)\times
\overline\cD_a^{+0+}\,\mathbf{S}\,\big|_{[\De,\ell,\bar\ell]},\\
\label{eq:commutation2b}
\mathbf{S}\,\overline\cD_a^{--0}\,\big|_{[\De,\ell,\bar\ell]} &=
-\mathcal{C}_2(\De,\bar\ell,\ell)\times
\overline\cD_a^{+0-}\,\mathbf{S}\,\big|_{[\De,\ell,\bar\ell]},\\
\label{eq:commutation3b}
\mathbf{S}\,\overline\cD_a^{+0+}\,\big|_{[\De,\ell,\bar\ell]} &=
-\mathcal{C}_3(\De,\bar\ell,\ell)\times
\overline\cD_a^{-+0}\,\mathbf{S}\,\big|_{[\De,\ell,\bar\ell]},\\
\label{eq:commutation4b}
\mathbf{S}\,\overline\cD_a^{+0-}\,\big|_{[\De,\ell,\bar\ell]} &=
-\mathcal{C}_4(\De,\bar\ell,\ell)\times
\overline\cD_a^{--0}\,\mathbf{S}\,\big|_{[\De,\ell,\bar\ell]}.
\end{align}
The coefficients read as follows
\begin{align}
\mathcal{C}_1(\De,\ell,\bar\ell)
&\equiv
+\frac{\mathcal{J}_2(4-\De+1/2, \ell, \bar\ell+1)}{1+\bar\ell}
=
\frac{-2i}{(6 - \ell + \bar\ell - 2 \Delta) (8 + \ell + \bar\ell - 2 \Delta)},\\
\mathcal{C}_2(\De,\ell,\bar\ell)
&\equiv
-\bar\ell\times\mathcal{J}_1(4-\De+1/2, \ell, \bar\ell-1)
=
\frac{2i}{(6+\ell-\bar\ell-2\Delta)(-4+\ell+\bar\ell+2\Delta)},\\
\mathcal{C}_3(\De,\ell,\bar\ell)
&\equiv
-\frac{\mathcal{J}_4(4-\De-1/2, \ell+1, \bar\ell)}{1+\ell}
=
-\frac{i}{2}\,(-4+\ell-\bar\ell+2\Delta)(-2+\ell+\bar\ell+2\Delta),\\
\mathcal{C}_4(\De,\ell,\bar\ell)
&\equiv
+\ell\times\mathcal{J}_3(4-\De-1/2, \ell-1, \bar\ell)
=
\frac{i}{2}\,(-4-\ell+\bar\ell+2\Delta)(6+\ell+\bar\ell-2\Delta).
\end{align}

As a simple application consider the coefficient of the square shadow transform~\eqref{eq:squareofshadow}. In 4d we denote it by $\mathcal{N}_\Delta^{(\ell,\bar{\ell})}$. One can write the following recursion relation
using the bubble operator~\eqref{eq:bubble_1}
\begin{align}
\nn
&\mathbf{S}^2\,
\left(\cD_{-0+}\cdot\overline{\cD}^{+0-}\right)
\cO_{\Delta}^{(\ell,\bar\ell)}
=\mathcal{B}_1(\De,\ell,\bar\ell)\,
\mathcal{N}_\Delta^{(\ell,\bar{\ell})}\,
\cO_{\Delta}^{(\ell,\bar\ell)}\\
\nn
&=
\mathcal{C}_1(\De+1/2,\ell,\bar\ell-1)\,
\mathcal{C}_3(\tl\De-1/2,\bar\ell-1,\ell)\,
\cD_{-0+}^a\mathbf{S}^2\overline{\cD}^{+0-}_a\cO_{\Delta}^{(\ell,\bar\ell)}\\
\nn
&=
\mathcal{C}_1(\De+1/2,\ell,\bar\ell-1)\,
\mathcal{C}_3(\tl\De-1/2,\bar\ell-1,\ell)\,
\mathcal{N}_{\Delta+1/2}^{(\ell,\bar\ell-1)}\,
\mathcal{B}_1(\De,\ell,\bar\ell)\cO_{\Delta}^{(\ell,\bar\ell)}.
\end{align}
As result we get a recursion relation on the coefficient $\mathcal{N}$.
Notice that the bubble coefficient cancels out.
Analogously one can derive a similar relation using the bubble operator~\eqref{eq:bubble_3}. Both relations read as
\begin{align}
\mathcal{N}_\Delta^{(\ell,\bar{\ell})}
&=
\mathcal{C}_1(\De+1/2,\ell,\bar\ell-1)\,
\mathcal{C}_3(\tl\De-1/2,\bar\ell-1,\ell)\,
\mathcal{N}_{\Delta+1/2}^{(\ell,\bar\ell-1)},\\
\mathcal{N}_\Delta^{(\ell,\bar{\ell})}
&=
\mathcal{C}_3(\De-1/2,\ell-1,\bar\ell)\,
\mathcal{C}_1(\tl\De+1/2,\bar\ell,\ell-1)\,
\mathcal{N}_{\Delta-1/2}^{(\ell-1,\bar\ell)}.
\end{align}
Using the base case~\eqref{eq:coefficient_square_shadow_transform} the solution reads as
\begin{equation}
\label{eq:shadow_square_4d}
\mathcal{N}_\Delta^{(\ell,\bar{\ell})}=\frac{(-1)^{\ell+\bar\ell}\times\pi^4}
{\left(\Delta-2+\frac{\ell-\bar\ell}{2}\right)
	\left(2-\Delta+\frac{\ell-\bar\ell}{2}\right)
	\left(\Delta+\frac{\ell+\bar\ell}{2}-1\right)
	\left(4-\Delta+\frac{\ell+\bar\ell}{2}-1\right)
}.
\end{equation}

\subsection{Two- and three-point pairings}
\label{sec:two_and_three_point_pairings}
We use~\eqref{eq:shadow_transform_4d} to define a conformally invariant pairing between $n$-point tensor structures with $n=2,3$ as the pairing applied to every point of the structure and its dual one as in~\eqref{eq:structurepairing}. It is totally symmetric under the exchange of structures
\begin{align}
\nn
&\p{
	\<\cO^{(\ell_1,\bar\ell_1)}_{\De_1}\ldots\cO^{(\ell_n,\bar\ell_n)}_{\De_n}\>^{(a)},
	\<\cO^{(\ell_1,\bar\ell_1)}_{\tl\De_1}\ldots\cO^{(\ell_n,\bar\ell_n)}_{\tl\De_n}\>^{(b)}
}=\\
&\p{
	\<\cO^{(\ell_1,\bar\ell_1)}_{\tl\De_1}\ldots\cO^{(\ell_n,\bar\ell_n)}_{\tl\De_n}\>^{(b)},
	\<\cO^{(\ell_1,\bar\ell_1)}_{\De_1}\ldots\cO^{(\ell_n,\bar\ell_n)}_{\De_n}\>^{(a)}
}.
\label{eq:property_correlator_4d}
\end{align}
More precisely according to~\eqref{eq:property_4d} the prefactor we get in the right-hand side of~\eqref{eq:property_correlator_4d} is
\begin{equation}
\label{eq:bosonic_condition}
(-1)^{\sum_{i=1}^n(\ell_i+\bar\ell_i)}=(-1)^{\text{even}}=+1.
\end{equation}
The relation~\eqref{eq:bosonic_condition} holds since there cannot be a non-zero vacuum expectation value of a fermionic quantity $\sum_{i=1}^n(\ell_i+\bar\ell_i)=\text{odd}$ in a Lorentz invariant theory. It becomes obvious in the index-free formalism where one can simply never construct an invariant tensor structure in such a case.

Before proceeding with computations of two- and three-point pairings let us rewrite the formula connecting the Plancherel measure with the square shadow coefficient~\eqref{eq:relationbetweenshadowsqandplancherel} in terms of the two-point pairing. In the index-free language the relation~\eqref{eq:plancherel_and_shadow} reads as
\begin{align}
\nn
&\cN(\De,\ell,\bar\ell) \frac{\mu(\De,\ell,\bar\ell)}{\vol\SO(1,1)}=
\int \frac{d^4 x d^4 y}{\vol \SO(5,1)}\\
\nn
&
\<\cO_{\De}^{(\ell,\bar\ell)}(x,s_3,\bar s_3)
\cO_{\De}^{(\bar\ell,\ell)}(y,s_2,\bar s_2)\>
\overleftrightarrow{G}_{s_2,\bar s_2}^{(\bar\ell,\ell)}
\<\cO_{\tl\De}^{(\bar\ell,\ell)}(y,s_2,\bar s_2)
\cO_{\tl\De}^{(\ell,\bar\ell)}(x,s_1,\bar s_1)\>
\overleftrightarrow{G}_{s_1,\bar s_1;\;s_3,\bar s_3}^{(\ell,\bar\ell)}\\
&=
\p{
	\<\cO^{(\ell,\bar\ell)}_{\Delta}
	\cO^{(\bar\ell,\ell)}_{\Delta}\>,\;
	\<\cO^{(\ell,\bar\ell)}_{\tl\Delta}
	\cO^{(\bar\ell,\ell)}_{\tl\Delta}\>
}.
\label{eq:plancherel_shadow_4d}
\end{align}
Here the ``gluing'' operator $\overleftrightarrow{G}_{s_1,\bar s_1;\;s_3,\bar s_3}^{(\ell,\bar\ell)}$ acts on spinor polarizations $s_1,\bar s_1$ to the left and $s_3,\bar s_3$ to the right. Moving it between the two-point structures we gain a $(-1)^{\ell+\bar\ell}$ factor. To obtain the final result we also flip the position of operators in the right two-point structure. It brings another  $(-1)^{\ell+\bar\ell}$ factor. Thus all the alternating signs cancel out.

\paragraph{Computation with a direct approach}
Let us provide here two simple examples of conformally invariant pairings defined in section~\ref{sec:pairings_between_structures} between two- and three-point tensor structures in 4d. For two- and three-point structures the pairing reduces only to contraction of indices.

Consider first a pairing of the two-point structure~\eqref{eq:2PointFunction}.
We evaluate it directly using \eqref{eq:inde-free_pairing} and~\eqref{eq:glueing_differential_operator}.  In the conformal frame it reads as
\begin{equation}
\nn
\<\cO^{(\ell,\bar\ell)}_{\Delta}(0,s_1,\bar s_1) 
\cO^{(\bar\ell,\ell)}_{\Delta}(\oo,s_2,\bar s_2)\>
=i^{\bar\ell-\ell}
\left(\bar s_{2}\,
\bar\sigma^{3}\,s_{1}\right)^\ell
\left(-\bar s_{1}\,
\bar\sigma^{3}\,s_{2}\right)^{\bar\ell}.
\end{equation}
It is simply equal to $+1$ for $\ell=0$. One has then
\begin{align}
\nn
&16\vol\SO(1,1)\vol\SO(4)\p{
	\<\cO^{(\ell,\bar\ell)}_{\Delta}
	\cO^{(\bar\ell,\ell)}_{\Delta}\>,\;
	\<\cO^{(\ell,\bar\ell)}_{\tl\Delta}
	\cO^{(\bar\ell,\ell)}_{\tl\Delta}\>
}\\
\nn
&=\<\cO^{(\ell,\bar\ell)}_{\Delta}(0,s_1,\bar s_1) 
\cO^{(\bar\ell,\ell)}_{\Delta}(\oo,s_2,\bar s_2)\>
\;
\overleftrightarrow{G}_{s_1,\bar s_1}^{(\ell,\bar\ell)}
\overleftrightarrow{G}_{s_2,\bar s_2}^{(\bar\ell,\ell)}
\;
\<\cO^{(\ell,\bar\ell)}_{\tl\Delta}(0,s_1,\bar s_1) 
\cO^{(\bar\ell,\ell)}_{\tl\Delta}(\oo,s_2,\bar s_2)\>\\
\nn
&=
\label{eq:two_point_pairing_4d}
\frac{(-1)^{\ell+\bar\ell}}{\ell!^2\bar\ell!^2}
\left(\partial_{\bar s_{2L}}\partial_{\bar s_{2R}}\right)^\ell
\left(\partial_{\bar s_{1L}}\partial_{\bar s_{1R}}\right)^{\bar\ell}\times
(\bar s_{2L}\bar s_{2R})^\ell\;
(\bar s_{1L}\bar s_{2R})^{\bar\ell}\\
&=(-1)^{\ell+\bar\ell}\times(1+\ell)(1+\bar\ell).
\end{align}
In the second equality we have inserted the labels $L$ and $R$ to distinguish contributions from the left and right part of the pairing.

Second, consider the pairing of the scalar-scalar-spin$(\ell,\ell)$ structure~\eqref{eq:tensor_structure_scalar_scalar_spin_L}.
In the conformal frame it reads as
\begin{equation}
\< \phi_{\Delta_1}(0) \phi_{\Delta_2}(e) \cO_{\Delta_3}^{(\ell,\ell)}(\infty,s,\bar s) \>=
\left(\bar s\bar\sigma^{3}s\right)^\ell.
\end{equation}
Again it is simply equal to $+1$ for $\ell=0$. We have then
\begin{align}
\nn
&16\vol\SO(3)\p{
	\<\phi_{\Delta_1}\phi_{\Delta_2}\cO_{\Delta_3}^{(\ell,\ell)}\>,\;
	\<\phi_{\tl\Delta_1}\phi_{\tl\Delta_2}\cO_{\tl\Delta_3}^{(\ell,\ell)}\>
}\\
\nn
&=
\left(\bar s\bar\sigma^{3}s\right)^\ell
\;
\overleftrightarrow{G}_{s,\bar s}^{(\ell,\bar\ell)}
\;
\left(\bar s\bar\sigma^{3}s\right)^\ell\\
\label{eq:scalar_three_point_pairing_4d}
&=\frac{1}{\ell!^2}(\partial_{\bar s_L}\partial_{\bar s_R})^\ell (\bar s_L \bar s_R)^\ell=(1+\ell).
\end{align}

It is interesting to compare the results~\eqref{eq:two_point_pairing_4d} and~\eqref{eq:scalar_three_point_pairing_4d} to their analogues obtained in the vector formalism by restricting our attention to traceless symmetric tensors $\ell=\bar\ell = J$. The two-point pairing in the vector formalism can be read off from~\eqref{eq:relationbetweenshadowsqandplancherel} together with~\eqref{eq:normforstts}. For $d=4$ it is identical to~\eqref{eq:two_point_pairing_4d}. We can now establish a connection between the 4d and the vector formalisms by requiring to preserve this condition or equivalently to demand that
\begin{equation}
\<\cO^{(\ell,\ell)}_{\Delta}(0,s_1,\bar s_1) 
\cO^{(\ell,\ell)}_{\Delta}(\oo,s_2,\bar s_2)\>_{\text{4d}}=
\<
\cO_{\De,\ell}(0,z_1)
\cO_{\De,\ell}(\infty,z_2)
\>_{\text{vector}}.
\end{equation}
We remind that the two-point function in the vector formalism was defined in~\eqref{eq:twoptconvnentionflatspace}. This leads to the following map
\begin{equation}
z^\mu = \pm 2^{-1/2}\times(\bar s \bar\sigma^\mu s),
\end{equation}
where $z^\mu$ is a polarization vector, see~\eqref{eq:index_free_vector_formalism}.
An overall sign remains ambiguous. Using this and the definition of the three-point structure in the vector formalism~\eqref{eq:scalarscalarspinjstructure} it is easy to show that
\begin{equation}
\label{eq:connection_between_three_point_structures_in_two_formalisms}
\<
\phi_{\Delta_1}(0)
\phi_{\Delta_2}(e)
\cO_{\Delta_3}^{(\ell,\ell)}(\infty,s,\bar s)
\>_{\text{4d}}=
(\mp)^\ell\, 2^{\ell/2}
\times
\<
\f_{\De_1}(0)
\f_{\De_2}(e)
\cO_{\De_3,\ell}(\infty,z)
\>_{\text{vector}}.
\end{equation}
Now from~\eqref{eq:connection_between_three_point_structures_in_two_formalisms} we expect that the three-point pairing~\eqref{eq:scalar_three_point_pairing_4d} is equal to~\eqref{eq:scalarscalarspinjpairing} times an overall $2^{\ell}$ factor for $d=4$. We then observe that this is indeed the case.

\paragraph{Computation with weight-shifting operators}
With the ability to integrate by parts introduced in section~\ref{eq:weightshiftingshadowcoeffs} we can also compute three-point pairings using weight-shifting operators. Although this is in principle simply a matter of contracting indices as demonstrated above, for higher spin structures this procedure becomes tedious whereas weight-shifting operators provide a simple algebraic way of organizing the computation. Let us illustrate how it works on several examples.

\subparagraph{Pairing of two-point structures}
As the simplest possible example we consider the most general two-point pairing defined in~\eqref{eq:2PointFunction}. It satisfies the following recursion relation
\begin{align}
\nn
&\p{
\<\cO^{(\ell,\bar\ell)}_{\Delta}
\cO^{(\bar\ell,\ell)}_{\Delta}\>,\;
\<\cO^{(\ell,\bar\ell)}_{\tl\Delta}
\cO^{(\bar\ell,\ell)}_{\tl\Delta}\>
}=\\
\nn
&=
i\;\p{
(D^1_{-0+}\cdot \overline D_2^{-+0})
\<\cO^{(\ell,\bar\ell-1)}_{\Delta+1/2}
\cO^{(\bar\ell-1,\ell)}_{\Delta+1/2}\>,\;
\<\cO^{(\ell,\bar\ell)}_{\tl\Delta}
\cO^{(\bar\ell,\ell)}_{\tl\Delta}\>
}\\
\nn
&=+\frac{i}{\bar\ell^2}
\p{
\<\cO^{(\ell,\bar\ell-1)}_{\Delta+1/2}
\cO^{(\bar\ell-1,\ell)}_{\Delta+1/2}\>,\;
(\overline D_2^{--0}\cdot D^1_{-0-})
\<\cO^{(\ell,\bar\ell)}_{\tl\Delta}
\cO^{(\bar\ell,\ell)}_{\tl\Delta}\>
}\\
\nn
&=-\left(1+\frac{1}{\bar\ell}\right)\times
\p{
	\<\cO^{(\ell,\bar\ell-1)}_{\Delta+1/2}
	\cO^{(\bar\ell-1,\ell)}_{\Delta+1/2}\>,\;
	\<\cO^{(\ell,\bar\ell-1)}_{\tl\Delta-1/2}
	\cO^{(\bar\ell-1,\ell)}_{\tl\Delta-1/2}\>
}
\end{align}
Analogously one can write a similar recursion relation
\begin{equation}\label{eq:2_point_pairing_recursion_2}
\nn
\p{
	\<\cO^{(\ell,\bar\ell)}_{\Delta}
	\cO^{(\bar\ell,\ell)}_{\Delta}\>,\;
	\<\cO^{(\ell,\bar\ell)}_{\tl\Delta}
	\cO^{(\bar\ell,\ell)}_{\tl\Delta}\>
}=
-\left(1+\frac{1}{\ell}\right)\times
\p{
	\<\cO^{(\ell-1,\bar\ell)}_{\Delta-1/2}
	\cO^{(\bar\ell,\ell-1)}_{\Delta-1/2}\>,\;
	\<\cO^{(\ell,\bar\ell)}_{\tl\Delta+1/2}
	\cO^{(\bar\ell,\ell)}_{\tl\Delta+1/2}\>
}.
\end{equation}
The solution to both recursion recursion relations reads as
\begin{equation}\label{eq:2_point_pairing_generic}
\p{
	\<\cO^{(\ell,\bar\ell)}_{\Delta}
	\cO^{(\bar\ell,\ell)}_{\Delta}\>,\;
	\<\cO^{(\ell,\bar\ell)}_{\tl\Delta}
	\cO^{(\bar\ell,\ell)}_{\tl\Delta}\>
}=
\frac{(-1)^{\ell+\bar\ell}\times(1+\ell)(1+\bar\ell)}{16\vol\SO(1,1)\vol\SO(4)}.
\end{equation}
The correct proportionality factors in~\eqref{eq:2_point_pairing_generic} are determined by considering $\ell=0$ case. The appearance of $16\vol\SO(1,1)\vol\SO(4)$ is explained in section~\ref{sec:pairings_between_structures}.
The result perfectly matches~\eqref{eq:two_point_pairing_4d}.

\subparagraph{Pairing of  scalar-scalar-spin$(\ell,\ell)$}
Consider a three-point pairing of the tensor structure~\eqref{eq:tensor_structure_scalar_scalar_spin_L}, it satisfies the following recursion
\begin{align}
\nn
&\p{
	\<\phi_{\De_1}\phi_{\De_2} \cO_{\De_3}^{(\ell,\ell)}\>,\, \<\phi_{\tl\Delta_1}\phi_{\tl\Delta_2} \cO_{\tl\Delta_3}^{(\ell,\ell)}\>
}\\
\nn
&=\mathcal{A}_3\times
\p{
	(D^2_{-0-}\cdot \overline D_3^{-+0})
	(D^3_{-0+}\cdot \overline D_2^{+0+})
	\<\phi_{\De_1}\phi_{\De_2} \cO_{\Delta_3+1}^{(\ell-1,\ell-1)}\>,\, \<\phi_{\tl\De_1}\phi_{\tl\De_2} \cO_{\tl\Delta_3}^{(\ell,\ell)}\>
}\\
\nn
&=\mathcal{A}_3/\ell^2\times
\p{
	\<\phi_{\De_1}\phi_{\De_2} O_{\Delta_3+1}^{(\ell-1,\ell-1)}\>,\,
	(\overline D_2^{+0-}\cdot D^3_{-0-})
	(\overline D_3^{--0}\cdot D^2_{-0+})
	\<\phi_{\tl\De_1}\phi_{\tl\De_2} \cO_{\tl\Delta_3}^{(\ell,\ell)}\>
}\\
\label{eq:pairing_scalar_scalar_spin_L}
&=\left(1+\frac{1}{\ell}\right) \times
\p{
	\<\phi_{\De_1}\phi_{\De_2} \cO_{\Delta_3+1}^{(\ell-1,\ell-1)}\>,\,
	\<\phi_{\tl\De_1}\phi_{\tl\De_2} \cO_{\tl\Delta_3-1}^{(\ell-1,\ell-1)}\>
},
\end{align}
where the coefficient $\mathcal{A}$ reads as
\begin{equation}
\mathcal{A}_3^{-1}\equiv(\De_2-1)(\De_1+\De_2-\De_3+\ell-2).
\end{equation}
The solution to the recursion relation~\eqref{eq:pairing_scalar_scalar_spin_L} is simply given by
\begin{equation}\label{eq:basic_pairing}
\p{
	\<\phi_{\De_1}\phi_{\De_2} \cO_{\De_3}^{(\ell,\ell)}\>,\, \<\phi_{\tl\Delta_1}\phi_{\tl\Delta_2} \cO_{\tl\Delta_3}^{(\ell,\ell)}\>
}=
\frac{1+\ell}{16\vol\SO(3)}.
\end{equation}
The correct proportionality factors in~\eqref{eq:basic_pairing} are determined by considering $\ell=0$ case. The appearance of $4\vol\SO(3)$ is explained in section~\ref{sec:pairings_between_structures}. The result matches perfectly~\eqref{eq:scalar_three_point_pairing_4d}.

\subparagraph{Pairing of scalar-spin$(p,0)$-spin$(\ell,\ell)$}
Next we consider the structure~\eqref{eq:seed_3pt_1}.
We use the relation~\eqref{eq:recursion_4D_second} and perform manipulations analogous to~\eqref{eq:pairing_scalar_scalar_spin_L} to obtain
\begin{align}
&\p{
	\<\phi_{\De_1}\,f^{(p,0)}_{\De_2}\,\cO^{(\ell,\ell+p)}_{\Delta_3}\>,\;
	\<\phi_{\tl\De_1}\,f^{(p,0)}_{\tl\De_2}\,\cO^{(\ell,\ell+p)}_{\tl\Delta_3}\>
}\\
&=
\frac{\ell+p+1}{\ell+p}\;
\p{
	\<\phi_{\De_1}\,f^{(p-1,0)}_{\De_2+1/2}\,\cO^{(\ell,\ell+p-1)}_{\Delta_3+1/2}\>,\;
	\<\phi_{\tl\De_1}\,f^{(p-1,0)}_{\tl\De_2-1/2}\,\cO^{(\ell,\ell+p-1)}_{\tl\Delta_3-1/2}\>
}\\
&=
\frac{\ell+p+1}{\ell+1}
\p{
	\<\phi_{\De_1}\,\phi_{\De_2+p/2}\,\cO^{(\ell,\ell)}_{\Delta_3+p/2}\>,\;
	\<\phi_{\tl\De_1}\,\phi_{\tl\De_2-p/2}\,\cO^{(\ell,\ell)}_{\tl\Delta_3-p/2}\>
}.
\end{align}
Using the result~\eqref{eq:basic_pairing} we write down the final solution
\begin{equation}\label{eq:seed_pairing_4d}
\p{
	\<\phi_{\De_1}\,f^{(p,0)}_{\De_2}\,\cO^{(\ell,\ell+p)}_{\Delta_3}\>,\;
	\<\phi_{\tl\De_1}\,f^{(p,0)}_{\tl\De_2}\,\cO^{(\ell,\ell+p)}_{\tl\Delta_3}\>
}
=
\frac{\ell+p+1}{16\vol\SO(3)}.
\end{equation}

\subparagraph{Pairing of fermion-fermion-spin$(\ell,\ell)$}
To conclude we study an example with multiple tensor structures~\eqref{eq:fermion_3pt_1}. Using the relation~\eqref{eq:fermion_bar_fermion_structures_via_differential_operators} we can write the following recursion
\begin{align}
\nn
&\p{
	\<\psi^\dag_{\De_1}\,\psi_{\De_2}\,\cO^{(\ell,\ell)}_{\De_3}\>^{(m)},\;
	\<\psi^\dag_{\tl\De_1}\,\psi_{\tl\De_2}\,\cO^{(\ell,\ell)}_{\tl\De_3}\>^{(n)}
} =\\
\nn
&\sum_{r=1}^2M^{m}_{12}{}_r
\p{D^{(r)}
	\<\phi_{\De_1+3/2-r}\phi_{\De_2+3/2-r}\cO^{(\ell,\ell)}_{\De_3}\>,\;
	\<\psi^\dag_{\tl\De_1}\,\psi_{\tl\De_2}\,\cO^{(\ell,\ell)}_{\tl\De_3}\>^{(n)}
}=\\
\nn
&\sum_{r=1}^2M^{m}_{12}{}_r
\p{
	\<\phi_{\De_1+3/2-r}\phi_{\De_2+3/2-r}\cO^{(\ell,\ell)}_{\De_3}\>,\;
	D^{\prime\prime (r)}
	\<\psi^\dag_{\tl\De_1}\,\psi_{\tl\De_2}\,\cO^{(\ell,\ell)}_{\tl\De_3}\>^{(n)}
}=\\
&\sum_{r=1}^2
M^{m}_{12}{}_r
U^{rn}
\p{
	\<\phi_{\De_1+3/2-r}\phi_{\De_2+3/2-r}\cO^{(\ell,\ell)}_{\De_3}\>,\;
	\<\phi_{\tl\De_1-3/2+r}\,\phi_{\tl\De_2-3/2+r}\,\cO^{(\ell,\ell)}_{\tl\De_3}\>
},
\end{align}
where the differential operator $D^{(n)}$ is defined in~\eqref{eq:fermionic_dif_operators} and $D''$ is obtained from $D$ using the integration by parts properties~\eqref{eq:integration_by_parts_1}-\eqref{eq:integration_by_parts_4_bar}, it is given by
\begin{equation}
D^{\prime\prime (r)}\equiv \left(
\overline \cD_2^{--0}\cdot \cD^1_{-0-},\quad
\cD^2_{+-0}\cdot \overline\cD_1^{+0-}
\right).
\end{equation}
The matrix $M_{12}$ is given by~\eqref{eq:fermions_matrix_differential_basis} and the components of the matrix $U$ are given by
\begin{align}\nn
U^{11} &=1/2\,U^{12}=2,\\\nn
U^{21} &=2(\De_1-3/2) (\De_2-3/2)
\Big((\De_1+\De_{2}-\De_{3}-1) (\De_1+\De_2+\De_3-5)-\ell(2+\ell)\Big),\\\nn
U^{22} &=(\Delta_1-3/2) (\Delta_2-3/2)
(\Delta_1+\Delta_2-\Delta_3-\ell-3)
(\Delta_1+\Delta_2+\Delta_3+\ell-3).
\end{align}
Taking into account~\eqref{eq:basic_pairing} we arrive at the rather simple final result
\begin{equation}\label{eq:three_point_pairing_fermions}
\p{
	\<\psi^\dag_{\De_1}\,\psi_{\De_2}\,\cO^{(\ell,\ell)}_{\De_3}\>^{(m)},\;
	\<\psi^\dag_{\tl\De_1}\,\psi_{\tl\De_2}\,\cO^{(\ell,\ell)}_{\tl\De_3}\>^{(n)}
}
=\begin{pmatrix}
2 & 1\\
1 & \frac{1+\ell}{\ell}
\end{pmatrix}
\times\frac{1+\ell}{16\vol\SO(3)}.
\end{equation}

\section{Conventions for 3d representations}
\label{app:3dconventions}

In this appendix we describe our conventions for 3d representation theory, which for the most parts is consistent with~\cite{Iliesiu:2015akf,Iliesiu:2015qra,Iliesiu:2017nrv,Kravchuk:2016qvl}. 

First, consider the Lorentz group $\SO(2,1)$.\footnote{We use the mostly plus metric with $\eta_{00}=-1$.} Its double cover is $\Spin(2,1)\simeq \Sp(2,\R)$. We define the Lorentz generators $M^{\mu\nu}$ to have
the commutation relations
\be
	[M^{\mu\nu},M^{\s\r}]=\eta^{\nu\s}M^{\mu\r}-\eta^{\mu\s}M^{\nu\r}+\eta^{\nu\r}M^{\s\mu}-\eta^{\mu\r}M^{\s\nu}.
\ee
These generators are anti-hermitian when acting on the physical Hilbert space, $M^\dagger=-M$. They act on local operators according to
\be
	[M^{\mu\nu},\cO^a(0)]=-(\cM^{\mu\nu})^a{}_b \cO^b(0),
\ee
where the matrices $\cM^{\mu\nu}$ define the representations and have the same commutation relations as $M$. In particular, for the spinor representation we have
\be
	(\cM^{\mu\nu})^{\a}{}_{\b}=\frac{1}{4}([\gamma^\mu,\gamma^\nu])^{\a}{}_{\b},
\ee
where we define the gamma-matrices $(\gamma^\mu)^\a{}_\b$ as
\be
	\gamma^0=\begin{pmatrix}
		0 & 1 \\ -1 & 0
	\end{pmatrix},\qquad
	\gamma^1=\begin{pmatrix}
	0 & 1 \\ 1 & 0
	\end{pmatrix},\qquad
	\gamma^2=\begin{pmatrix}
	1 & 0 \\ 0 & -1
	\end{pmatrix}.
\ee
Note that since the gamma-matrices are real so are the generators $\cM$. This is natural since the spinor representation is simply the fundamental of $\Sp(2,\R)$. For convenience we will also define an abstract representation with basis elements $e^\a$ and the action of Lorentz group
\be
	M^{\mu\nu}\.e^\a\equiv -(\cM^{\mu\nu})^\a{}_\b e^\b.
\ee
The spin-$j$ representation of $\Spin(2,1)$ can then be realized in the space with basis elements
\be
	e^{\a_1\ldots \a_{2j}}=e^{(\a_1\ldots \a_{2j})},
\ee
symmetric in all indices. We define the basis of the dual representation as
\be
	e_{\a_1\ldots \a_{2j}}(e^{\b_1\ldots \b_{2j}})=\delta^{(\b_1}_{\a_1}\cdots \delta^{\b_{2j})}_{\a_{2j}}.
\ee
Note that this implies
\be
	e_{11\ldots 2}(e^{11\ldots 2})=\binom{2j}{k}^{-1},
\ee
where in both index sets $1$ occurs $k$ times. The spinor representation and its dual are equivalent, with the equivalence defined as
\be
	e_{\a_1\ldots \a_{2j}}=\O_{\a_1\b_1}\cdots \O_{\a_{2j}\b_{2j}} e^{\b_1\ldots \b_{2j}},
\ee
with $\O_{\a\b}=\O^{\a\b}=(\gamma^0)^\a{}_\b$. 

We will now identify these basis elements with the standard $|j,m\>$ basis from the theory of angular momentum. To that end, we first define Euclidean generators by
\be
	M^{0\mu}_E=-M^{\mu0}_E=iM^{0\mu},
\ee
and $M^{\mu\nu}_E=M^{\mu\nu}$ for the other components. We then introduce
\be
	J_\mu = \frac{-i}{2}\e_{\mu\nu\l} M^{\mu\l}_E,
\ee
and $J_\pm = J_0\pm i J_1$. The basis vectors $|j,m\>$ are then defined to satisfy the usual relations
\be
	J_2|j,m\> &= m|j,m\>,\\
	J_+|j,m\> &= \sqrt{(j-m)(j+m+1)}|j,m+1\>,\\
	J_-|j,m\> &= \sqrt{(j+m)(j-m+1)}|j,m-1\>.
\ee
The vectors $\<j,m|$ of the dual representation satisfy
\be
	\<j,m|j,m'\>\equiv\<j,m|(|j,m'\>)\equiv\delta_{mm'}.
\ee
We can again establish the isomorphism between the two representations by
\be\label{eq:jmduality}
	\<j,m|=i^{-2m}|j,-m\>.
\ee
Finally, we can find the correspondence (parentheses denote the binomial coefficients)
\be\label{eq:ejmisomoprhism}
	e^{1\ldots 2}&=i^{+j+m}\binom{2j}{j+m}^{-1/2}|j,m\>,\nn\\
	e_{1\ldots 2}&=i^{-j-m}\binom{2j}{j+m}^{-1/2}\<j,m|,
\ee
where in both index sets $1$ occurs $j-m$ times.

For integer $j$ we can also establish the isomorphism of the above representations with traceless symmetric tensors. We write
\be\label{eq:vectorspinorisomorphism}
	e^{\mu_1\ldots \mu_j}=\frac{(-i)^j}{2^{j/2}}\gamma^{\mu_1}_{\a_1\a_2}\cdots \gamma^{\mu_j}_{\a_{2j-1}\a_{2j}}e^{\a_1\ldots \a_{2j}}.
\ee
With this convention the pairing between $e^{\mu_1\ldots \mu_j}$ and $e_{\mu_1\ldots \mu_j}$ is equivalent to the pairing between $e^{\a_1\ldots \a_{2j}}$ and $e_{\a_1\ldots \a_{2j}}$. However, all the representation that we have studied so far are real and it is somewhat unnatural to use $i$ in defining this isomorphism.\footnote{This is the rare case when the metric signature matters -- had we chosen to work in mostly minus signature, we wouldn't need these $i$ factors, while in the mostly plus signature $i$ necessarily has to appear somewhere. This is a reflection of the fact that the real Clifford algebras $Cl(2,1)$ and $Cl(1,2)$ are not isomorphic.} This is most problematic when a hermitian $\cO$ operator is used in place of $e$ --- $\cO^{\mu_1\ldots \mu_j}$ and $\cO^{\a_1\ldots \a_{2j}}$ cannot be both hermitian for odd $j$. We thus declare that the operators in spinor notation have phase $i^{-j}$. This leads to the hermiticity conditions
\be\label{eq:3dconjugationrules}
	(\cO^{\a_1\ldots \a_{2j}})^\dagger &= e^{i\pi j} \cO^{\a_1\ldots \a_{2j}},\\
	(\cO^{\mu_1\ldots \mu_j})^\dagger &= \cO^{\mu_1\ldots \mu_j}.
\ee

Finally, the Wick rotation to the Euclidean signature is obtained by setting for all vector indices $x_E^0=ix_E$. The spinor indices are not altered, and thus in Euclidean signature the isomorphism~\eqref{eq:vectorspinorisomorphism} is established by Euclidean gamma-matrices with $\gamma^0_E=i\gamma^0$.

\subsection{Tensor structures in different formalisms}
\label{app:3dconventions:tensorstructures}
To keep track of tensor structures we contract the local operators with polarization\footnote{In~\cite{Kravchuk:2016qvl} $s_{\a_1\ldots \a_{2j}}\equiv s_{\a_1}\cdots s_{\a_{2j}}$.} $s_{\a_1\ldots \a_{2j}}$, which transforms in the same way as $e_{\a_1\ldots \a_{2j}}$,
\be
	\cO(x,s)\equiv s_{\a_1\ldots \a_{2j}} \cO^{\a_1\ldots \a_{2j}}(x).
\ee
Occasionally we also write
\be
	s_{\a_1\ldots \a_{2j}}=s_{\a_1}\cdots s_{\a_{2j}}
\ee
for a spinor polarization $s_\a$.

Then naturally the tensor structure $\<\cO(s_1,x_1)\cdots \cO(s_n,x_n)\>$ is an element of $j_1\otimes\cdots\otimes j_n$. In particular, by using the isomorphism described in this section it can be expanded in the basis
\be
	|j_1,m_1\>\otimes \cdots \otimes |j_n,m_n\>.
\ee

For example, in~\cite{Kravchuk:2016qvl} the structures $[q_1q_2q_3]$ are introduced for 3-point functions, and by unwinding the definitions we find (taking into account $m_1+m_2+m_3=0$)
\be
	[m_1m_2m_3]=&i^{-j_1-j_2-j_3}\left[\binom{2j_1}{j_1+m_1}\binom{2j_2}{j_2+m_2}\binom{2j_3}{j_3+m_3}\right]^{-1/2}\nn\\ &\qquad\times|j_1,m_1\>\otimes|j_2,m_2\>\otimes|j_3,m_3\>.
\ee

\subsection{The two-point function}
\label{app:3dconventions:twopt}
The two-point function of spin-$j$ operators can be written as~\cite{Iliesiu:2015qra}
\be\label{eq:3dtwopt}
	\<\cO(x_1,s_1)\cO(x_2,s_2)\>=\frac{(s_1^\a \gamma^\mu_{\a\b} s_2^\b x_{12,\mu})^{2j}}{x_{12}^{2\De+2j}}.
\ee
In order to relate this to the normalization in~\cite{Iliesiu:2015qra} one has to recall that due to our reality conditions we essentially have $\cO_{here}(x,s)=i^{-j}\cO_{there}(x,s)$. For integer spin operators we can use the isomorphism~\eqref{eq:vectorspinorisomorphism}, and defining
\be
	\cO(x,z) = z_{\mu_1}\cdots z_{\mu_j} \cO^{\mu_1\ldots \mu_j}(z,x)
\ee
for a null polarization $z$, we find
\be
	\<\cO(x_1,z_1)\cO(x_2,z_2)\>=\frac{(z_1^\mu I_{\mu\nu}(x_{12}) z_2^\nu)^j}{x_{12}^{2\De}},
\ee
where
\be
	I_{\mu\nu}(x)=\eta_{\mu\nu}-2\frac{x_\mu x_\nu}{x^2}.
\ee

Let us first evaluate the two point function in the configuration
\be
	\<\cO(\hat e_2,s_1)\cO(0,s_2)\>=(s_{1,1} s_{2,2}+s_{1,2} s_{2,1})^{2j},
\ee
where $\hat e_2^\mu = (0,0,1)$. By using the isomorphism~\eqref{eq:ejmisomoprhism} we find
\be
	\<\cO(\hat e_2,s_1)\cO(0,s_2)\>=\sum_{m=-j}^j i^{-2j} |j,m\>\otimes |j,-m\>.
\ee
Using~\eqref{eq:jmduality} we can dualize the right operator to obtain the expression for the shadow kernel,
\be\label{eq:Kstandard}
	K(\hat e_2) = \sum_{m=-j}^j (-1)^{j-m} |j,m\>\<j,m|.
\ee
This is just the reflection operator in the direction of $\hat e_2$. Any given Euclidean $x$ can be represented in spherical coordinates as
\be
	x=r e^{-i J_2\phi} e^{-i J_1 \theta} \hat e_2.
\ee
Using the invariance condition $UK(x)U^{-1}=K(Ux)$ we find
\be
	K(x) = r^{-2\De} e^{-i J_2\phi} e^{-i J_1 \theta} K(\hat e_2) e^{i J_1 \theta} e^{i J_2\phi}=r^{-2\De} e^{-i J_2\phi} e^{-2i J_1 \theta} e^{i J_2\phi} K(\hat e_2)
\ee
where we used that $K(\hat e_2)$ is a reflection and $J$ is a pseudo vector. We thus find
\be\label{eq:3dkernel}
	K(x) = r^{-2\De}\sum_{m,m'=-j}^j (-1)^{j-m} e^{i(m-m')\phi} d^j_{m',m}(-2\theta) |j,m'\>\<j,m|,
\ee
where $d^j_{m',m}(\theta)=\<j,m'|e^{iJ_1\theta}|j,m\>$ is the small Wigner $d$-function.\footnote{Our convention for the $d$-function is consistent with \texttt{Mathematica}'s \texttt{WignerD[\{j,m',m\},$\theta$]}}

\section{Fourier transform of 3d two-point function}
\label{app:twoptfourier}

In this section we compute the Fourier transform of the 3d two-point function~\eqref{eq:3dkernel}, which is defined by the Euclidean integral\footnote{Note that in this section we compute the Fourier transform of an operator with dimension $\De$, while in the main text we are mainly interested in the shadow operator with dimension $\tl\De=3-\De$.}
\be
	K(p)=\int d^3 x e^{-ip\.x}K(x).
\ee
Because of rotation and scaling symmetry it suffices to evaluate it at $p=(0,0,1)$, which leads to the integral
\be
	\sum_{m,m'=-j}^j\int d\phi\, d\!\cos\theta \,dr\, r^{2-2\De} e^{-i r\cos\theta}(-1)^{j-m} e^{i(m-m')\phi} d^j_{m',m}(-2\theta) |j,m'\>\<j,m|=
\ee
The $\phi$-integral is trivial and sets $m'=m$,
\be
	=2\pi \sum_{m=-j}^j\int d\!\cos\theta \,dr\, r^{2-2\De} e^{-i r\cos\theta}(-1)^{j-m} d^j_{m,m}(-2\theta) |j,m\>\<j,m|.
\ee
Let us write
\be
	\cI_{j,m}(r)=\int d\!\cos\theta e^{-i r\cos\theta} d^j_{m,m}(-2\theta).
\ee
We have $d^{\half}_{\half,\half}(-2\theta) = \cos\theta$, and thus by using the CG decomposition we can find
\be
	i\ptl_r \cI_{j,m}(r) = \frac{j+m+1}{2j+1}\cI_{j+\half,m+\half}(r)+\frac{j-m}{2j+1}\cI_{j-\half,m+\half}(r).
\ee
Together with the base case
\be
	\cI_{0,0}(r) = 2\frac{\sin r}{r}
\ee
and identity $\cI_{j,m}=\cI_{j,-m}$ this recursion relation determines all $\cI_{j,m}$. We then need to compute
\be
	\cA_{j,m}(\De)=2\pi (-1)^{j-m} \int dr\, r^{2-2\De} \cI_{j,m}(r).
\ee
Substituting the recursion relation for $\cI$ into this definition and integrating by parts, we find
\be
	2i(\De-1)\cA_{j,m}(\De+\half)=\frac{j+m+1}{2j+1}\cA_{j+\half,m+\half}(\De)-\frac{j-m}{2j+1}\cA_{j-\half,m+\half}(\De),
\ee
with the base case
\be
	\cA_{0,0}(\De)=4\pi\Gamma(2-2\De)\sin\pi\De.
\ee
This recursion relation is solved by
\be\label{eq:Ajm}
	\cA_{j,m}(\De)=(-1)^{j-m}\frac{\Gamma(\De+m-1)\Gamma(\De-m-1)}{\Gamma(\De+j-1)\Gamma(\De-j-1)}\cA_{j,j}(\De)
\ee
and
\be
	\cA_{j,j}(\De)=(-i)^{2j}\frac{4\pi\sin\pi(\De+j)\Gamma(2-2\De)(\De-1)}{\De+j-1}.
\ee
In terms of $\cA$ the Fourier transform for $p=(0,0,1)$ is
\be
	K(p)=\sum_{m=-j}^j \cA_{j,m}(\De)|j,m\>\<j,m|.
\ee

\section{Numerical 3d $\<TTTT\>$ OPE coefficients}
\label{app:TTTTMFTcoefficients}
In the tables below we record the OPE matrices $P_{ab}$ for $\<TTTT\>$ MFT four-point function in 3d. For parity-even operators we have
\begin{center}
	\begin{tabular}{r||c|c|c|c|c}
		$\De$ & $j=0$& $j=2$ & $j=4$ & $j=6$ & $j=8$ \\
		\hline\hline
		$0$ &1 &  &  & & \\
		\hline
		$6$ &$\frac{1}{3240}$&$\frac{1}{9676800}$&$\begin{pmatrix} \frac{1}{12700800} & 0 \\ 0 & 0 \end{pmatrix}$&& \\
		\hline
		$8$ &$\frac{1}{92400}$&$\frac{1}{1303948800}$&$\begin{pmatrix} \frac{23}{8298702720} & \frac{-1}{377213760} \\ \frac{-1}{377213760} & \frac{1}{8573040} \end{pmatrix}$&$\begin{pmatrix} \frac{1}{1886068800} & 0 \\ 0 & 0 \end{pmatrix}$& \\
		\hline
		$10$ & \!\!$\frac{1}{1842750}$\!\!\!&\!\!\!$\frac{1}{78460462080}$\!\!\!&\!\!\!$\begin{pmatrix} \frac{31}{229572743400} & \frac{-1}{35318883600} \\ \frac{-1}{35318883600} & \frac{1}{339604650} \end{pmatrix}$\!\!\!&\!\!\!$\begin{pmatrix} \frac{1}{51602265000} & \frac{-1}{10320453000} \\ \frac{-1}{10320453000} & \frac{1}{172007550} \end{pmatrix}$\!\!\!&\!\!\!$\begin{pmatrix} \frac{1}{72243171000} & 0 \\ 0 & 0 \end{pmatrix}$\!\!\!\\
		\hline
	\end{tabular}
\end{center}
One can check that all the matrices are positive-semidefinite. For parity-odd operators we find
\begin{center}
	\begin{tabular}{r||c|c|c|c|c|c}
		$\De$& $j=0$& $j=2$ & $j=4$ & $j=5$ & $j=6$ & $j=7$ \\
		\hline\hline
		$7$ &$\frac{-1}{1920}$&$\frac{-1}{7902720}$&$\frac{-1}{10534551552}$&&& \\
		\hline 
		$8$ &&&&$\frac{-1}{7097654108160}$&&\\
		\hline
		$9$ & $\frac{-1}{9100}$ & $\frac{-1}{237758976}$ & $\frac{-1}{1607401635840} $ && $\frac{-1}{469695283200}$ &\\
		\hline
		$10$ & &&&$\frac{-1}{179850593894400}$&&$\frac{-1}{998689595904000}$
	\end{tabular}
\end{center}
Note that the OPE matrices of parity-odd operators are negative because unitarity requires the OPE coefficients of parity-odd structures to be pure imaginary.

\bibliographystyle{JHEP}
\bibliography{refs}

\providecommand{\href}[2]{#2}\begingroup\raggedright\begin{thebibliography}{10}

\bibitem{Maldacena:1997re}
J.~M. Maldacena, \emph{{The Large N limit of superconformal field theories and
  supergravity}}, \href{http://dx.doi.org/10.1023/A:1026654312961,
  10.4310/ATMP.1998.v2.n2.a1}{\emph{Int. J. Theor. Phys.} {\bf 38} (1999)
  1113--1133}, [\href{https://arxiv.org/abs/hep-th/9711200}{{\tt
  hep-th/9711200}}].

\bibitem{Gubser:1998bc}
S.~S. Gubser, I.~R. Klebanov and A.~M. Polyakov, \emph{{Gauge theory
  correlators from noncritical string theory}},
  \href{http://dx.doi.org/10.1016/S0370-2693(98)00377-3}{\emph{Phys. Lett.}
  {\bf B428} (1998) 105--114},
  [\href{https://arxiv.org/abs/hep-th/9802109}{{\tt hep-th/9802109}}].

\bibitem{Witten:1998qj}
E.~Witten, \emph{{Anti-de Sitter space and holography}},
  \href{http://dx.doi.org/10.4310/ATMP.1998.v2.n2.a2}{\emph{Adv. Theor. Math.
  Phys.} {\bf 2} (1998) 253--291},
  [\href{https://arxiv.org/abs/hep-th/9802150}{{\tt hep-th/9802150}}].

\bibitem{Alday:2007mf}
L.~F. Alday and J.~M. Maldacena, \emph{{Comments on operators with large
  spin}}, \href{http://dx.doi.org/10.1088/1126-6708/2007/11/019}{\emph{JHEP}
  {\bf 11} (2007) 019}, [\href{https://arxiv.org/abs/0708.0672}{{\tt
  0708.0672}}].

\bibitem{Fitzpatrick:2012yx}
A.~L. Fitzpatrick, J.~Kaplan, D.~Poland and D.~Simmons-Duffin, \emph{{The
  Analytic Bootstrap and AdS Superhorizon Locality}},
  \href{http://dx.doi.org/10.1007/JHEP12(2013)004}{\emph{JHEP} {\bf 1312}
  (2013) 004}, [\href{https://arxiv.org/abs/1212.3616}{{\tt 1212.3616}}].

\bibitem{Komargodski:2012ek}
Z.~Komargodski and A.~Zhiboedov, \emph{{Convexity and Liberation at Large
  Spin}}, \href{http://dx.doi.org/10.1007/JHEP11(2013)140}{\emph{JHEP} {\bf
  1311} (2013) 140}, [\href{https://arxiv.org/abs/1212.4103}{{\tt 1212.4103}}].

\bibitem{Alday:2015eya}
L.~F. Alday, A.~Bissi and T.~Lukowski, \emph{{Large spin systematics in CFT}},
  \href{http://dx.doi.org/10.1007/JHEP11(2015)101}{\emph{JHEP} {\bf 11} (2015)
  101}, [\href{https://arxiv.org/abs/1502.07707}{{\tt 1502.07707}}].

\bibitem{Alday:2015ewa}
L.~F. Alday and A.~Zhiboedov, \emph{{An Algebraic Approach to the Analytic
  Bootstrap}}, \href{http://dx.doi.org/10.1007/JHEP04(2017)157}{\emph{JHEP}
  {\bf 04} (2017) 157}, [\href{https://arxiv.org/abs/1510.08091}{{\tt
  1510.08091}}].

\bibitem{Alday:2016njk}
L.~F. Alday, \emph{{Large Spin Perturbation Theory for Conformal Field
  Theories}},
  \href{http://dx.doi.org/10.1103/PhysRevLett.119.111601}{\emph{Phys. Rev.
  Lett.} {\bf 119} (2017) 111601},
  [\href{https://arxiv.org/abs/1611.01500}{{\tt 1611.01500}}].

\bibitem{Simmons-Duffin:2016wlq}
D.~Simmons-Duffin, \emph{{The Lightcone Bootstrap and the Spectrum of the 3d
  Ising CFT}}, \href{http://dx.doi.org/10.1007/JHEP03(2017)086}{\emph{JHEP}
  {\bf 03} (2017) 086}, [\href{https://arxiv.org/abs/1612.08471}{{\tt
  1612.08471}}].

\bibitem{Rattazzi:2008pe}
R.~Rattazzi, V.~S. Rychkov, E.~Tonni and A.~Vichi, \emph{{Bounding scalar
  operator dimensions in 4D CFT}},
  \href{http://dx.doi.org/10.1088/1126-6708/2008/12/031}{\emph{JHEP} {\bf 12}
  (2008) 031}, [\href{https://arxiv.org/abs/0807.0004}{{\tt 0807.0004}}].

\bibitem{Iliesiu:2015qra}
L.~Iliesiu, F.~Kos, D.~Poland, S.~S. Pufu, D.~Simmons-Duffin and R.~Yacoby,
  \emph{{Bootstrapping 3D Fermions}},
  \href{http://dx.doi.org/10.1007/JHEP03(2016)120}{\emph{JHEP} {\bf 03} (2016)
  120}, [\href{https://arxiv.org/abs/1508.00012}{{\tt 1508.00012}}].

\bibitem{Iliesiu:2017nrv}
L.~Iliesiu, F.~Kos, D.~Poland, S.~S. Pufu and D.~Simmons-Duffin,
  \emph{{Bootstrapping 3D Fermions with Global Symmetries}},
  \href{https://arxiv.org/abs/1705.03484}{{\tt 1705.03484}}.

\bibitem{Dymarsky:2017xzb}
A.~Dymarsky, J.~Penedones, E.~Trevisani and A.~Vichi, \emph{{Charting the space
  of 3D CFTs with a continuous global symmetry}},
  \href{https://arxiv.org/abs/1705.04278}{{\tt 1705.04278}}.

\bibitem{Dymarsky:2017yzx}
A.~Dymarsky, F.~Kos, P.~Kravchuk, D.~Poland and D.~Simmons-Duffin, \emph{{The
  3d Stress-Tensor Bootstrap}},
  \href{http://dx.doi.org/10.1007/JHEP02(2018)164}{\emph{JHEP} {\bf 02} (2018)
  164}, [\href{https://arxiv.org/abs/1708.05718}{{\tt 1708.05718}}].

\bibitem{Karateev:2017}
D.~Karateev, P.~Kravchuk, M.~Serone and A.~Vichi, \emph{{to appear..}}, .

\bibitem{Kitaev}
A.~Kitaev, ``{A simple model of quantum holography, Talks at KITP, April 7 and
  May 27}.''
  \href{http://online.kitp.ucsb.edu/online/entangled15/kitaev/}{http://online.kitp.ucsb.edu/online/entangled15/kitaev/}
  and
  \href{http://online.kitp.ucsb.edu/online/entangled15/kitaev2/}{http://online.kitp.ucsb.edu/online/entangled15/kitaev2/},
  2015.

\bibitem{Maldacena:2016hyu}
J.~Maldacena and D.~Stanford, \emph{{Remarks on the Sachdev-Ye-Kitaev model}},
  \href{http://dx.doi.org/10.1103/PhysRevD.94.106002}{\emph{Phys. Rev.} {\bf
  D94} (2016) 106002}, [\href{https://arxiv.org/abs/1604.07818}{{\tt
  1604.07818}}].

\bibitem{PR}
J.~Polchinski and V.~Rosenhaus, \emph{{The Spectrum in the Sachdev-Ye-Kitaev
  Model}}, \href{http://dx.doi.org/10.1007/JHEP04(2016)001}{\emph{JHEP} {\bf
  04} (2016) 001}, [\href{https://arxiv.org/abs/1601.06768}{{\tt 1601.06768}}].

\bibitem{GR}
D.~J. Gross and V.~Rosenhaus, \emph{{A Generalization of Sachdev-Ye-Kitaev}},
  \href{http://dx.doi.org/10.1007/JHEP02(2017)093}{\emph{JHEP} {\bf 02} (2017)
  093}, [\href{https://arxiv.org/abs/1610.01569}{{\tt 1610.01569}}].

\bibitem{Murugan:2017eto}
J.~Murugan, D.~Stanford and E.~Witten, \emph{{More on Supersymmetric and 2d
  Analogs of the SYK Model}},
  \href{http://dx.doi.org/10.1007/JHEP08(2017)146}{\emph{JHEP} {\bf 08} (2017)
  146}, [\href{https://arxiv.org/abs/1706.05362}{{\tt 1706.05362}}].

\bibitem{Giombi:2017dtl}
S.~Giombi, I.~R. Klebanov and G.~Tarnopolsky, \emph{{Bosonic tensor models at
  large $N$ and small $\epsilon$}},
  \href{http://dx.doi.org/10.1103/PhysRevD.96.106014}{\emph{Phys. Rev.} {\bf
  D96} (2017) 106014}, [\href{https://arxiv.org/abs/1707.03866}{{\tt
  1707.03866}}].

\bibitem{Liu:2018jhs}
J.~Liu, E.~Perlmutter, V.~Rosenhaus and D.~Simmons-Duffin,
  \emph{{$d$-dimensional SYK, AdS Loops, and $6j$ Symbols}},
  \href{https://arxiv.org/abs/1808.00612}{{\tt 1808.00612}}.

\bibitem{Heemskerk:2009pn}
I.~Heemskerk, J.~Penedones, J.~Polchinski and J.~Sully, \emph{{Holography from
  Conformal Field Theory}},
  \href{http://dx.doi.org/10.1088/1126-6708/2009/10/079}{\emph{JHEP} {\bf 0910}
  (2009) 079}, [\href{https://arxiv.org/abs/0907.0151}{{\tt 0907.0151}}].

\bibitem{Fitzpatrick:2011dm}
A.~L. Fitzpatrick and J.~Kaplan, \emph{{Unitarity and the Holographic
  S-Matrix}}, \href{http://dx.doi.org/10.1007/JHEP10(2012)032}{\emph{JHEP} {\bf
  1210} (2012) 032}, [\href{https://arxiv.org/abs/1112.4845}{{\tt 1112.4845}}].

\bibitem{Dobrev:1977qv}
V.~K. Dobrev, G.~Mack, V.~B. Petkova, S.~G. Petrova and I.~T. Todorov,
  \emph{{Harmonic Analysis on the n-Dimensional Lorentz Group and Its
  Application to Conformal Quantum Field Theory}},
  \href{http://dx.doi.org/10.1007/BFb0009678}{\emph{Lect. Notes Phys.} {\bf 63}
  (1977) 1--280}.

\bibitem{Gadde:2016fbj}
A.~Gadde, \emph{{Conformal constraints on defects}},
  \href{https://arxiv.org/abs/1602.06354}{{\tt 1602.06354}}.

\bibitem{Caron-Huot:2017vep}
S.~Caron-Huot, \emph{{Analyticity in Spin in Conformal Theories}},
  \href{https://arxiv.org/abs/1703.00278}{{\tt 1703.00278}}.

\bibitem{Gromov:2017cja}
N.~Gromov, V.~Kazakov, G.~Korchemsky, S.~Negro and G.~Sizov,
  \emph{{Integrability of Conformal Fishnet Theory}},
  \href{http://dx.doi.org/10.1007/JHEP01(2018)095}{\emph{JHEP} {\bf 01} (2018)
  095}, [\href{https://arxiv.org/abs/1706.04167}{{\tt 1706.04167}}].

\bibitem{Gromov:2018hut}
N.~Gromov, V.~Kazakov and G.~Korchemsky, \emph{{Exact Correlation Functions in
  Conformal Fishnet Theory}},  \href{https://arxiv.org/abs/1808.02688}{{\tt
  1808.02688}}.

\bibitem{Kravchuk:2018htv}
P.~Kravchuk and D.~Simmons-Duffin, \emph{{Light-ray operators in conformal
  field theory}},  \href{https://arxiv.org/abs/1805.00098}{{\tt 1805.00098}}.

\bibitem{Ferrara1972}
S.~Ferrara, A.~F. Grillo, G.~Parisi and R.~Gatto, \emph{The shadow operator
  formalism for conformal algebra. vacuum expectation values and operator
  products}, \href{http://dx.doi.org/10.1007/BF02907130}{\emph{Lettere al Nuovo
  Cimento (1971-1985)} {\bf 4} (May, 1972) 115--120}.

\bibitem{SimmonsDuffin:2012uy}
D.~Simmons-Duffin, \emph{{Projectors, Shadows, and Conformal Blocks}},
  \href{http://dx.doi.org/10.1007/JHEP04(2014)146}{\emph{JHEP} {\bf 04} (2014)
  146}, [\href{https://arxiv.org/abs/1204.3894}{{\tt 1204.3894}}].

\bibitem{Fradkin:1978pp}
E.~S. Fradkin and M.~{\relax Ya}. Palchik, \emph{{Recent Developments in
  Conformal Invariant Quantum Field Theory}},
  \href{http://dx.doi.org/10.1016/0370-1573(78)90172-2}{\emph{Phys. Rept.} {\bf
  44} (1978) 249--349}.

\bibitem{Vasiliev:1981dg}
A.~N. Vasiliev, {\relax Yu}.~M. Pismak and {\relax Yu}.~R. Khonkonen,
  \emph{{1/$N$ Expansion: Calculation of the Exponents $\eta$ and Nu in the
  Order 1/$N^2$ for Arbitrary Number of Dimensions}},
  \href{http://dx.doi.org/10.1007/BF01019296}{\emph{Theor. Math. Phys.} {\bf
  47} (1981) 465--475}.

\bibitem{Karateev:2017jgd}
D.~Karateev, P.~Kravchuk and D.~Simmons-Duffin, \emph{{Weight Shifting
  Operators and Conformal Blocks}},
  \href{http://dx.doi.org/10.1007/JHEP02(2018)081}{\emph{JHEP} {\bf 02} (2018)
  081}, [\href{https://arxiv.org/abs/1706.07813}{{\tt 1706.07813}}].

\bibitem{Cuomo:2017wme}
G.~F. Cuomo, D.~Karateev and P.~Kravchuk, \emph{{General Bootstrap Equations in
  4D CFTs}},  \href{https://arxiv.org/abs/1705.05401}{{\tt 1705.05401}}.

\bibitem{KnappStein1}
A.~W. Knapp and E.~M. Stein, \emph{Intertwining operators for semisimple
  groups}, \href{http://dx.doi.org/10.2307/1970887}{\emph{Ann. of Math. (2)}
  {\bf 93} (1971) 489--578}.

\bibitem{Simmons-Duffin:2017nub}
D.~Simmons-Duffin, D.~Stanford and E.~Witten, \emph{{A spacetime derivation of
  the Lorentzian OPE inversion formula}},
  \href{http://dx.doi.org/10.1007/JHEP07(2018)085}{\emph{JHEP} {\bf 07} (2018)
  085}, [\href{https://arxiv.org/abs/1711.03816}{{\tt 1711.03816}}].

\bibitem{Kravchuk:2016qvl}
P.~Kravchuk and D.~Simmons-Duffin, \emph{{Counting Conformal Correlators}},
  \href{http://dx.doi.org/10.1007/JHEP02(2018)096}{\emph{JHEP} {\bf 02} (2018)
  096}, [\href{https://arxiv.org/abs/1612.08987}{{\tt 1612.08987}}].

\bibitem{RepresentationsAndSpecialFunctions}
N.~J. Vilenkin and A.~U. Klimyk, \emph{Representation of Lie Groups and Special
  Functions}, vol.~3.
\newblock Springer Netherlands, 1992.

\bibitem{Dolan:2000ut}
F.~A. Dolan and H.~Osborn, \emph{{Conformal four point functions and the
  operator product expansion}},
  \href{http://dx.doi.org/10.1016/S0550-3213(01)00013-X}{\emph{Nucl. Phys.}
  {\bf B599} (2001) 459--496},
  [\href{https://arxiv.org/abs/hep-th/0011040}{{\tt hep-th/0011040}}].

\bibitem{Dolan:2003hv}
F.~A. Dolan and H.~Osborn, \emph{{Conformal partial waves and the operator
  product expansion}},
  \href{http://dx.doi.org/10.1016/j.nuclphysb.2003.11.016}{\emph{Nucl. Phys.}
  {\bf B678} (2004) 491--507},
  [\href{https://arxiv.org/abs/hep-th/0309180}{{\tt hep-th/0309180}}].

\bibitem{Dolan:2004iy}
F.~Dolan and H.~Osborn, \emph{{Conformal partial wave expansions for
  $\mathcal{N}=4$ chiral four point functions}},
  \href{http://dx.doi.org/10.1016/j.aop.2005.07.005}{\emph{Annals Phys.} {\bf
  321} (2006) 581--626}, [\href{https://arxiv.org/abs/hep-th/0412335}{{\tt
  hep-th/0412335}}].

\bibitem{DO2}
F.~Dolan and H.~Osborn, \emph{{Conformal partial waves and the operator product
  expansion}},
  \href{http://dx.doi.org/10.1016/j.nuclphysb.2003.11.016}{\emph{Nucl.Phys.}
  {\bf B678} (2004) 491--507},
  [\href{https://arxiv.org/abs/hep-th/0309180}{{\tt hep-th/0309180}}].

\bibitem{Costa:2011mg}
M.~S. Costa, J.~Penedones, D.~Poland and S.~Rychkov, \emph{{Spinning Conformal
  Correlators}}, \href{http://dx.doi.org/10.1007/JHEP11(2011)071}{\emph{JHEP}
  {\bf 11} (2011) 071}, [\href{https://arxiv.org/abs/1107.3554}{{\tt
  1107.3554}}].

\bibitem{Weinberg:2010fx}
S.~Weinberg, \emph{{Six-dimensional Methods for Four-dimensional Conformal
  Field Theories}},
  \href{http://dx.doi.org/10.1103/PhysRevD.82.045031}{\emph{Phys. Rev.} {\bf
  D82} (2010) 045031}, [\href{https://arxiv.org/abs/1006.3480}{{\tt
  1006.3480}}].

\bibitem{Elkhidir:2014woa}
E.~Elkhidir, D.~Karateev and M.~Serone, \emph{{General Three-Point Functions in
  4D CFT}}, \href{http://dx.doi.org/10.1007/JHEP01(2015)133}{\emph{JHEP} {\bf
  01} (2015) 133}, [\href{https://arxiv.org/abs/1412.1796}{{\tt 1412.1796}}].

\bibitem{Kravchuk:2017dzd}
P.~Kravchuk, \emph{{Casimir recursion relations for general conformal blocks}},
  \href{http://dx.doi.org/10.1007/JHEP02(2018)011}{\emph{JHEP} {\bf 02} (2018)
  011}, [\href{https://arxiv.org/abs/1709.05347}{{\tt 1709.05347}}].

\bibitem{Kitaev:2017hnr}
A.~Kitaev, \emph{{Notes on $\widetilde{\mathrm{SL}}(2,\mathbb{R})$
  representations}},  \href{https://arxiv.org/abs/1711.08169}{{\tt
  1711.08169}}.

\bibitem{Goncalves:2014rfa}
V.~Goncalves, J.~Penedones and E.~Trevisani, \emph{{Factorization of Mellin
  amplitudes}}, \href{http://dx.doi.org/10.1007/JHEP10(2015)040}{\emph{JHEP}
  {\bf 10} (2015) 040}, [\href{https://arxiv.org/abs/1410.4185}{{\tt
  1410.4185}}].

\bibitem{Costa:2014rya}
M.~S. Costa and T.~Hansen, \emph{{Conformal correlators of mixed-symmetry
  tensors}}, \href{http://dx.doi.org/10.1007/JHEP02(2015)151}{\emph{JHEP} {\bf
  02} (2015) 151}, [\href{https://arxiv.org/abs/1411.7351}{{\tt 1411.7351}}].

\bibitem{Costa:2011dw}
M.~S. Costa, J.~Penedones, D.~Poland and S.~Rychkov, \emph{{Spinning Conformal
  Blocks}}, \href{http://dx.doi.org/10.1007/JHEP11(2011)154}{\emph{JHEP} {\bf
  11} (2011) 154}, [\href{https://arxiv.org/abs/1109.6321}{{\tt 1109.6321}}].

\bibitem{Echeverri:2015rwa}
A.~Castedo~Echeverri, E.~Elkhidir, D.~Karateev and M.~Serone,
  \emph{{Deconstructing Conformal Blocks in 4D CFT}},
  \href{http://dx.doi.org/10.1007/JHEP08(2015)101}{\emph{JHEP} {\bf 08} (2015)
  101}, [\href{https://arxiv.org/abs/1505.03750}{{\tt 1505.03750}}].

\bibitem{Echeverri:2016dun}
A.~Castedo~Echeverri, E.~Elkhidir, D.~Karateev and M.~Serone, \emph{{Seed
  Conformal Blocks in 4D CFT}},
  \href{http://dx.doi.org/10.1007/JHEP02(2016)183}{\emph{JHEP} {\bf 02} (2016)
  183}, [\href{https://arxiv.org/abs/1601.05325}{{\tt 1601.05325}}].

\bibitem{Elkhidir:2017iov}
E.~Elkhidir and D.~Karateev, \emph{{Scalar-Fermion Analytic Bootstrap in 4D}},
  \href{https://arxiv.org/abs/1712.01554}{{\tt 1712.01554}}.

\bibitem{Bzowski:2013sza}
A.~Bzowski, P.~McFadden and K.~Skenderis, \emph{{Implications of conformal
  invariance in momentum space}},
  \href{http://dx.doi.org/10.1007/JHEP03(2014)111}{\emph{JHEP} {\bf 03} (2014)
  111}, [\href{https://arxiv.org/abs/1304.7760}{{\tt 1304.7760}}].

\bibitem{Bzowski:2015pba}
A.~Bzowski, P.~McFadden and K.~Skenderis, \emph{{Scalar 3-point functions in
  CFT: renormalisation, beta functions and anomalies}},
  \href{http://dx.doi.org/10.1007/JHEP03(2016)066}{\emph{JHEP} {\bf 03} (2016)
  066}, [\href{https://arxiv.org/abs/1510.08442}{{\tt 1510.08442}}].

\bibitem{Bzowski:2017poo}
A.~Bzowski, P.~McFadden and K.~Skenderis, \emph{{Renormalised 3-point functions
  of stress tensors and conserved currents in CFT}},
  \href{https://arxiv.org/abs/1711.09105}{{\tt 1711.09105}}.

\bibitem{Bzowski:2018fql}
A.~Bzowski, P.~McFadden and K.~Skenderis, \emph{{Renormalised CFT 3-point
  functions of scalars, currents and stress tensors}},
  \href{https://arxiv.org/abs/1805.12100}{{\tt 1805.12100}}.

\bibitem{Mack:1975je}
G.~Mack, \emph{{All Unitary Ray Representations of the Conformal Group SU(2,2)
  with Positive Energy}},
  \href{http://dx.doi.org/10.1007/BF01613145}{\emph{Commun.Math.Phys.} {\bf 55}
  (1977) 1}.

\bibitem{Mack:1976pa}
G.~Mack, \emph{{Convergence of Operator Product Expansions on the Vacuum in
  Conformal Invariant Quantum Field Theory}},
  \href{http://dx.doi.org/10.1007/BF01609130}{\emph{Commun. Math. Phys.} {\bf
  53} (1977) 155}.

\bibitem{Iliesiu:2015akf}
L.~Iliesiu, F.~Kos, D.~Poland, S.~S. Pufu, D.~Simmons-Duffin and R.~Yacoby,
  \emph{{Fermion-Scalar Conformal Blocks}},
  \href{http://dx.doi.org/10.1007/JHEP04(2016)074}{\emph{JHEP} {\bf 04} (2016)
  074}, [\href{https://arxiv.org/abs/1511.01497}{{\tt 1511.01497}}].

\end{thebibliography}\endgroup

\end{document}